\documentclass{aa}  
\usepackage{graphicx}
\usepackage[varg]{txfonts}
\usepackage{longtable}
\usepackage{natbib}
\usepackage{url}
\usepackage{color}
\usepackage{multirow,bigdelim}
\usepackage{cases}
\usepackage{rotating}
\usepackage[colorinlistoftodos]{todonotes}
\bibpunct{(}{)}{;}{a}{}{,} % to follow the A&A style
\newcommand\kms{{\rm\,km\,s^{-1}}}

\usepackage{hyperref} %[draft]
\hypersetup{
     breaklinks,
     colorlinks   = true,
     citecolor    = blue,
     urlcolor     = blue,
     linkcolor    = blue
}

\begin{document} 

\title{Disentangling the Arcturus stream}
%\subtitle{Clues to the merger history of the Milky Way}

\titlerunning{Disentangling the Arcturus stream} 
%\subtitle{}

\author{
Iryna Kushniruk
\and 
Thomas Bensby
}

\institute{
Lund Observatory, Department of Astronomy and Theoretical Physics, Box 43, SE-221\,00 Lund, Sweden\\
\email{[iryna; tbensby]@astro.lu.se}
}

\date{Received 8 February 2019 / Accepted 9 September 2019}

% \abstract{}{}{}{}{} 
% 5 {} token are mandatory
 
\abstract
% context heading (optional)
% {} leave it empty if necessary  
{The Arcturus stream is an over-density of stars in velocity space and its origin has been much debated recently without any clear conclusion. The (classical) dissolved open cluster origin is essentially refuted, instead the discussions try to distinguish between an accretion, a resonant, or an external-perturbation origin for the stream. As kinematic structures are observational footprints of ongoing and past dynamical processes in disk galaxies, resolving the nature of the Arcturus stream can provide clues to the formation history of the Milky Way and its stellar populations.
}
% aims heading (mandatory)
{
We aim to characterise the kinematical and chemical properties of the Arcturus stream in order to resolve its origin.
}
% methods heading (mandatory)
{
The space velocities, angular momenta and actions for a sample of more than 5.8 million stars, composed from {\it Gaia} DR2, are analysed with a wavelet transform method to characterise kinematic over-densities in the Galactic disk. The kinematic characteristics of each identified group is used to select possible members of the groups from the GALAH and APOGEE spectroscopic surveys to further study and constrain their chemical properties.   
}
% results heading (mandatory)
{
In the velocity and angular momentum spaces the already known Sirius, Pleiades, Hyades, Hercules, AF06, Arcturus and KFR08 streams are clearly identified. The Hercules stream appears to be a mixture of thin and thick disk stars. The Arcturus stream, as well as the AF06 and KFR08 streams, are low-velocity and low-angular momentum structures with chemical compositions similar to the thick disk. These three groups extend further from the Galactic plane compared to the Hercules stream. The detections of all the groups were spaced by approximately $20-30 \kms$ in azimuthal velocity. 
}
% conclusions heading (optional), leave it empty if necessary 
{A wide spread of chemical abundances within the Arcturus stream indicates that the group is not a dissolved open cluster. Instead the Arcturus stream, together with the AF06 and KFR08 streams, are more likely to be part of a phase-space wave, that could have been caused by an ancient merger event. This conclusion is based on that the different structures are detected in steps of $20-30\,\kms$ in azimuthal velocity, that the kinematic and chemical features are different from what is expected for bar-originated structures, and that the lower-velocity streams extend further from the disk than bar-originated structures. 
}

\keywords{
stars: kinematics and dynamics -- galaxy: formation -- galaxy: evolution -- galaxy: kinematics and dynamics
}

\maketitle
%

%===================================================================
%===================================================================
\section{Introduction}
%===================================================================
%===================================================================
How large spiral galaxies form and evolve into the complicated structures that are observed today is an active area of research, and presents many challenges, both theoretically and observationally. As the Milky Way is the only galaxy where stars and structures can be studied in great detail, it may serve as a benchmark galaxy when constraining models of galaxy formation. It is therefore utterly important to understand what the Milky Way looks like, and where the observed stellar populations and structures come from. Currently the Milky Way contains a plethora of structures, both physical and kinematic, whose nature and origins are unclear.

%To better understand the mechanisms that played a key role in the formation of the Galaxy this study focus on kinematic structures, groups of stars that share similar velocities. 
Many studies have shown that the velocity distribution of stars in the Milky Way disk is clumpy \citep[e.g.][]{_dehnen98, _skuljan99, _famaey05, _antoja08, _antoja12, _kushniruk17, _ramos18}. The kinematic and chemical properties of such structures can be used to constrain the properties and the formation history of the Milky Way. For example, the Hercules stream has been widely used to probe the pattern speed and the length of the Galactic bar \citep[e.g.][]{_dehnen00, _minchev07, _antoja14, _wegg15, _perez17}. Kinematic structures can be used to study the spiral structure of the Milky Way \citep[e.g.][]{_chakrabarty07, _sellwood18, _quillen18}. The studies of kinematic streams especially in the Galactic halo can tell us about the merger history of the Milky Way \citep[e.g.][]{_navarro04, _helmi06, _helmi17, _koppelman18, _helmi18}. The analysis of the {\it Gaia} DR2 \citep{_brown18, _katz18k} revealed that the kinematic over-densities are a part of a much more complicated structure that is seen as arches and ridges across velocity space and as clumps in action space \citep{_trick18}. This structure is possibly caused by spiral arms or is a result of a phase-mixing due to a past merger event \citep[e.g.][]{_antoja18, _ramos18, _quillen18}. As these studies have shown, learning more about the nature of kinematic structures can improve our understanding of the Milky Way's evolution. In this paper we will investigate the properties and origin of the Arcturus stream. 

A set of  about 50 stars, including the star Arcturus ($\alpha$ Bootis), was discovered by \citet{_eggen71} to have very similar V space velocity component of $V\simeq-100 \kms$. \citet{_eggen71} proposed that this over-density in velocity space is composed of stars that escaped from an open cluster, and was therefore named the Arcturus {\it moving group}. Nowadays the hypothesis of the Arcturus over-density being a moving group is almost refuted as there is no chemical homogeneity within the group \citep[e.g.][]{_williams09, _ramya12, _bensby14}, that should be if the stars originated from the same open cluster \citep[e.g.][]{_desilva07, _bovy16}. We have therefore chosen to adopt the `stream' nomenclature when referring to this Arcturus over-density of stars in velocity space. 

Two other possible origins of the Arcturus stream are now favoured and are widely discussed. The first is an accretion event scenario, where a small satellite galaxy merged with the Milky Way and caused this dynamical structure \citep[e.g.][]{_navarro04, _helmi06}. The second possibility is that it has originated due to resonances with the Galactic bar or spiral arms that cause kinematic over-densities \citep[e.g.][]{_gardner10, _monari13}. %Assuming both scenario it is possible to numerically simulate a structure with the properties similar to the Arcturus stream. 
The chemical properties of the stream does not show any chemical peculiarities that should be in the case of the extra-galactic origin \citep[e.g.][]{_ramya12, _bensby14}. At the same time the low angular momentum and the low velocity of the stream indicate that it could be another tidal debris sub-structure in the Galactic halo \citep[e.g.][]{_arifyanto06, _klement08, _zhao14}. Despite numerous approaches to study the origin of the Arcturus stream (for example, numerical simulations, kinematic analysis or studies of elemental abundances), there is no consensus on its origin.

The aim of this paper is to characterise the nature of the Arcturus stream and constrain its origin. We start  detecting and characterising the velocities of the Arcturus stream using a large stellar sample constructed from the {\it Gaia} DR2 catalogue (see Sect.~\ref{_sec_sample}). Then we search for over-densities in the velocity, angular momentum and action spaces to obtain the kinematic characteristics of the stream (see Sects.~\ref{_sec_method}, \ref{_sec_results} and \ref{_sec_lowvel}). After that we investigate the chemical characteristics of the group using the data from the GALAH \citep{_buder18} and APOGEE \citep{_holzman18} spectroscopic surveys (see Sect.~\ref{_sec_chem}). We conclude by discussing possible origins for the Arcturus stream based on the kinematic and spectroscopic findings (see Sects.~\ref{_sec_arcturus} and \ref{_sec_the_end}).

%===================================================================
%===================================================================
\section{Stellar sample}\label{_sec_sample}
%===================================================================
%===================================================================

To search for the Arcturus stream a wavelet analysis was applied for a stellar sample defined by velocities, angular momentum and action components. To calculate these parameters positions on the sky, proper motions, parallaxes, radial velocities, and the corresponding uncertainties for these properties are needed.

The size of the stellar sample and the quality of the astrometric data play a key role when hunting for kinematic structures. The larger the sample of stars with available high-precision astrometric data, in the greater detail it is possible to study kinematical structures of the Galaxy. The currently best data source is the {\it Gaia} satellite, which is an on-going full-sky mission that aims to provide high-precision astrometric parameters for more than a billion targets over the whole sky. The most recent data release, {\it Gaia} DR2 \citep{_brown18} contains astrometric data for almost 1.7 billion targets, and for a small subsample of about 7 million targets, also radial velocities.  

A stellar sample of 5\,844\,487 stars was constructed from the {\it Gaia} DR2 catalogue in the following way:

\begin{itemize}
\item 7\,173\,615 stars were obtained from \citet{_mcmillan18}, who estimated distances for {\it Gaia} DR2 stars with measured radial velocities. 
%(X = ÃÂ¢ÃÂÃÂ7.8kpc,Y = 0.kpc (second model in table 2 of McMillan  Binney 2010) and Z = 0.014 kpc (Binney, Gerhard  Spergel 1997))

\item Stars with bad fits of {\it Gaia} DR2 astrometric parameters were filtered out to avoid possible systematic errors in the stellar sample. Following the procedure suggested in \citet{_lindegren18n} a re-normalised unit weight error ($RUWE$) is used to estimate goodness of astrometric fits. Selecting those targets with  $RUWE < 1.4$ leaves us with 6\,692\,285 targets. Photometric filtering that cleans the sample from star with bad astrometric solutions (see Eq.~2 in \citet{_arenou18}) was also applied. This cut leaves us with 6\,683\,408 stars.

\item Space velocities $U, V, W$\footnote{$U$ points towards Galactic centre, $V$ velocity defines the direction of the Galactic rotation, $W$ points at the North Galactic Pole.} together with angular momenta and actions that will used later in this paper were computed using {\it galpy}\footnote{Available at \url{http://github.com/jobovy/galpy}} package \citep{_bovy15}. For action estimates we used MWPotential2014 axisymmetric gravitational potential model pre-defined in {\it galpy}. Velocity uncertainties $\sigma_U$, $\sigma_V, \sigma_W$ were computed following equations from \citet{_johnson87}. The velocities are given relative to the Local Standard of Rest: ($U_{\odot}$, $V_{\odot}$, $W_{\odot}$) = (11.1, 12.24, 7.25) $\kms$ \citep{_schonrich10}. 

Taking into account the results from, for example, \citet{_zhao14}, that the typical size of kinematical structures is around 20\,$\kms$. Therefore we need to cut stars with $\sigma_U, \sigma_V > 20 \kms$, because such large velocity uncertainties will influence the precision of the results, that is the position in velocity space of the structures. This leaves us with 6\,002\,514 stars.

\item Next, the sample was constrained to stars that are located within a distance of 5\,kpc from the Sun. This filters out stars that are located in the outskirts or very inner parts of the Galaxy, and thus, cannot be a part of any of the local kinematic structures. The limit of 5\,kpc was chosen to avoid regions in direct contact with for example the Galactic bar, whose half-length is about 3\,kpc \citep[e.g.][]{_dehnen00, _minchev10,_monari17}.
% when assuming the fast-rotating bar. 
According to \citep{_bailerjones15} distance estimates should not be dominated by using pre-{\it Gaia} information or so-called priors if fractional parallax uncertainty does not exceed 20\%. Typical parallax uncertainty for bright sources in {\it Gaia} DR2 is about 0.4 $\mu$as \citet{_lindegren18n}. Converting 5 kpc cut into $\mu$as and calculating fractional parallax uncertainty we obtain 20\% meaning that distance estimates in the sample should not be effected by priors. After this cut there are 5\,844\,487 stars left that will go into our analysis.

\end{itemize}

%--------------------------------------------------------------------
\begin{figure*}
   \centering
   \resizebox{0.75\hsize}{!}{
   \includegraphics[viewport = 70  20 770 600,clip]{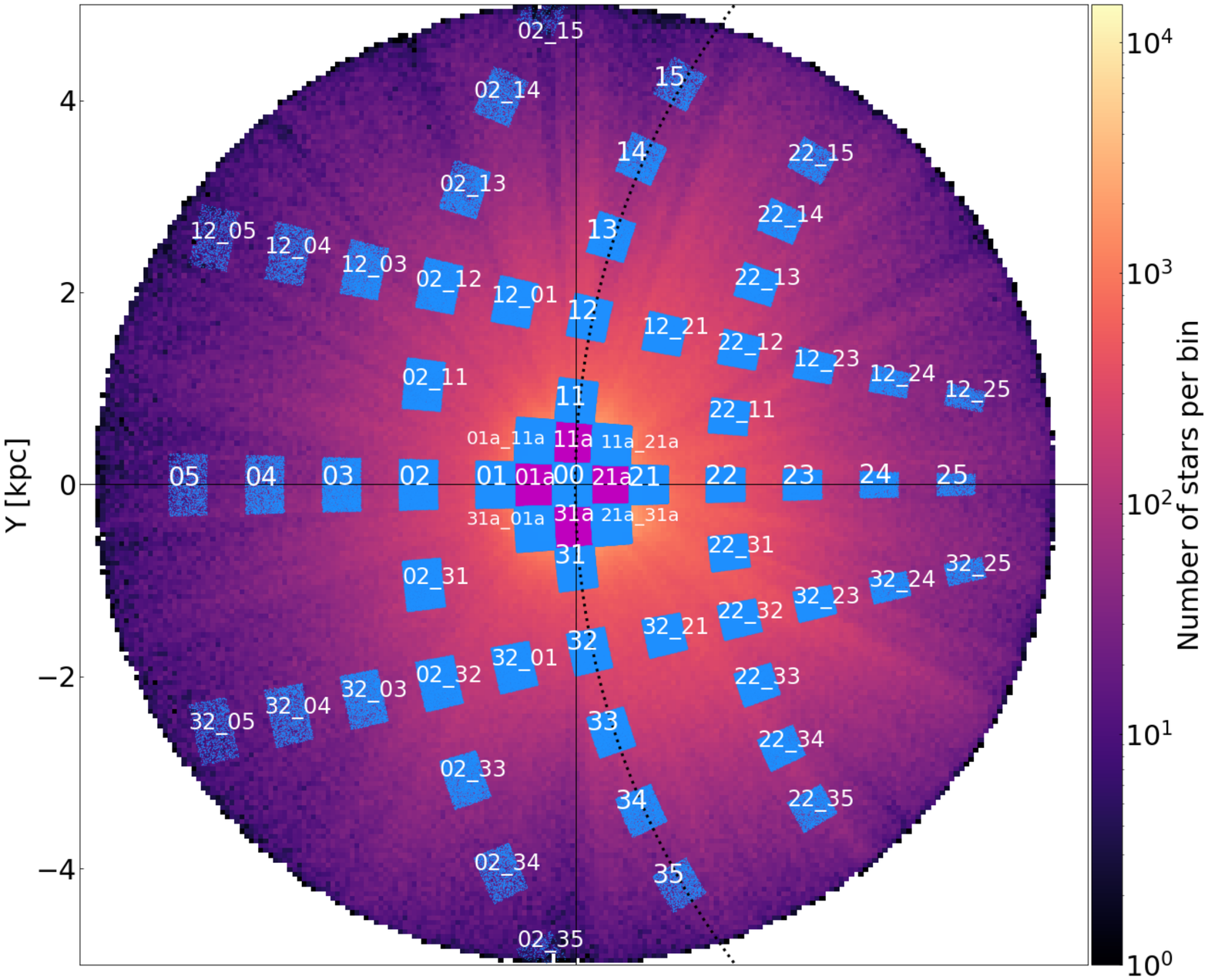}
   }
   \resizebox{0.75\hsize}{!}{
   \includegraphics[viewport = 0  150 835 430,clip]{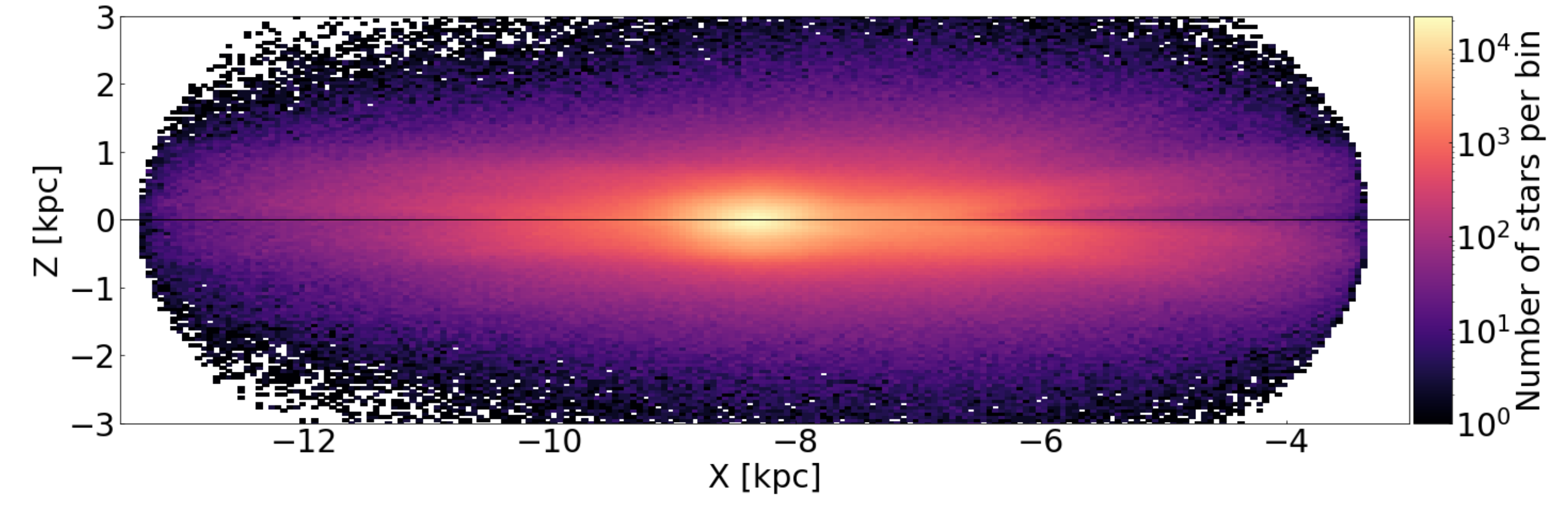}
   }
   \caption{
   Distribution of 5\,844\,487 stars with $\sigma_U$ and $\sigma_V \,<\,20\,\kms$ and $R\le 5$ kpc in $X$ and $Y$ (top), and $X$ and $Z$ (bottom) Cartesian Galactic coordinates. Blue and magenta boxes show 66 small regions investigated in this work and their names are whiten on top in white color. The dashed line in the top plot shows the Solar circle. The black line in the bottom plot shows $Z=0$ kpc. The bin size is 0.05\,kpc for both plots.
   \label{_fig_regions}
   }
\end{figure*}
%--------------------------------------------------------------------

Since kinematic structures are local phenomena \citep[e.g.][]{_antoja12, _ramos18, _trick18} and the stellar sample covers a wide range in $X$ and $Y$, it was divided it into 66 smaller volumes that were investigated separately. Each box is 0.4\,kpc in radial coordinate and $3^{\circ}$ in azimuthal angle\footnote{R is the radial coordinate pointing towards the Galactic anti-centre, and $\phi$ is the azimuthal angle following the direction opposite to the Galactic rotation.}. The top plot in Fig.~\ref{_fig_regions} shows the distribution of the 5\,844\,487 stars in the Galactic Cartesian $X-Y$ plane and how it is divided into small volumes. The name of each region, the number of stars, the median distance from the Sun, and median distance uncertainty are given in Table~\ref{_tab_info} for each of the 65 volumes. The bottom plot in  Fig.~\ref{_fig_regions} shows the sample in Cartesian $X$ and $Z$ coordinates, where $Z$ is a vertical component of Galactocentric coordinate system (points towards North Galactic Pole).

%===================================================================
%===================================================================
\section{Method}\label{_sec_method}
%===================================================================
%===================================================================

%--------------------------------------------------------------------
\begin{figure*}
   \centering
   \resizebox{\hsize}{!}{
   \includegraphics[viewport = 0  0  1550 980,clip]{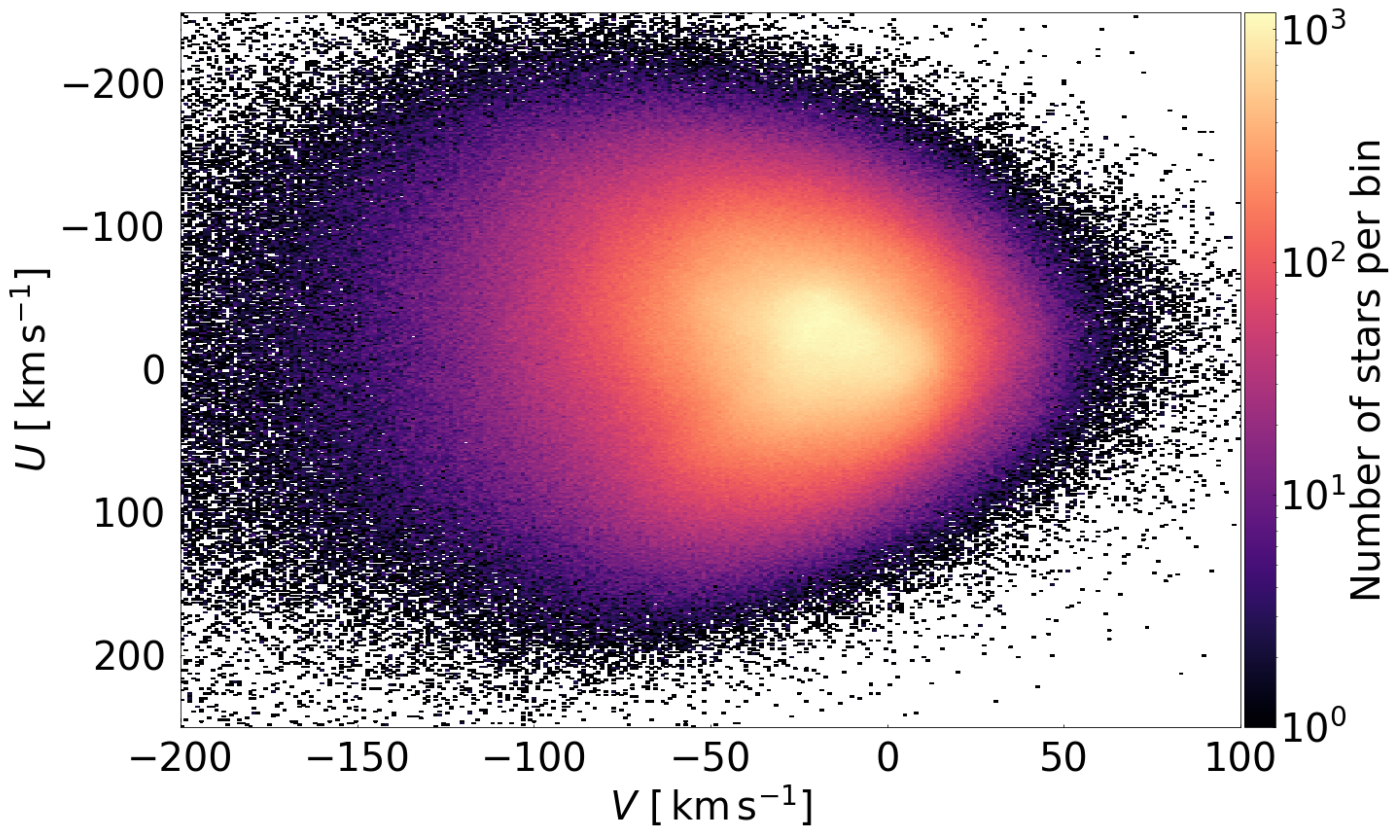}
   \includegraphics[viewport = 0  0  1550 980,clip]{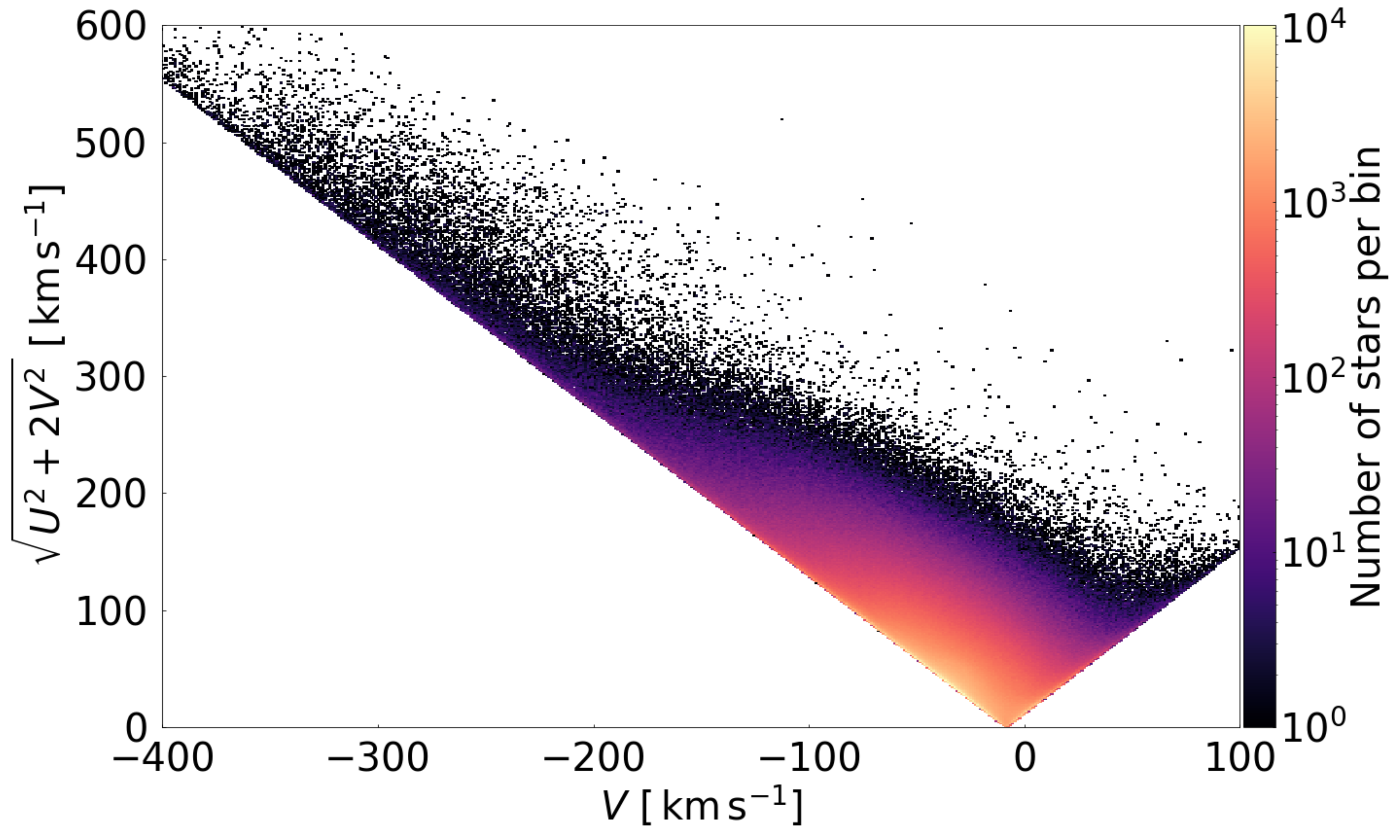}
   }
   \resizebox{\hsize}{!}{
   \includegraphics[viewport = 0  0  1550 1000,clip]{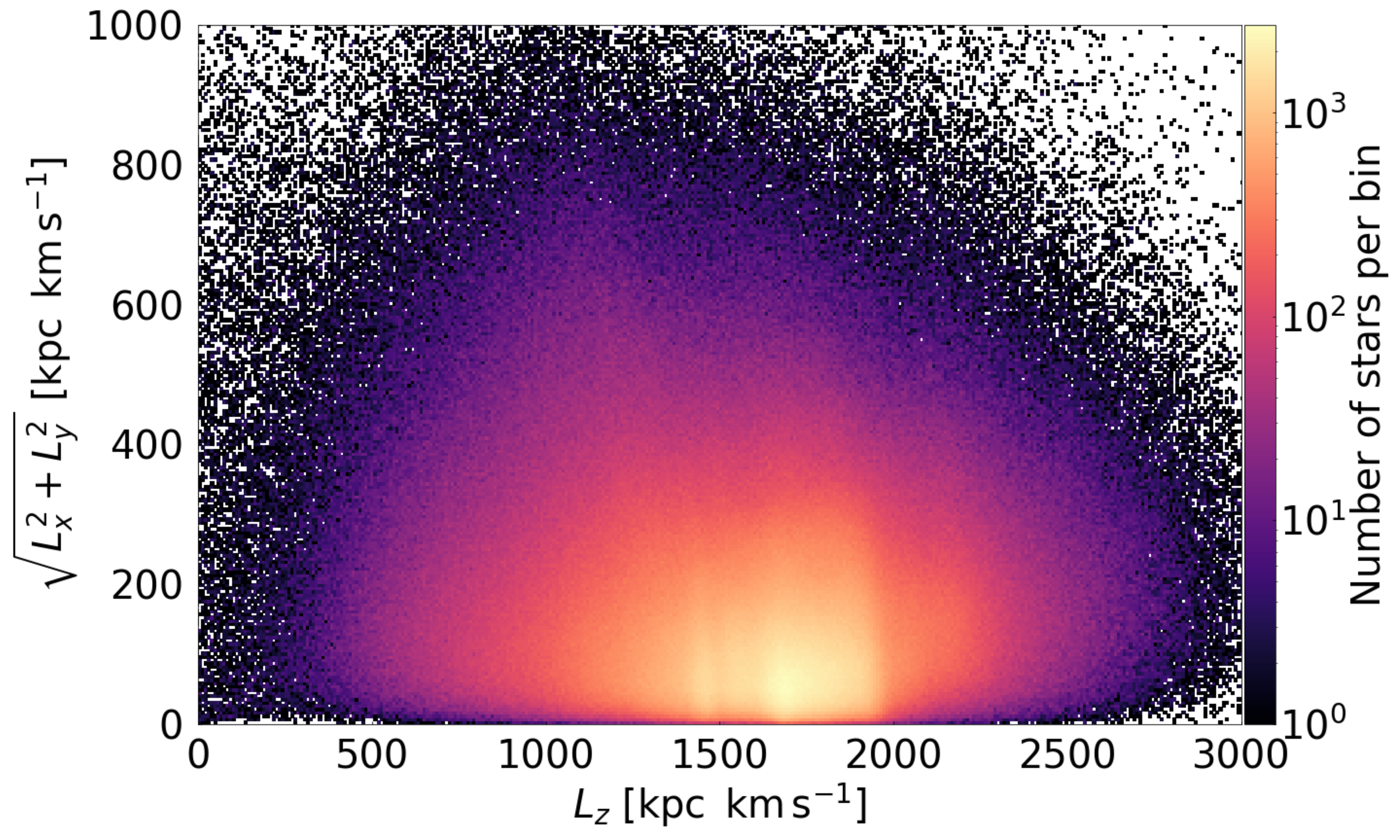}
   \includegraphics[viewport = 0  0  1550 1000,clip]{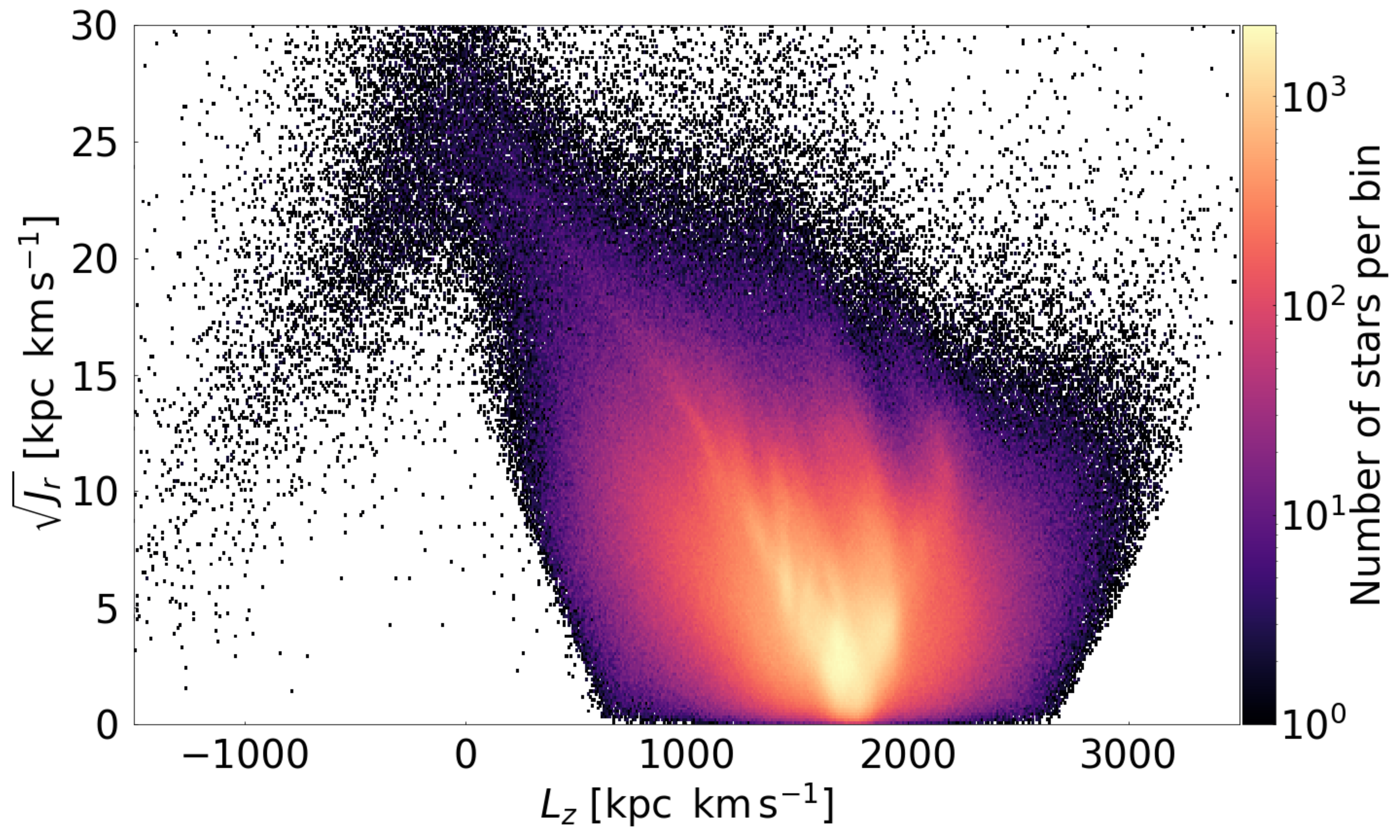}
   }
   \caption{
 Density map of the selected sample of 5\,844\,487 stars in $U-V$ space (top left), in $V-\sqrt{U^2+2V^2}$ space (top right), in $L_z-\sqrt{L_x^2+L_y^2}$ space (bottom left) and in $L_z-\sqrt{J_{R}}$ space (bottom right). The bin sizes are 1 $\kms$, 1 $\kms$, 10 kpc $\kms$ and $10 \times 0.1$ kpc $\kms$, respectively. 
   \label{_fig_sample}
   }
\end{figure*}
%--------------------------------------------------------------------

Our current knowledge about the Arcturus stream is based on observations within a small region of about 500\,pc around the Sun. Its origin is unknown, mainly due to that there are only rough estimates of its kinematic characteristics, and as it is chemical properties are not well studied. Our strategy is therefore to search for streams in four different planes defined by combinations of velocity, angular momentum and action components respectively: the $U-V$ plane, the $V - \sqrt{U^2+2V^2}$ plane, the $L_z - \sqrt{L_x^2+L_y^2}$ plane, and the $L_z - \sqrt{J_r}$ plane. %The choice of the investigated planes is based on previous works where the Arcturus stream was mentioned. 
%In this study we analyse the stars in 65 small regions in all four spaces. 
This will allow us to characterise structures in terms of velocities, angular momenta and actions, and obtain stronger criteria on how to select stars-members of kinematic structures.

%\subsection{Calculation of space velocities, orbits, and actions}

\subsection{Investigated planes}

The distributions of stars in all four planes are shown Fig.~\ref{_fig_sample}. A majority of stars have negative $V$ velocities between $V\simeq0$ and $-200 \kms$ and angular momentum $L_z$ between 0 and 2500\,kpc\,$\kms$. The disk stars are located at $L_z\simeq1800$ kpc $\kms$, the halo stars are expected at $L_z\simeq0$ kpc $\kms$.

\subsubsection{The $U-V$ plane}\label{_sec_sec_uv}
The $U-V$ plane is widely used to search for kinematic structures \citep[e.g.][]{_dehnen98, _antoja08, _antoja12, _kushniruk17, _antoja18, _ramos18, _katz18}. It allows to trace kinematic over-densities of different origin without making any assumptions on orbital parameters of stars as well as no assumptions on the Galactic potential. The only limitation of this method is that stellar volumes must be relatively small (around $0.1-0.5$\,pc in $X$ and $Y$, see \citealt{_trick18}), since kinematic structures in the $U-V$ plane are local. On the other hand, this limitation is an advantage, since it allows us to follow how the structures move in physical space \citep{_ramos18}. The Arcturus stream is expected to be one of the arches in the $U-V$ plane localised around $V\simeq-100 \kms$ in the nearby sample. The stream is likely to cover a wide range of $U$ velocities \citep{_williams09}. 

\subsubsection{The $V-\sqrt{U^2+2V^2}$ plane}\label{_sec_sec_uuvv}
Examining the distribution of stars in the $V - \sqrt{U^2+2V^2}$ plane was proposed by \citet{_arifyanto06}. $V$ is proportional to $L_z$, a vertical component of the angular momentum and is an integral of motion in axisymmetric potentials. $\sqrt{U^2+2V^2}$ is a measure of eccentricity in the Dekker's approximation \citep{_dekker76}. This means that we search for structures that share similar orbital eccentricity. This approach is applicable only for planar orbits in axisymmetric potentials. The method shows reliable results for nearby stars with eccentricities up to 0.5 \citep{_arifyanto06}. The $V-\sqrt{U^2+2V^2}$ plane was used by \citet{_klement08} and \citet{_zhao14} to search for kinematic structures and allowed to reveal several structures including the Arcturus stream. We expect to detect the Arcturus steam at the velocities around $V~-100 \kms$ in the solar neighbourhood \citep[e.g.][]{_williams09}.

\subsubsection{The $L_z-\sqrt{L_x^2+L_y^2}$ plane}\label{_sec_sec_lplz}
Another approach to search for kinematic groups was proposed by \citet{_helmi99} who suggested to examine distribution of stars in the plane characterised by the $L_z$ and $\sqrt{L_x^2+L_y^2}$ integrals of motion, where $L_x$, $L_y$, and $L_z$ are angular momentum components in $X$, $Y$ and $Z$ directions. The method is used to search for phase-mixed stars on similar orbits. The disadvantage of this method is that $\sqrt{L_x^2+L_y^2}$ is not fully conserved in axisymmetric potentials, but still allows to reveal dynamical structures \citep[e.g.][]{_helmi99, _klement08, _zhao14}. The Arcturus stream is expected at $L_z$ in the range between 700 and 1100 $\kms$ kpc \citep[e.g.][]{_navarro04}.

\subsubsection{The $L_z-\sqrt{J_r}$ plane}\label{_sec_sec_act}
The most general method to search for kinematic structures is to investigate action space. Actions are conserved quantities that characterise stellar orbits. In this work we will use radial and azimuthal actions $J_r$ and $L_z$ that are a measure of orbital eccentricity and orbital angular momentum. As suggested in \citet{_trick18} taking the square root of radial action will make the final plots more clear. The action space was investigated by, for instance, \citet{_sellwood10} and \citet{_trick18} and is rich on kinematic over-densities as. We expect to detect the Arcturus stream at $L_z$ in the range between 700 and 1100\,kpc\,$\kms$ \citep[e.g.][]{_navarro04}.

%-------------------------------------------------
\subsection{Wavelet transform}\label{_sec_sec_waw_tr}
%-------------------------------------------------

To search for kinematic structures the methodology described in \citet{_kushniruk17} was used with some additions. To detect over-densities a wavelet transform was applied to the stellar sample in the $U-V$, $V - \sqrt{U^2+2V^2}$, $L_z - \sqrt{L_x^2+L_y^2}$, and $L_z - \sqrt{J_r}$ planes. Then the noise from the wavelet maps was filtered and Monte Carlo simulations were used to verify whether the detected structures are real or not.  

The data was analysed by the wavelet transform with the `{\it a trous}' algorithm \citep{_starck98} applied to the stars in all 65 regions in the four different planes separately. The input data is a binned stellar density map in the velocity, angular momentum and action planes. The bin sizes $\Delta$ were set to 1\,$\kms$ for the $U-V$, and the $V - \sqrt{U^2+2V^2}$ planes, to 2\,$\kms$ kpc for $the \sqrt{L_x^2+L_y^2} - L_z$ plane, and to $0.1\times 10$\,kpc\,$\kms$ for the $L_z - \sqrt{J_r}$ plane. Due to the limitations of the usage of the $V - \sqrt{U^2+2V^2}$ plane, as discussed in Sect.~\ref{_sec_sec_uuvv}, the stars that have orbits with eccentricities $e>0.5$ were cut out. The output data is a set of wavelet coefficients at different scales that contain information about the presence of substructures. A higher wavelet coefficient means a higher probability that the structure is real. The scale $J$ is proportional to the size of the detectable structures $s$. Scales $J=1$, 2, 3 and 4 were investigated for all maps. The relation between scale and bin size $s_J=2^J\Delta$ characterises typical sizes of detectable structures. Then the wavelet coefficient maps were filtered from Poisson noise. The wavelet transform part as well as noise filtering from the output wavelet maps were performed in MR software\footnote{Available at \url{http://www.multiresolutions.com/mr/}} developed by CEA (Saclay, France) and Nice Observatory. More details on the algorithm itself can be found in \citet{starck_astronomical_2002}, and more details on the methods used to search for over-densities and structures can be found in \citet{_kushniruk17}.

%--------------------------------------------------------------------
\begin{figure*}
   \centering
   \resizebox{\hsize}{!}{
   \includegraphics[viewport = 0  0 430 280,clip]{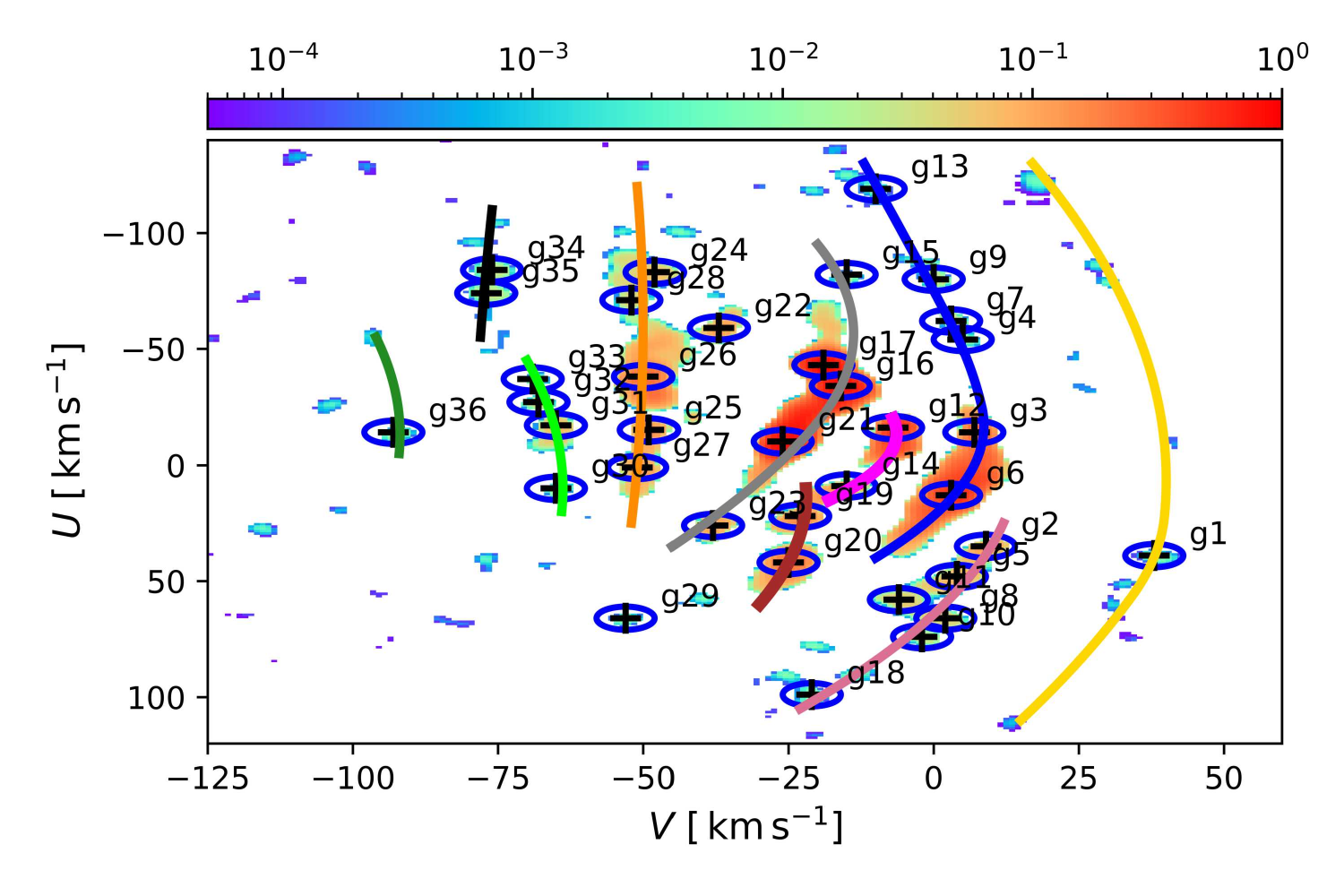}
   \includegraphics[viewport = 0  0 430 280,clip]{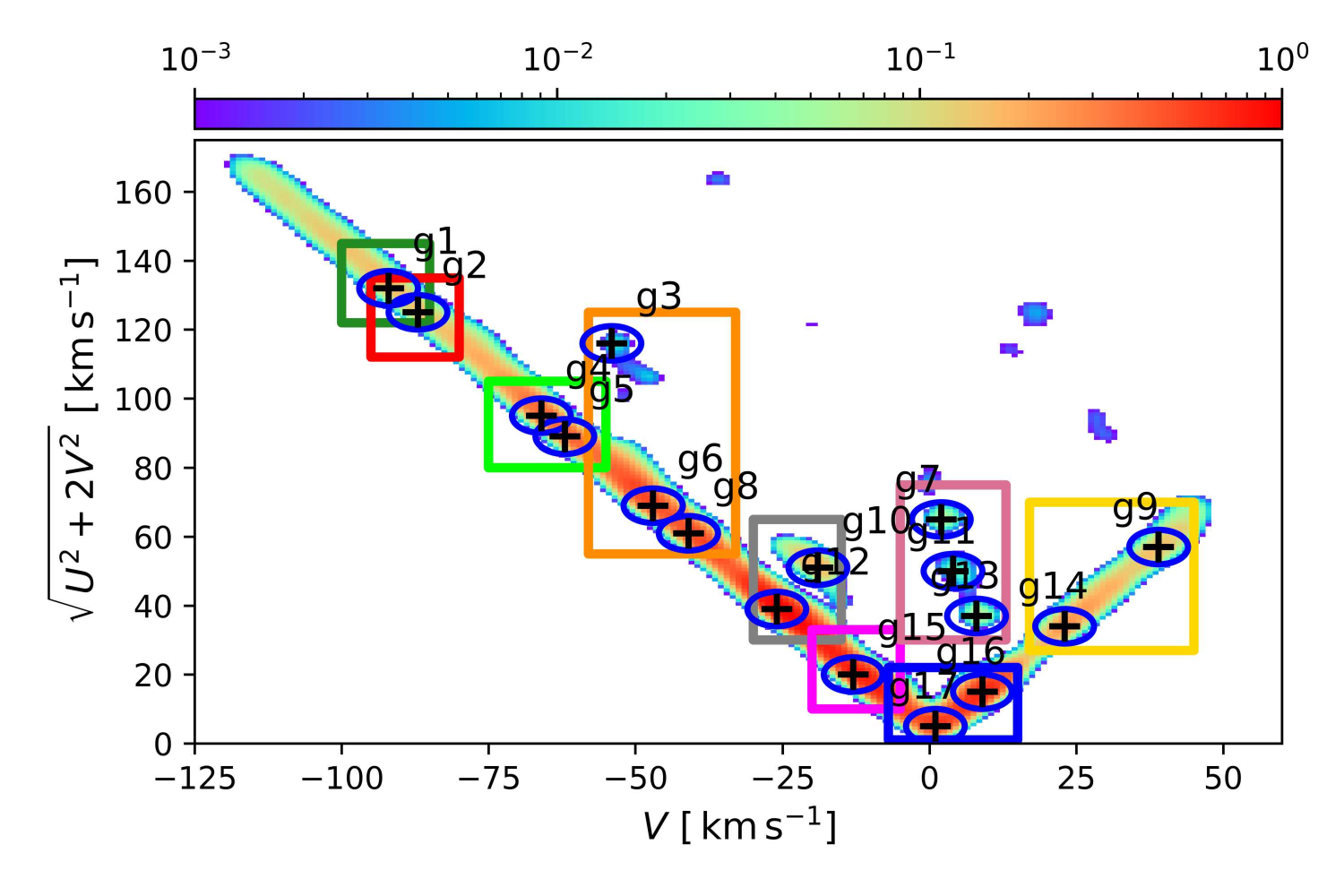}
   }
   \resizebox{\hsize}{!}{
   \includegraphics[viewport = 0  0 430 280,clip]{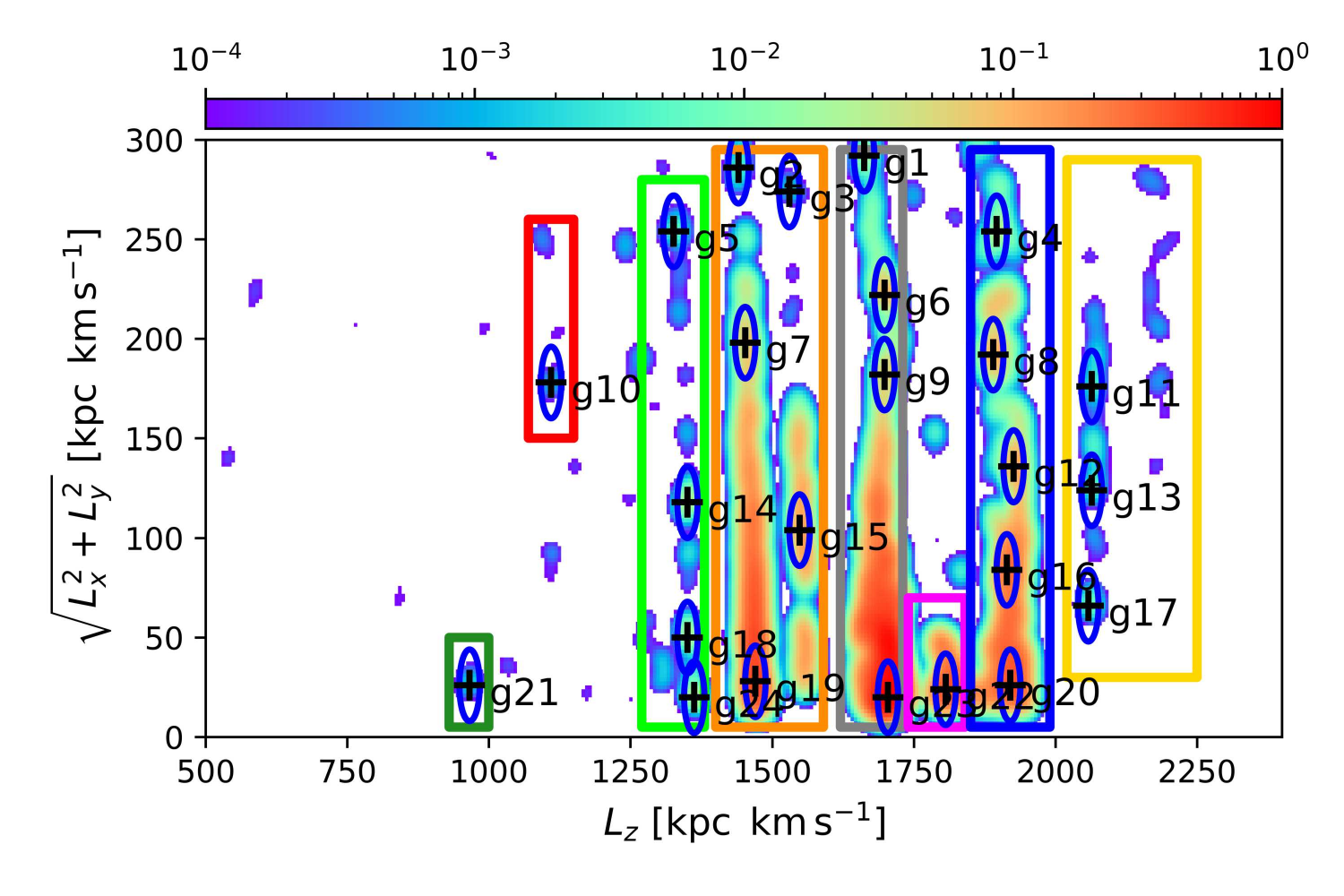}
   \includegraphics[viewport = 0  0 430 280,clip]{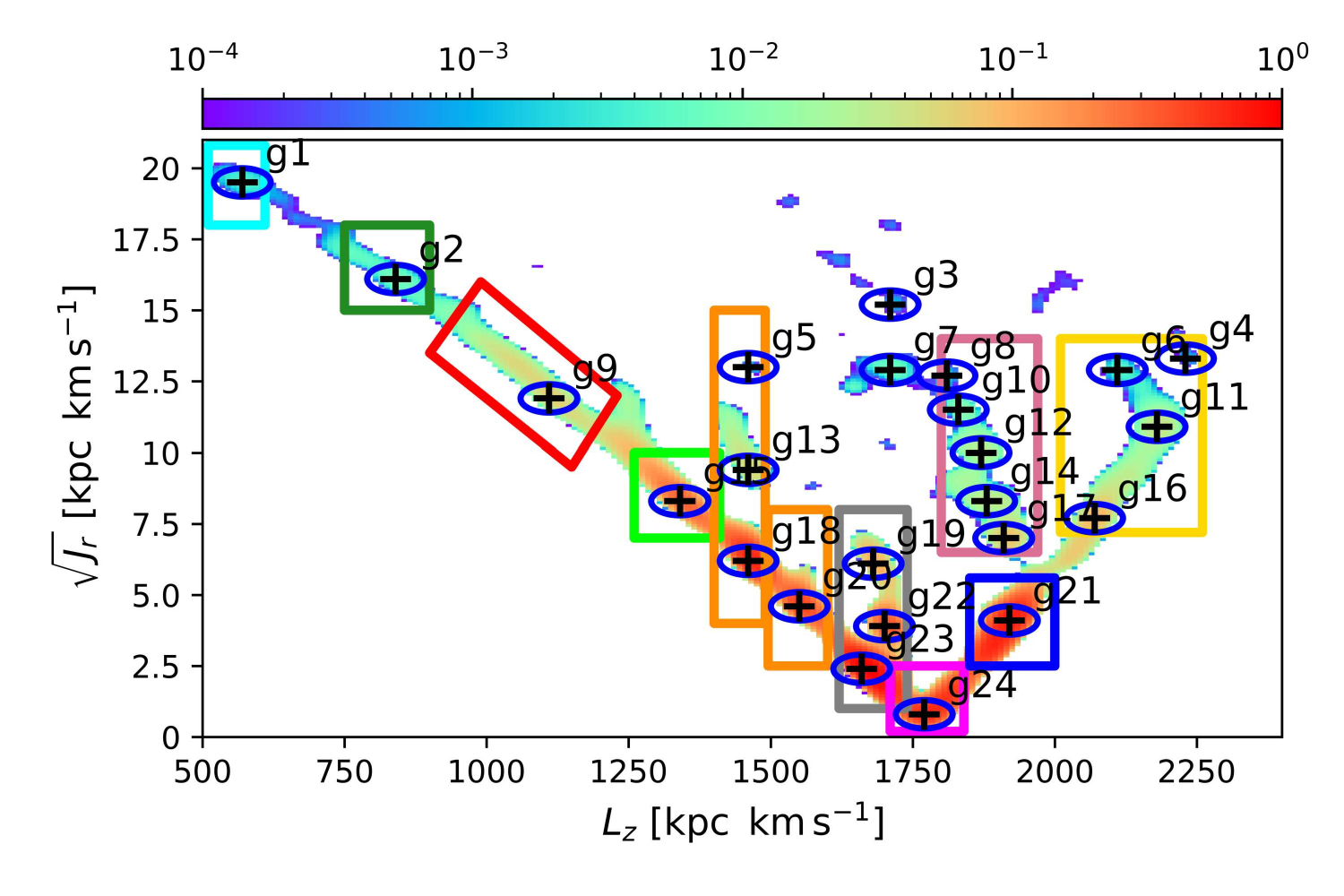}
   }
   \resizebox{\hsize}{!}{
   \includegraphics[viewport = 0  40 430 150,clip]{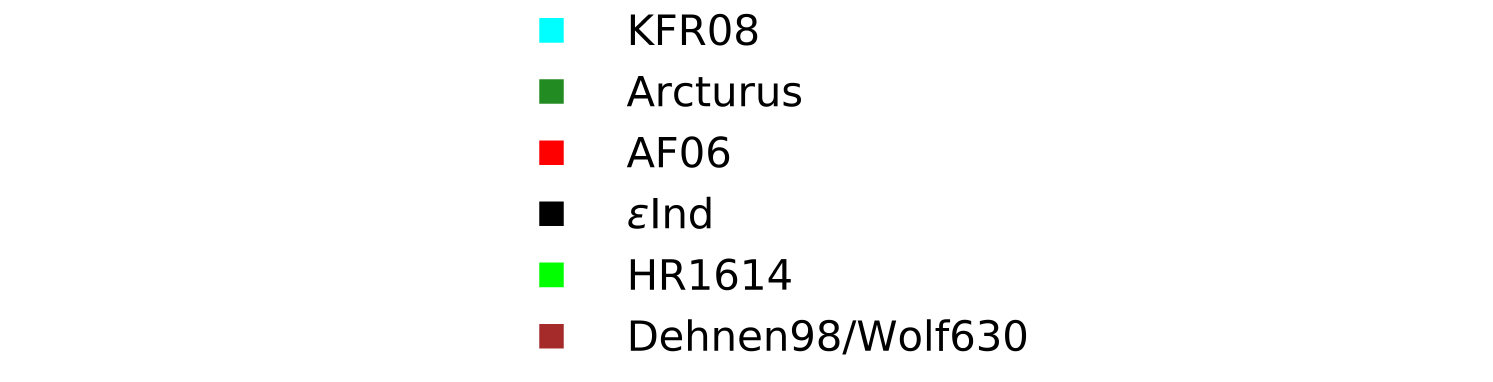}
   \includegraphics[viewport = 0  40 430 150,clip]{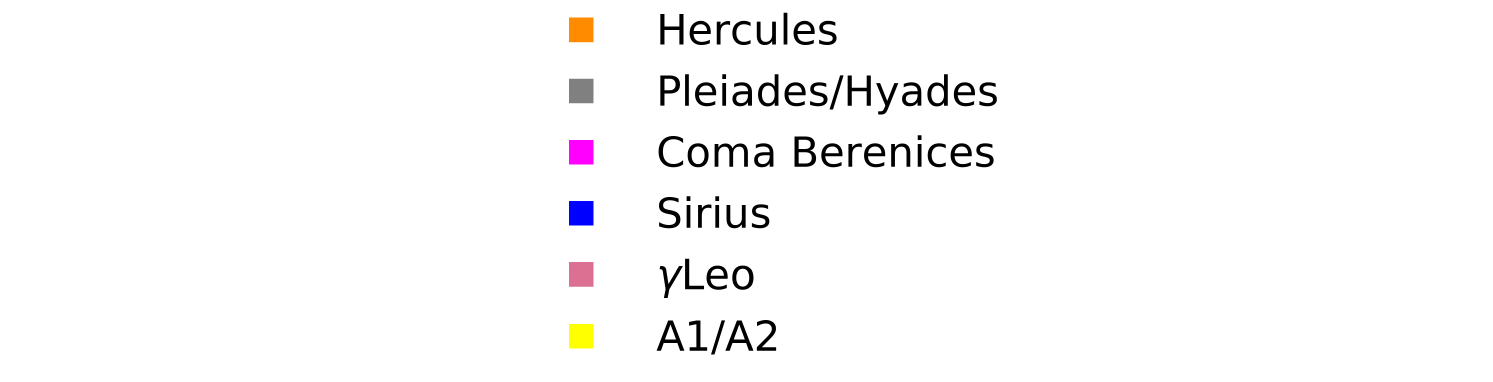}
   }
   \caption{Wavelet coefficient maps of central region 00 retrieved in $U-V$ (top left), $V-\sqrt{U^2+2V^2}$ (top right), $L_z-\sqrt{L_x^2+L_y^2}$ (bottom left) and $L_z-\sqrt{J_{R}}$ (bottom right) space for scale $J=2$. Colour bars show normalised wavelet coefficients. Kinematic structures are shown as blue circles with radius 5 $\kms$ or 5 kpc $\kms$ and their centres are shown with black crosses. Lines and boxes of different colours correspond to the group names as listed in the legend. Note that numbers assigned to the groups do not match between the planes.
   \label{_fig_00_2}
   }
\end{figure*}
%--------------------------------------------------------------------

%--------------------------------------------------------------------
\begin{figure*}
   \centering
   \resizebox{\hsize}{!}{
   \includegraphics[viewport = 0  0 430 280,clip]{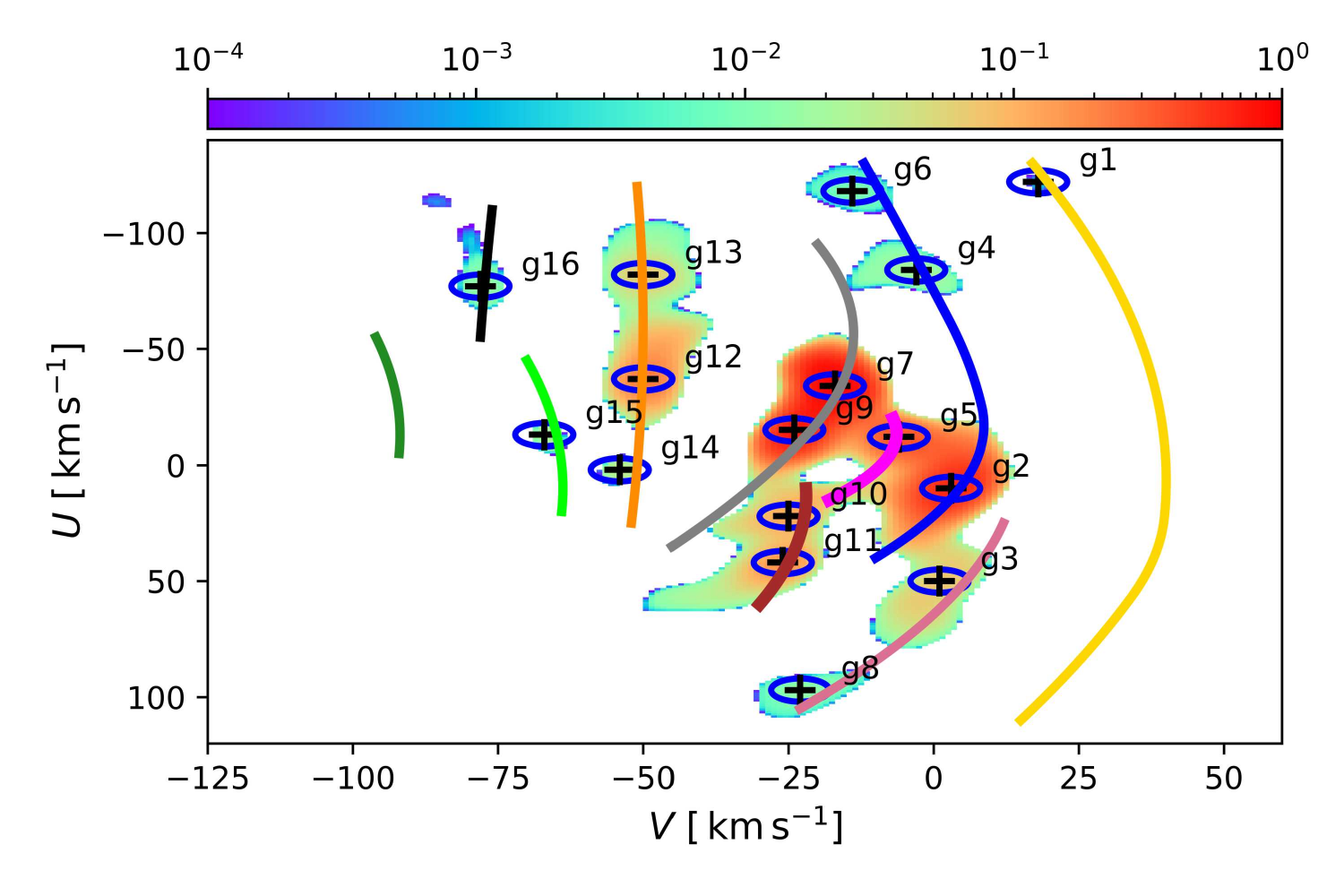}
   \includegraphics[viewport = 0  0 430 280,clip]{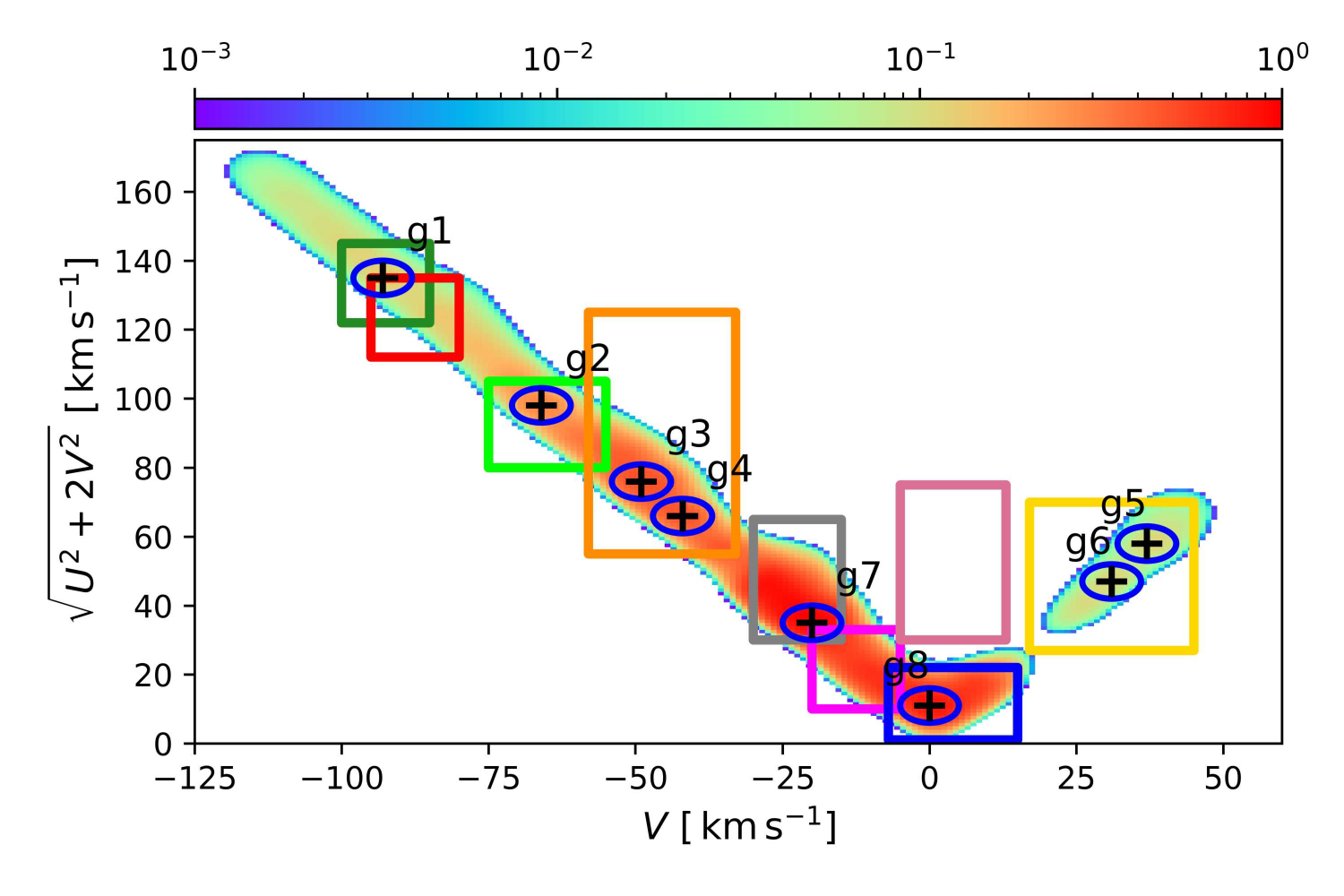}
   }
   \resizebox{\hsize}{!}{
   \includegraphics[viewport = 0  0 430 280,clip]{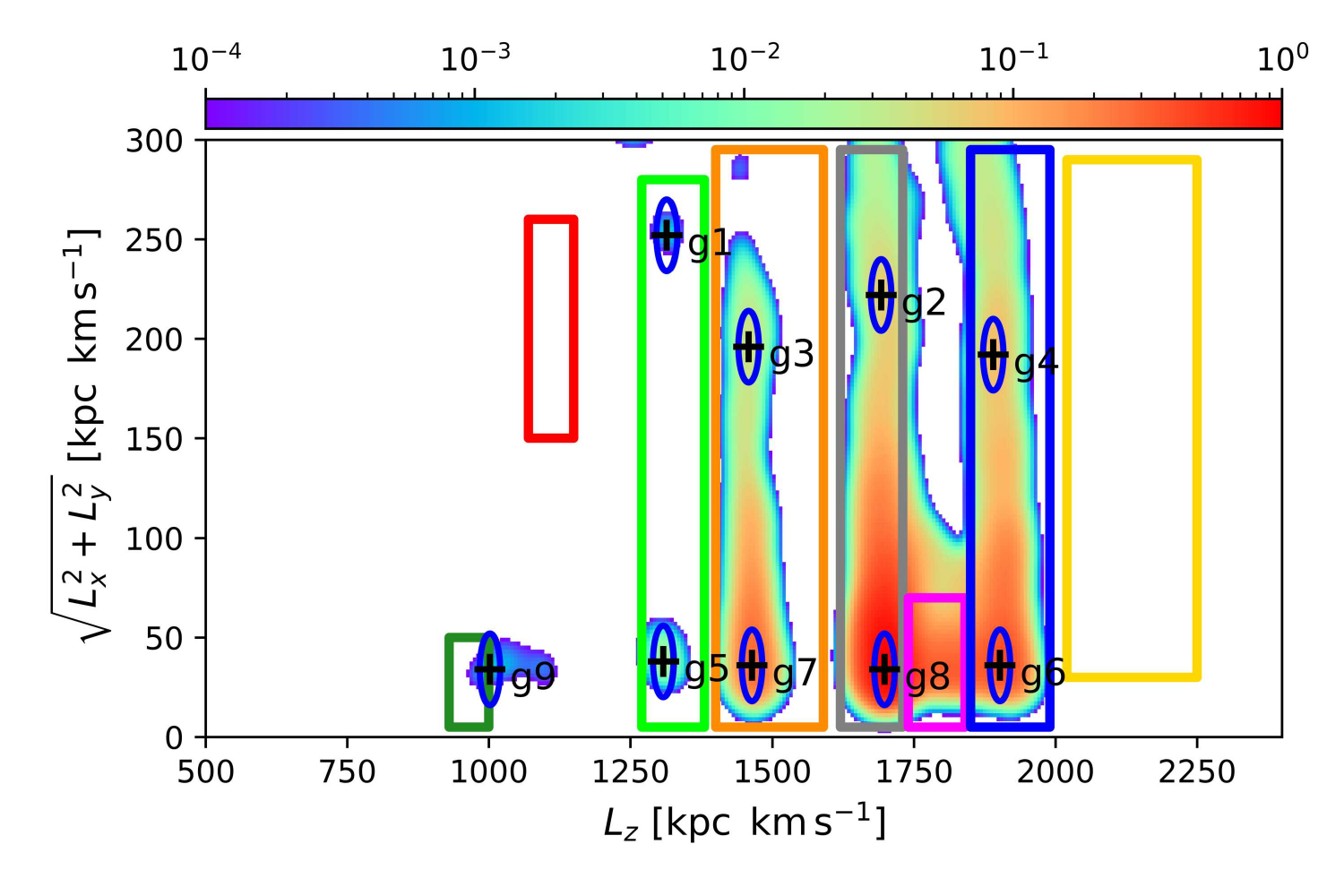}
   \includegraphics[viewport = 0  0 430 280,clip]{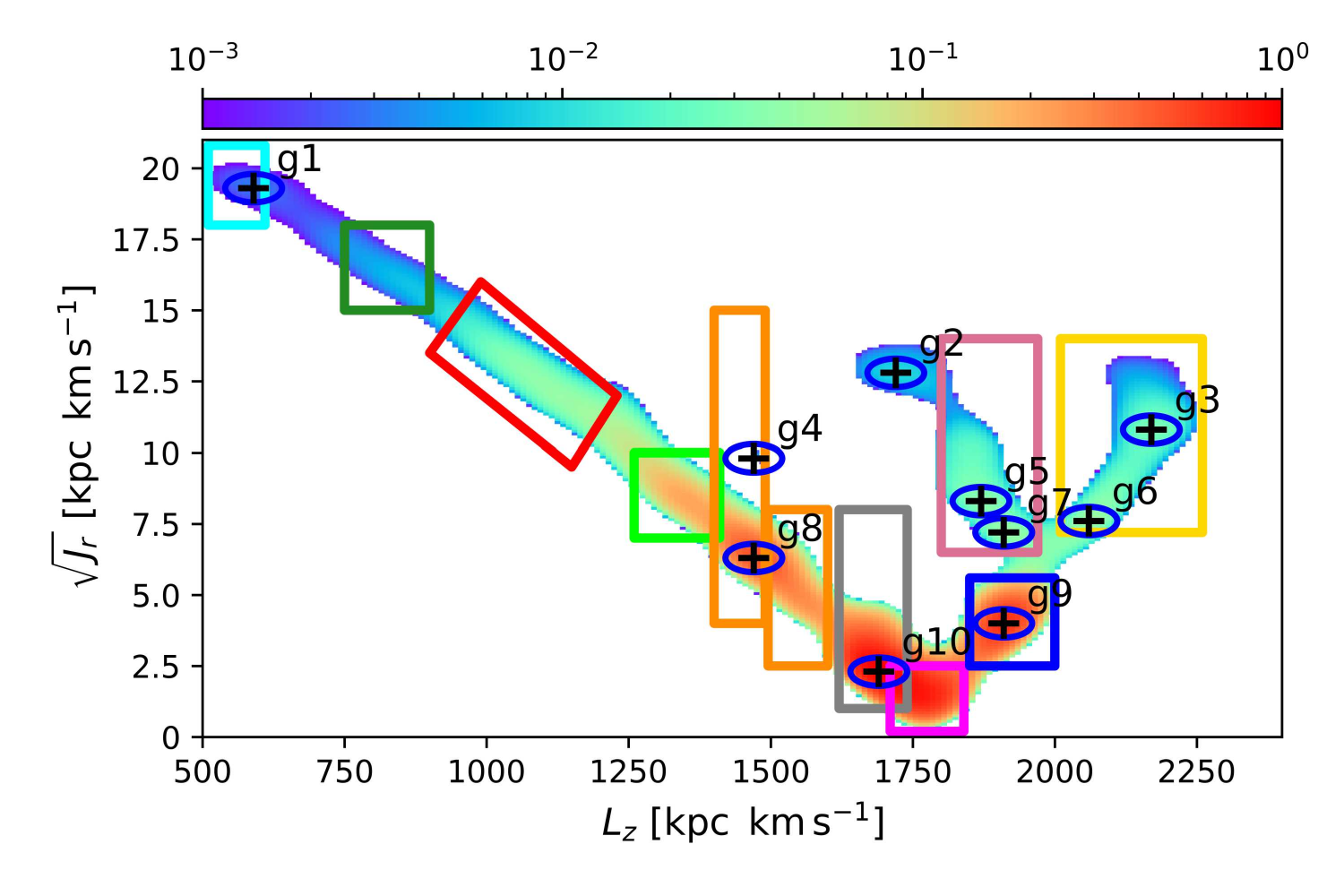}
   }
   \caption{Same as Figure \ref{_fig_00_2} but for scale $J=3$. Lines and boxes correspond to the structures detected at scale $J=2$. Note that numbers assigned to the groups do not match between the planes.
   \label{_fig_00_3}
   }
\end{figure*}
%--------------------------------------------------------------------

%--------------------------------------------------------------------
\begin{figure}
   \centering
   \resizebox{\hsize}{!}{
   \includegraphics[viewport = -8  8 435 280,clip]{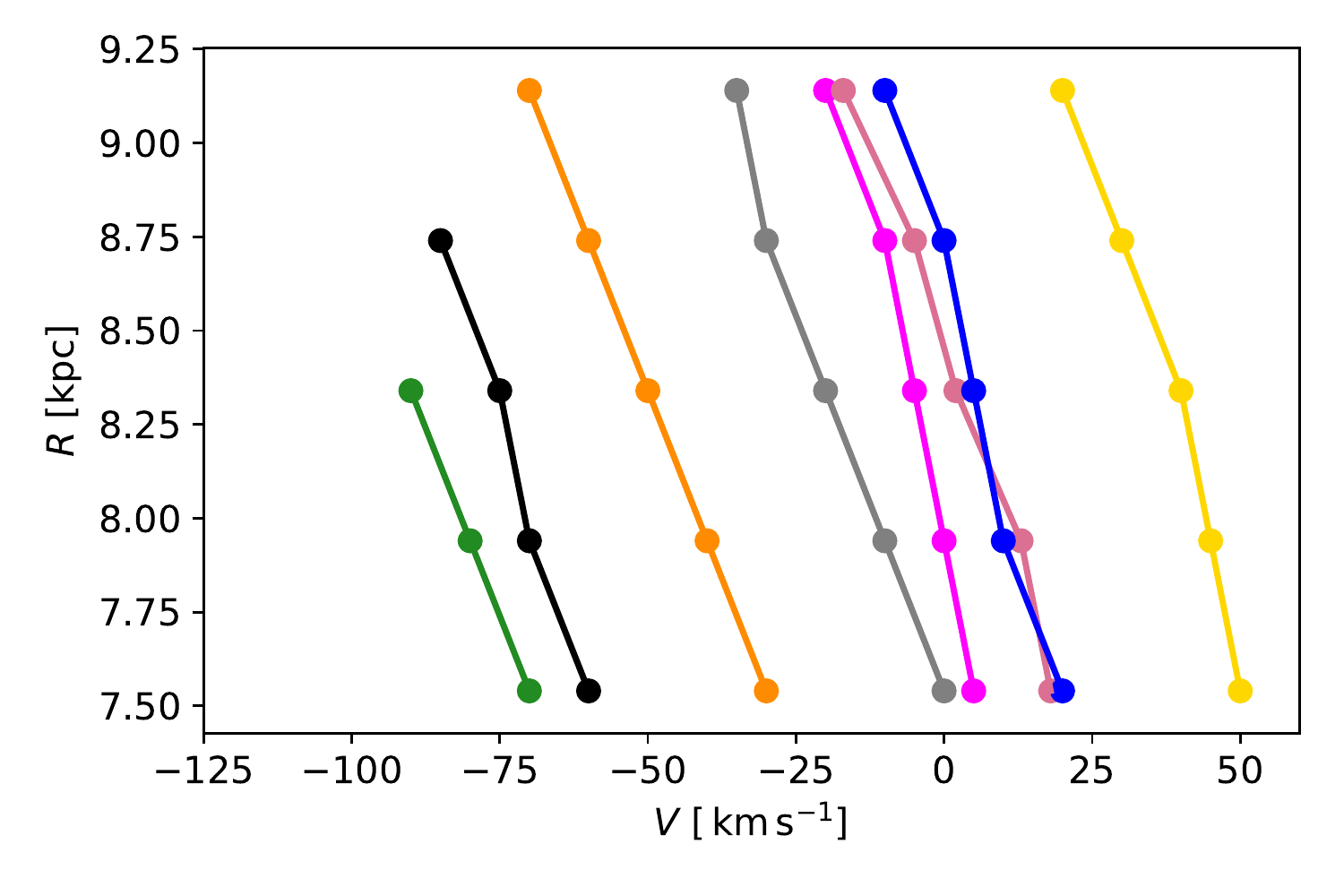}}
   \resizebox{\hsize}{!}{
   \includegraphics[viewport = 0  50 425 280,clip]{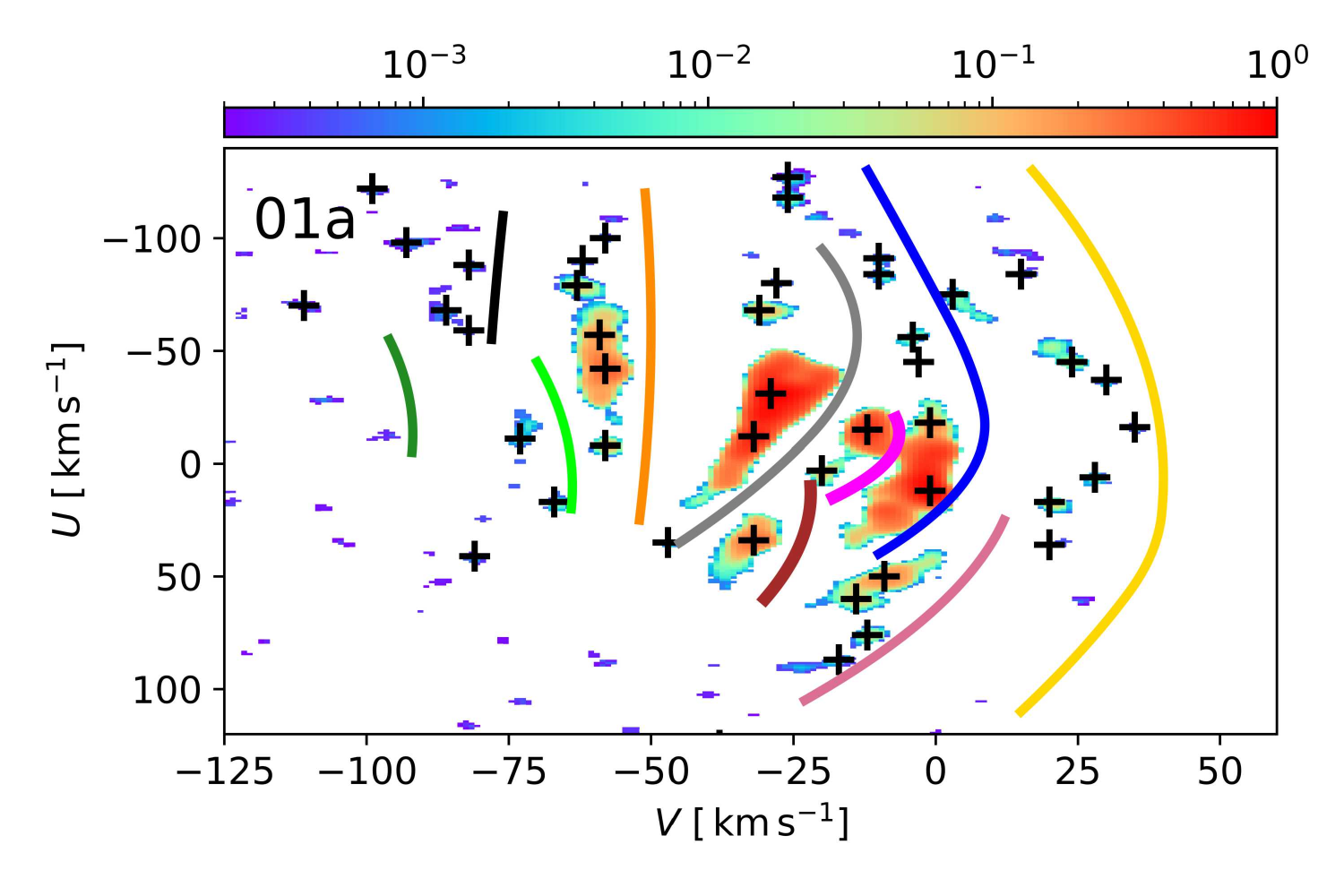}}
   \resizebox{\hsize}{!}{
   \includegraphics[viewport = 0  5 422 242,clip]{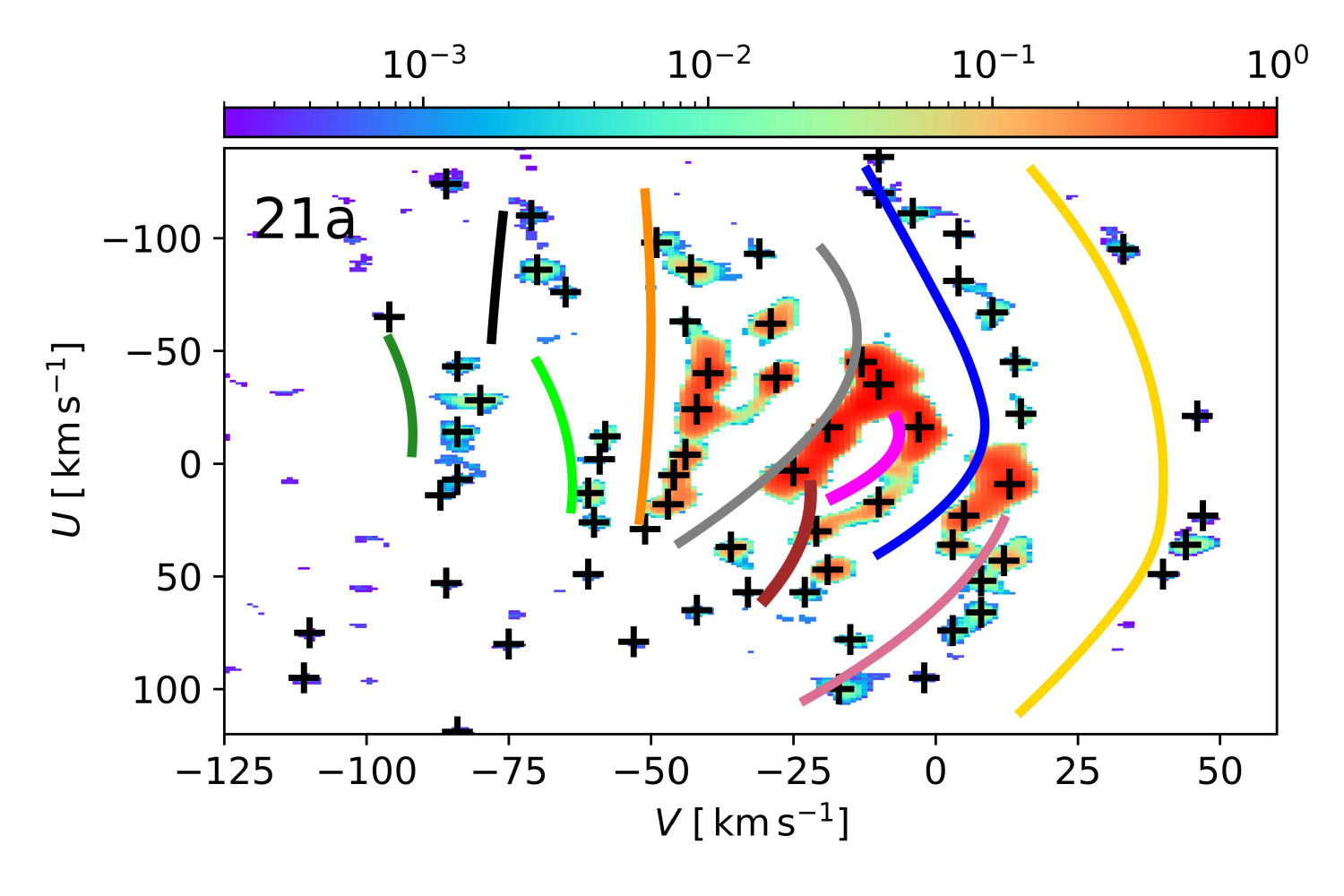}
   }
   \caption{{\it Top plot:} Change of central $V$ velocities of kinematic structures as a function of Galactocentric radii $R$. Each dot corresponds to the centre of the structure determined by the wavelet transform in volumes 01, 01a, 00, 21 and 21a. Names of the groups associated with the lines are listed in the legend of Figure \ref{_fig_00_2}. {\it Middle and bottom plots:} Example of wavelet maps in the $U-V$ space in regions 01a and 21a. Colour bar shows normalised wavelet coefficients. Black crosses show centres of the detected structures in regions 01a and 21a. Lines show the location of kinematic structures detected in region 00 (see legend of Fig. \ref{_fig_00_2}).
   \label{_fig_shift}
   }
\end{figure}
%--------------------------------------------------------------------

%--------------------------------------------------------------------
\begin{figure*}
   \centering
   \resizebox{\hsize}{!}{
   \includegraphics[viewport = 0  50 425 280,clip]{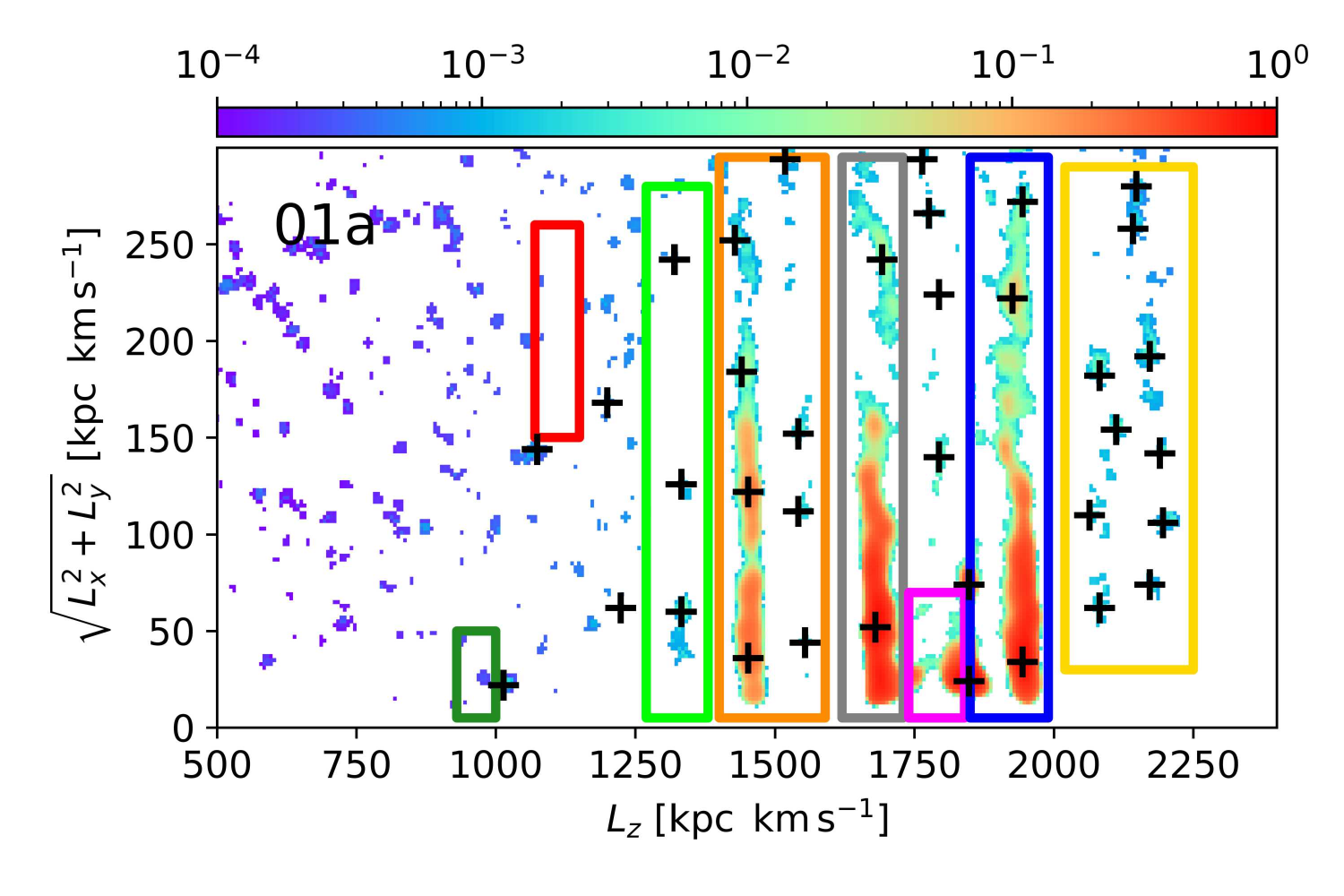}
   \includegraphics[viewport = 0  50 425 280,clip]{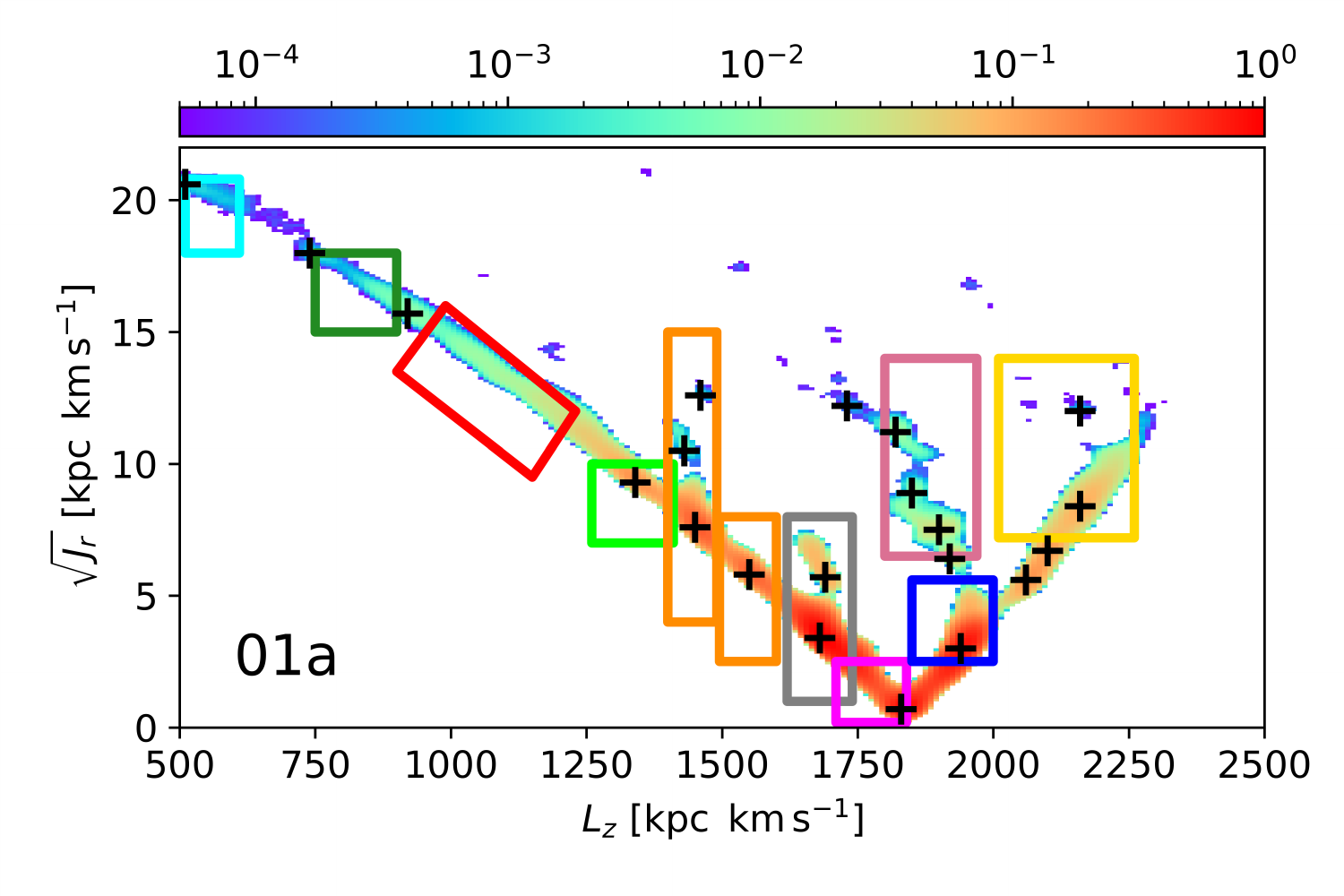}
   }
   \resizebox{\hsize}{!}{
   \includegraphics[viewport = 0  50 425 242,clip]{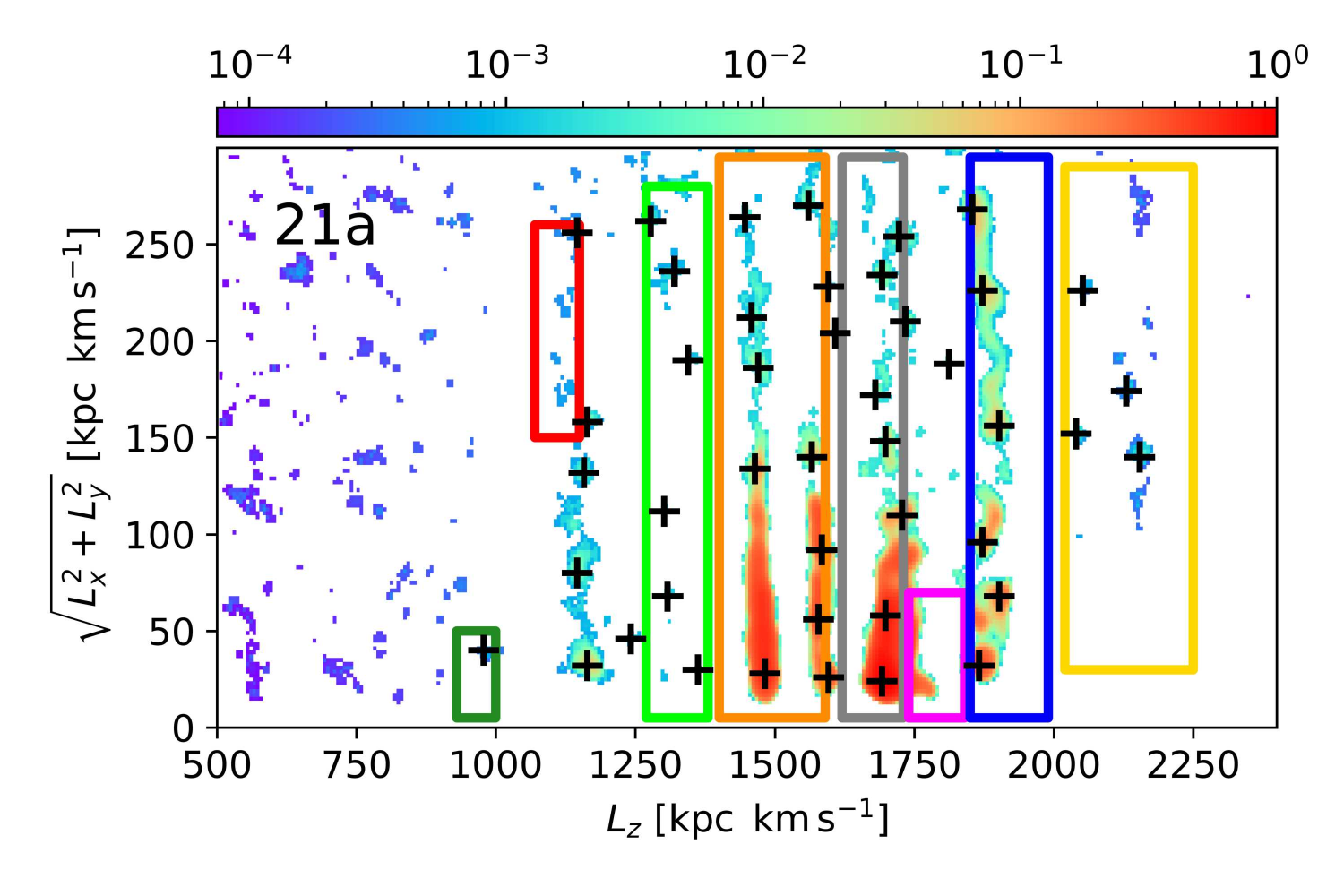}
   \includegraphics[viewport = 0  50 425 242,clip]{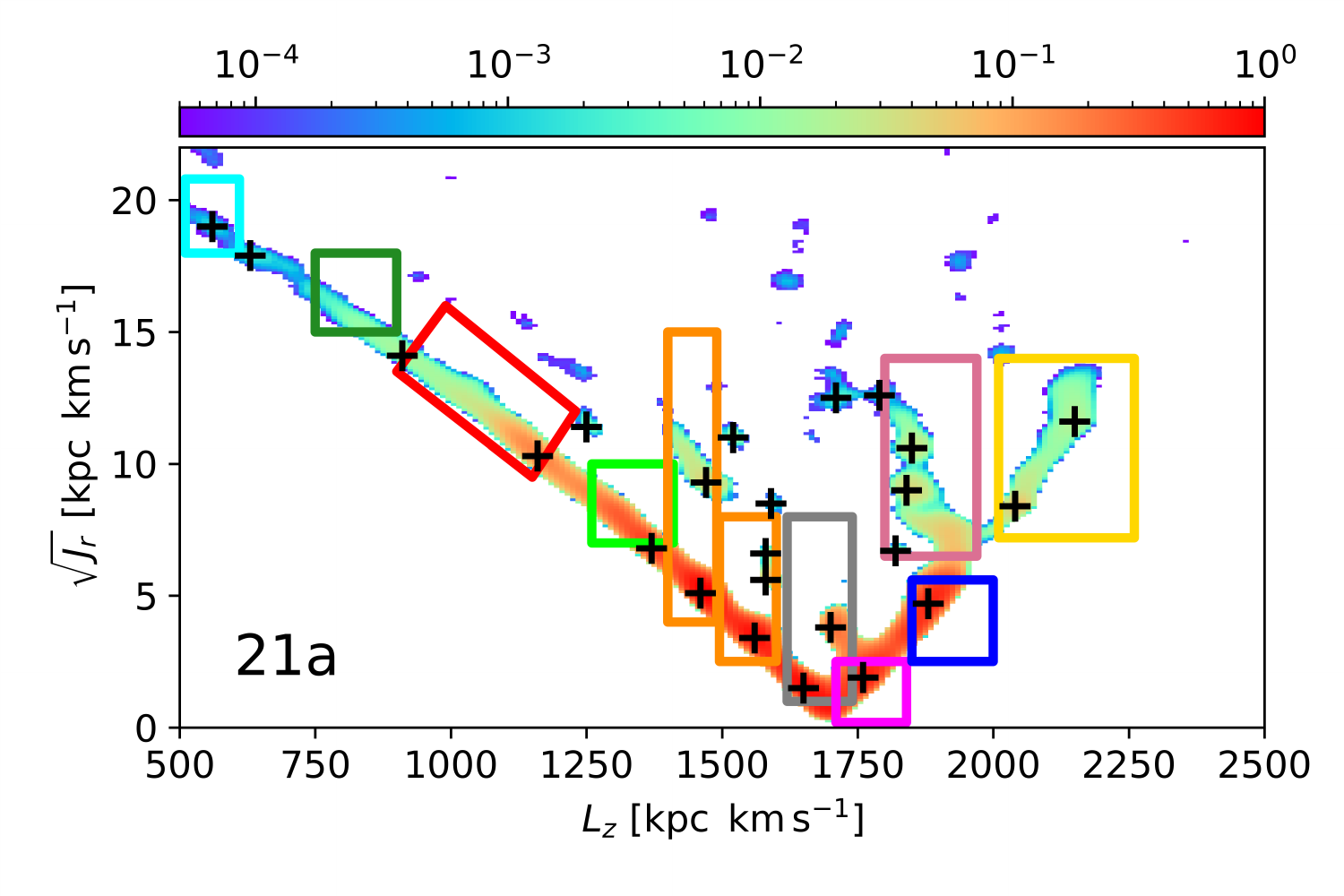}
   }  
   \resizebox{\hsize}{!}{
   \includegraphics[viewport = 0  5 425 242,clip]{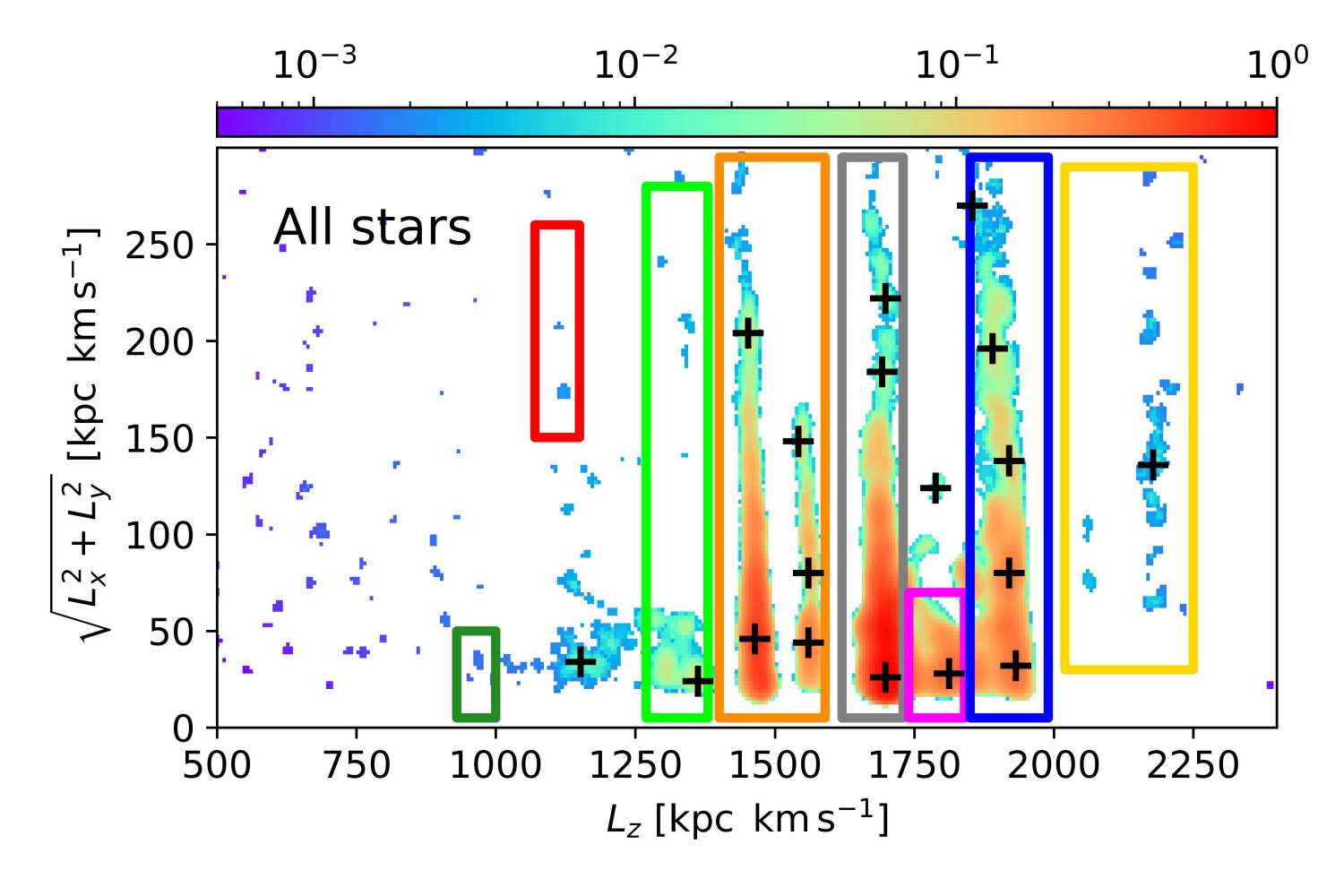}
   \includegraphics[viewport = 0  5 425 242,clip]{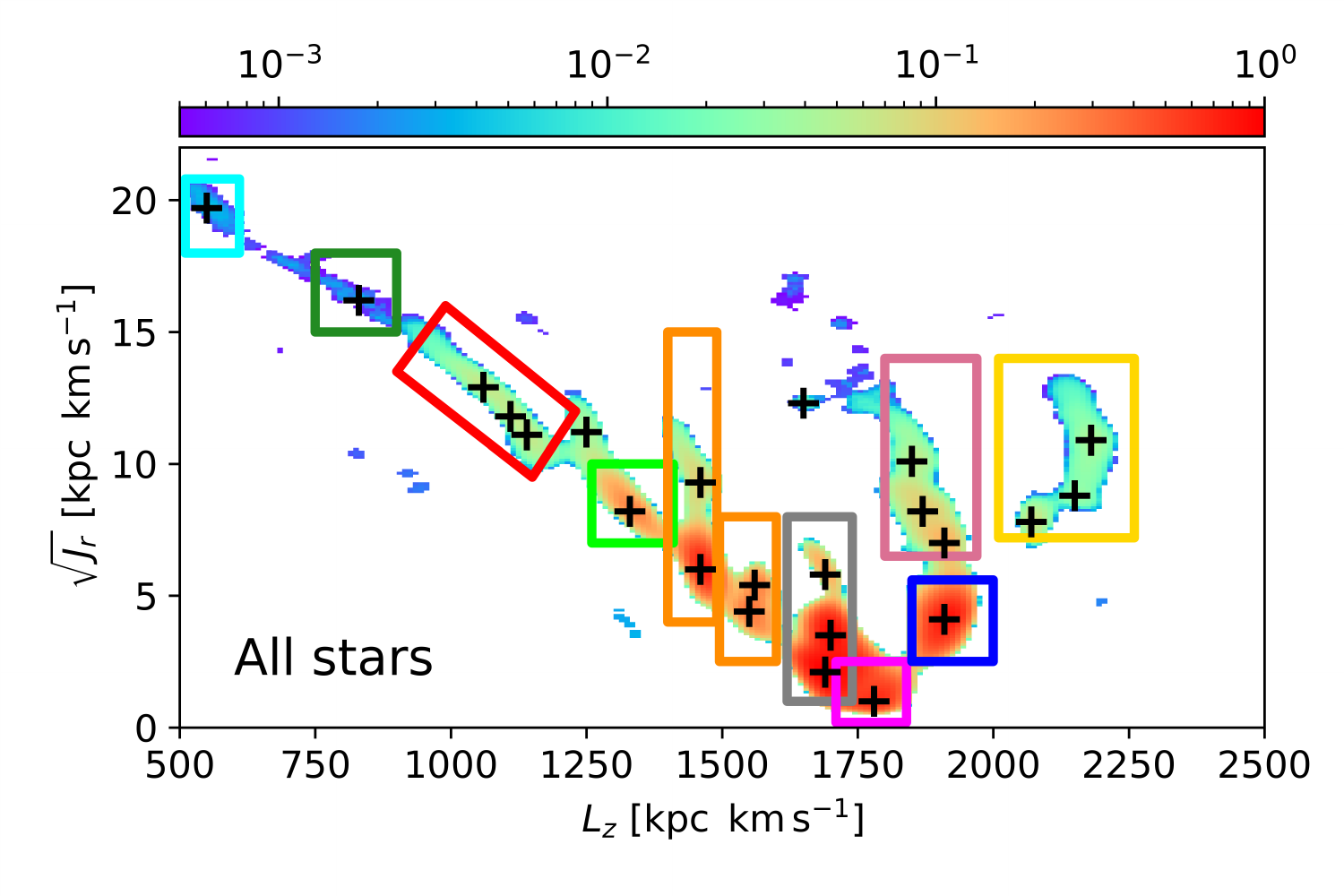}
   }
\caption{Wavelet coefficient maps of regions 01a (top), 21a (middle) and all stars studied in this work (bottom) retrieved in $L_z-\sqrt{L_x^2+L_y^2}$ (left column) and $L_z-\sqrt{J_{R}}$ (right column) space for scale $J=2$. Colour bars show normalised wavelet coefficients. Black crosses show centres of the structures. Boxes around the crosses show location of kinematic structures based on the nearby sample 00 (see Figure \ref{_fig_00_2}). Colours of the boxes correspond to different names of the structures listed in the legend of Figure \ref{_fig_00_2}.
   \label{_fig_shift2}
   }
\end{figure*}
%--------------------------------------------------------------------

%-------------------------------------------------
\subsection{Acquiring positions of the detected peaks} 
\label{_sec_sec_acquire_peaks}
%-------------------------------------------------
To get precise positions of the peaks Monte Carlo (MC) simulations were performed. MC samples were created assuming that each star can be represented as a Gaussian velocity distribution with $\mu = (U, V)$ and $\sigma = (\sigma_U, \sigma_V)$ for the two velocity components. To generate MC samples in angular momentum and action space orbits of stars were computed assuming that positions, proper motions and radial velocities can be represented as Gaussians, similarly as velocities. Here it is assumed that Gaussians are independent and do not consider correlations between astrometric parameters. These MC samples are then analysed in the same manner as the original data. Convergence is reached when the number of structures and their positions do not change as more simulations are added. Typically, results converge after about 30 simulations, but to be on the safe side, 100 MC samples were created for all regions. Then MC wavelet maps for different scales were over-plot used to search for peaks applying {\it peak\_local\_max} feature from {\it scikit-image}\footnote{https://scikit-image.org/} Python package \citep{_sc_image}. In this work we focus on the $J=2$ and $J=3$ scales as they allow us to detect most structures.

%===================================================================
%===================================================================
\section{Results}\label{_sec_results}
%===================================================================
%===================================================================
%In this section we present the detections of kinematic structures in the $U$--$V$, $V$--$\sqrt{U^2+2V^2}$ and $L_z$--$\sqrt{L_x^2+L_y^2}$ and $L_z$--$\sqrt{J_r}$ planes for 65 small volumes. 

%-----------------------------------------------------------
\subsection{Stellar streams in the nearby sample}
%-----------------------------------------------------------
Figures~\ref{_fig_00_2} and \ref{_fig_00_3} show 100 over-plotted wavelet maps for the central region 00 in four different planes for scales 2 and 3, respectively. Both scales show richness of kinematic structures for the nearby sample. The list of the centres of the peaks and the corresponding uncertainties are given in Table \ref{_tab_groups_2} for scale $J=2$ and in Table \ref{_tab_groups_3} for scale $J=3$. The fact that well-known groups like Sirius, Coma Berenices, Hyades, Pleiades and Hercules were identified at the expected positions shows that our method is sound (see Tables~3, 5, 6 and 7 in \citealt{_kushniruk17} that summarise literature values for the $U$ and $V$ velocities of Sirius, Coma Berenices, Hyades, Pleiades and Hercules). These groups are detected in all four planes. The detection of other groups varies between the planes. Figure~\ref{_fig_00_2} for scale $J=2$ shows the same structures as in Fig.~\ref{_fig_00_3} for scale $J=3$ but in greater details. It was decided to focus on scale $J=2$ since it is more sensitive to smaller structures. By comparing our results in the nearby region `00' (centred around the Sun, see Fig.~\ref{_fig_sample}) for scale $J=2$ with what has previously been found in the literature \citep[e.g.][]{_eggen98,_navarro04,_arifyanto06,_klement08,_williams09,_antoja12,_zhao14,_kushniruk17,_trick18,_antoja18,_ramos18}) we assign names to the structures. Curved lines and boxes of different colours in Figs.~\ref{_fig_00_2} and \ref{_fig_00_3} correspond to the names of the groups listed in the legend of Fig.~\ref{_fig_00_2}. The structures found in the nearby region are discussed below:

\begin{itemize}

\item {\bf A1/A2:} Groups with $V > 15 \kms$ and $L_z > 2000$\,kpc\,$\kms$ we link to arches A1 and A2 detected by \citet{_ramos18} (see their Table~2). A1/A2 is shown with yellow lines and boxes on the plots. 

\item {\bf Sirius:} Blue line and boxes correspond to the Sirius stream. Group 13 in the $U-V$ space is potentially Bobylev16 \citep[see][]{_bobylev16} and could be a continuation of Sirius.

\item {\bf $\gamma$Leo:} Pink line slightly above Sirius in $V$ is $\gamma$Leo stream \citep[see][]{_antoja12}. Unlike a big majority of the groups, $\gamma$Leo is located at positive $U$ velocities. The stream could be a continuation of Sirius arch since both have similar angular momenta. 
 
\item {\bf Coma Berenices:} Magenta line just below Sirius is Coma Berenices stream. Unlike arch-like neighbouring Sirius and Pleiades/Hyades, Coma Berenices is a clump in the $U-V$ plane and, consequently, is a shorter line in the angular momentum space.

\item {\bf Dehnen98/Wolf630:} Wolf630 and Dehnen98 \citep[see][]{_antoja12, _dehnen98} are two small groups in between Coma Berenices and Pleiades/Hyades streams. They are shown with brown colour and could be a continuation of Coma Berenices. 

\item {\bf Pleiades/Hyades:} A grey arch in the $U-V$ plane is associated with Pleiades/Hyades stream. Group 29 in the $U-V$ space linked to Antoja12(15) \citep[see][]{_antoja12} could be a continuation of stream. 

\item {\bf Hercules:} Orange lines and boxes correspond to the Hercules stream. It is likely to be composed of a few substructures that are visible in the angular momenta and action spaces.

\item {\bf HR1614:} The HR1614 moving group \citep[see][]{_feltzing00, _desilva07} we connect to the clumps just below Hercules in $V$. The group is shown with a lime colour. 

\item {\bf $\epsilon$ Ind:} Groups g34 and g35 in the $U-V$ plane are linked to a group called $\epsilon$Ind \citep[see][]{_antoja12}. The structure is marked with a black colour.

\item {\bf AF06:} AF06 stream was first found by \citet{_arifyanto06} in the range between $V \simeq -70$ and $-100\,\kms$. We did not find it in the $U-V$ space, but the group is detected in other three spaces and is shown with red boxes. 

\item {\bf Arcturus:} Group g36 in the $U-V$ plane could be the Arcturus stream. Median $V$ velocity and angular momentum of g36 are $V\simeq- 92 \kms$ and $L_z\sim1118$ kpc $\kms$. These values are a bit higher compared to, for example, values from \citet{_navarro04}, but are within the uncertainties. In the $V-\sqrt{U^2 + 2V^2}$ plane the nearest to Arcturus are groups g1 and g2 and have the same angular momentum and radial action as g36 in the $U-V$ plane. In the angular momentum space g21 has the parameters closest to Arcturus. In action space there are two candidates: g9 and g2. The first group is consistent with the groups detected in velocity spaces, the second one has lower angular momentum and $V$ velocity. Taking into account works by \citet[e.g.][]{_klement08, _zhao14} we link g2 to the Arcturus stream and group g9 to the AF06 stream. Arcturus is shown as green lines and boxes on the wavelet maps. 

\item {\bf KFR08:} Among the detected groups we assign one weak over-density in action space at $L_z \simeq 575$ kpc $\kms$ to the group called KFR08. The structure was first detected by \citet{_klement08} at $V\simeq-160$\,kpc\,$\kms$. Group g1 in detected in action space has exactly the same median $V$ velocity. KFR08 is shown with cyan colour on the wavelet maps.
\end{itemize}

Overall 36 groups at scale $J=2$ and 16 groups at scale $J=3$ were discovered in the $U-V$ plane that form larger-scale arches as discussed in \citet{_ramos18, _antoja18, _katz18k}. We also conclude that these arches correspond to the lines in the $L_z-\sqrt{L_x^2+L_y^2}$ plane and to clumps in the $V-\sqrt{U^2+2V^2}$ and $L_z-\sqrt{J_r}$ due to very similar properties of the groups (see Tables \ref{_tab_groups_2} and \ref{_tab_groups_3}). In the $V-\sqrt{U^2+2V^2}$ 17 and 8 groups were detected respectively. Action space mimics $V-\sqrt{U^2+2V^2}$ plane very much but allows to detect the structures in greater detail. In the angular momentum and action spaces 24 groups were found in each space at scale $J=2$ and 9 and 10 groups at scale $J=3$, respectively.

%-----------------------------------------------------------
\subsection{Stellar streams outside the solar neighbourhood}
%-----------------------------------------------------------
The solar neighbourhood volume is well-studied and thus it is relatively easy to match the detected groups with groups that have been identified by other studies in the literature. The behaviour of the groups outside the solar volume was studied by \citep[][e.g.]{_antoja12, _ramos18}. Both papers found a decreasing trend for $V$ velocity when moving to the volumes at larger $R$. In this work we also investigated the trends of the structures depending on the position in the Galaxy with a focus on the Arcturus stream.

If one looks at the small volumes from the left-hand side of Fig.~\ref{_fig_sample} and move towards positive $X$ values an increase in $V$ velocity can be observed. As an example of this trend Fig.~\ref{_fig_shift} shows how the $V$ velocity of the groups evolves with Galactocentric distance $R$ in regions 01, 01a, 00, 21 and 21a based on the analysis of the $U-V$ plane. For example, the Hercules stream in region 01 is detected at $V\simeq- 70 \kms$, and for comparison, in region 21a the Hercules stream shifts towards positive $V$ and is located at $V\simeq- 40 \kms$. Similar behaviours is observed for most of the major streams and is shown in the top plot of Fig.~\ref{_fig_shift}. On the wavelet maps for regions 01a and 21a that are shown in the middle and bottom plots of Fig.~\ref{_fig_shift} we draw the lines from the top left plot in Figure~\ref{_fig_00_2}. Taking volumes at larger R shifts the groups towards lower $V$ values and vice verse.    

If one fixes the galactocentric distance and starts exploring the regions at high $\phi$ moving down towards negative $\phi$ (for example, start at region 15 and go down to region 35), the streams are observed at the same position in $V$. Major streams including Sirius, Pleiades, Hyades and Hercules have the same angular momentum when fixing $R$ and looking at different $\phi$. This is different from results in \citet{_monari19} who found that the Hercules angular momentum changes with azimuth at solar radius. We also do not observe this change when fixing $R$ inside and outside the Solar circle. 

The shape of the angular momentum and action spaces change a bit at different $R$, but almost all main structures detected in the solar volume remain at the same positions within a box defined by volumes 02\_12, 22\_12, 22\_32 and 02\_32. Figure~\ref{_fig_shift2} shows wavelet transform maps for regions 01a and 21a (top and middle rows). Additionally a wavelet transform was applied to all the stars in the sample in the angular momentum and action spaces (bottom row) and compare to results in volumes 01a and 21a. If there are any groups with constant actions detected in the total sample then it possible to observe them in smaller volumes. Boxes of different colours mark kinematic structures detected in the region 00. The same boxes are plotted on top of maps for regions 01a and 21a. There is a small shift in action space when changing $R$, but generally main groups are located at the same positions.

The kinematic structures are mainly detected in the central regions within the rectangle defined by regions 02\_12, 22\_12, 22\_32 and 02\_32. The rest of the regions contain less stars and also have larger distance uncertainties. We tested if the structures really exist only inside the mentioned above box or the lack of the groups in the outer regions is a consequence of larger distances and smaller number of stars in sub-samples. To check if the latest is true 10\,000 stars were randomly selected in the central region 00 and repeated the wavelet analysis. Then, the results were compared with \citet{_dehnen98} who used a sample of 14\,000 stars in total. For our 10\,000 sample a similar result as \citet{_dehnen98} was received. The main conclusion of the test is that with the small samples it is possible to detect only main big structures like Sirius, Hyades, Pleiades and Hercules. The more stars are in the volume, the higher is the probability to detect low-velocity structures. Due to this limitation, it is not possible claim that there is such a radius where some of the groups stop existing.

There are many tiny groups detected in the $U-V$ plane that could potentially be a part of the Arcturus stream. We do not observe a clear arch that we can connect with the structure. Unlike the Arcturus stream, an arch at $V\simeq-80 \kms$ is clearly visible inside a box defined by regions 01, 11, 21, 31 for the AF06 stream. In the $V-\sqrt{U^2+2V^2}$ plane the structures are better resolved at scale $J=3$. In the angular momentum and action space there are strong detections of the low-velocity groups clearly visible at scale $J=3$. The Arcturus and KFR08 streams appear stronger at the smaller Galactocentric radii. Based on the analysis of all stars in action space, the Arcturus stream is an elongated structure in $L_z$. These means it covers a wider range of orbits, unlike, for example, the Hercules stream.

%--------------------------------------------------------------------
\begin{figure*}
   \centering
   \resizebox{\hsize}{!}{
   \includegraphics[viewport = 0  0 450 332,clip]{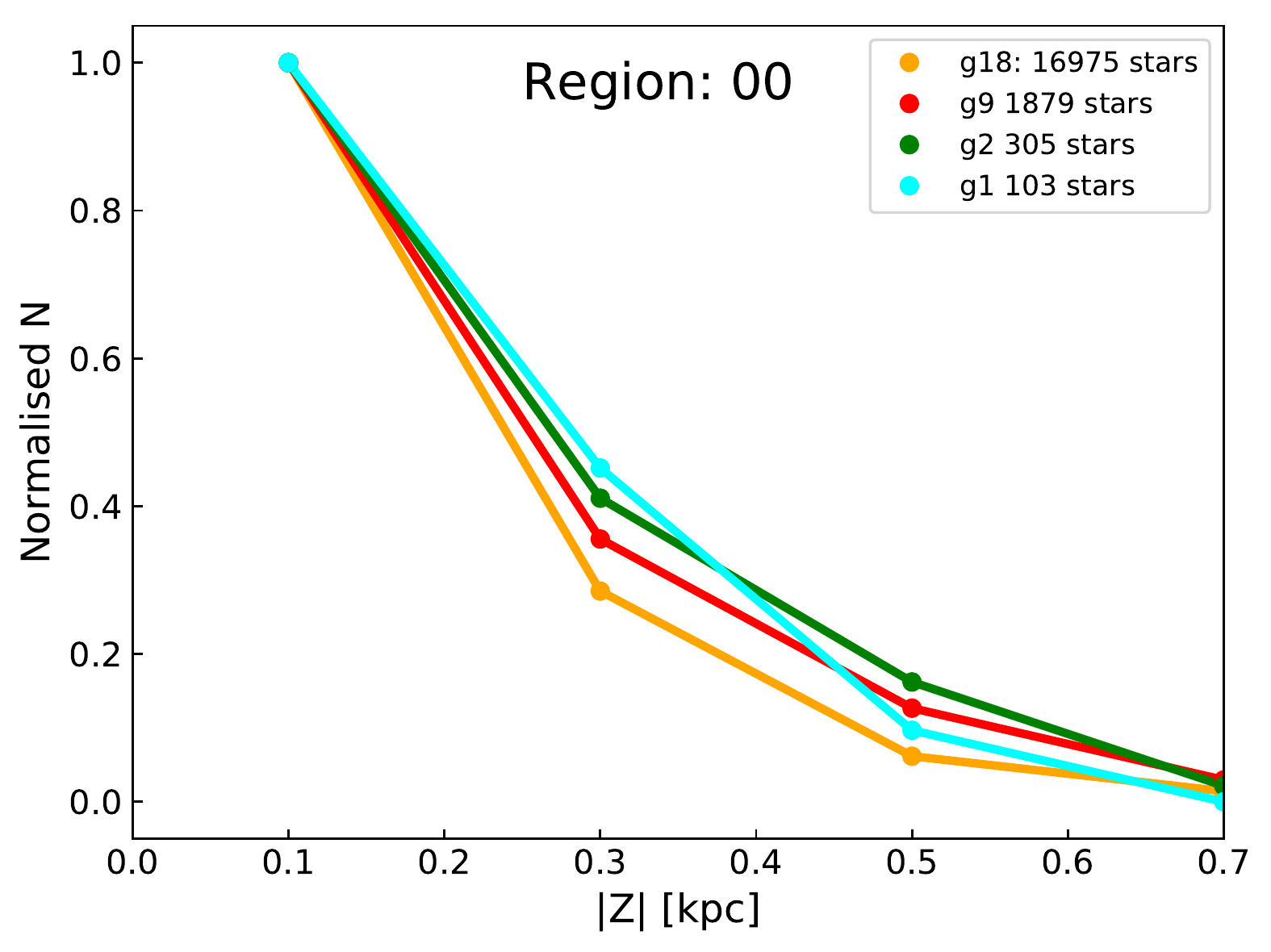}
   \includegraphics[viewport = 0  0 450 332,clip]{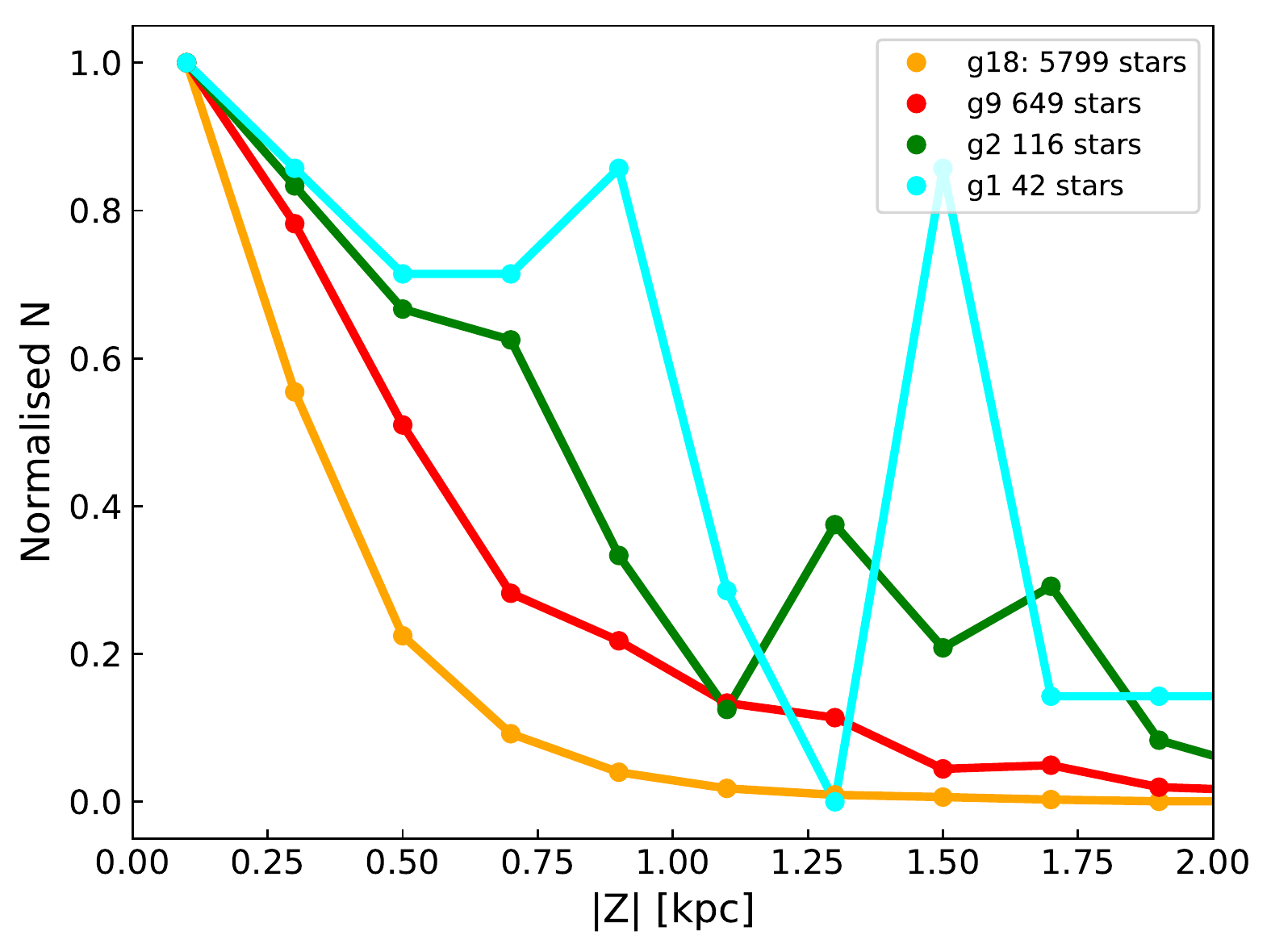}
   }
   \caption{A number of stars in Hercules, AF06, Arcturus and KFR08 versus module of distance from the Galactic disk $\left|Z\right|$ in region 00 (left) and in regions 01, 11, 21 and 31 (right).
   \label{_fig_gaia_z}
   }
\end{figure*}
%--------------------------------------------------------------------

%-----------------------------------------------------------
\section{The vertical extent of Arcturus and associated streams}\label{_sec_lowvel}
%-----------------------------------------------------------

We will focus on three low-velocity structures (g1, g2 and g9) detected in region 00 in action space between $V\simeq-70$ to $-160 \kms$. The groups that we associate with these velocities in the solar region are the AF06, Arcturus, and KFR08 streams. The question is if they are related, are they elongations of each other, and how different they are compared to the Hercules stream? The Hercules stream is chosen as a reference as it is one of the most studied kinematic structures and is a relatively metal-rich disk structure with the dynamical origin with the Galactic bar \citep[e.g.][]{_bensby07, _ramya16, _perez17}. One of the main peaks of the Hercules stream is group g18 detected in action space. We study properties of this group for a comparison with the low-velocity structures.

To further distinguish the three streams we investigate how the number density of stars in the Hercules, Arcturus, AF06, and KFR08 streams vary with vertical distance from the Galactic plane. Candidate member stars of the three streams were selected from the stellar sample constructed as described in Sect.~\ref{_sec_sample} using the characteristic velocities of the streams that found for region 00 (see  Table~\ref{_tab_groups_2}). From now on, it is assumed that kinematic groups are defined as stars on similar orbits. We assume that a star belongs to a group if its radial action and angular momentum falls into an ellipse around the centre of the group as shown in Fig.~\ref{_fig_00_2}.

The plot on the left-hand side in Fig.~\ref{_fig_gaia_z} shows the variation of the normalised number of stars in g1, g2, g9 and g18 streams as defined in action space (see Fig.~\ref{_tab_groups_2}) with module of the distance from the Galactic plane $|Z|$ for region 00. The Hercules stream is slightly more concentrated towards the Galactic plane compared to the three low-velocity structures. To check if this is valid in the regions outside the solar neighbourhood stars in regions 01, 11, 21 and 31 (see Fig.~\ref{_fig_sample}) that are members of groups g1, g2, g9 and g18 were selected. A star is defined as a member of a group if it has $J_r$ and $L_z$ values located within an ellipse around a group in action space as shown in Fig.~\ref{_fig_sample}. Since actions are conserved quantities along orbits of stars in static potentials it is expected that the structures will show up at the same positions in $L_z$ and $J_r$. The right-hand side plot in Fig.~\ref{_fig_gaia_z} is the same as the one on the left, but for regions 01, 11, 21 and 31. The Hercules stream is strongly concentrated to the Galactic plane and gets rapidly weaker with distance from the plane. At distances above $|Z|\gtrsim 0.7$\,kpc the density of Hercules stream stars has essentially dropped to zero. In comparison, the g1, g2 and g9 structures reach larger heights from the plane. The disappearance of the Hercules stream after about $0.7$\,kpc is consistent with the results from \citet{_antoja12} that detected the Hercules stream at lower confidence level at higher $Z$. To further probe the origins of the detected kinematic structures and how they relate to each other we will make use of the detailed elemental abundance data from recent spectroscopic surveys.

%===================================================================
%===================================================================
\section{Chemical properties of the kinematic streams}
\label{_sec_chem}
%===================================================================
%===================================================================

In this section we will investigate whether the detected streams show distinct elemental abundance patterns. To our aim the detailed abundance data from large spectroscopic surveys such as GALAH DR2 \citep{_buder18} and APOGEE DR 14 \citep{_holzman18} were used. GALAH DR2 accounts over 340\,000 stars and APOGEE DR14 accounts around 263\,000 stars. Both GALAH and APOGEE have determined radial velocities for all their targets, and those stars does only to a very limited amount overlap with the subsample of stars in Gaia that comes with measured radial velocities in Gaia DR2. This means there will be just a few stars in each kinematic group when cross-matching GALAH and APOGEE with our sample that is constructed from the Gaia DR2. 

Therefore, to increase the number of stars that can be associated with the streams that were detected and that have elemental abundances in the GALAH and APOGEE data releases, we compute space velocities $U,V,W$, angular momenta $L_x, L_y, L_z$ and radial action $J_r$ for all APOGEE and GALAH stars using astrometric data from {\it Gaia} DR2 and radial velocities from GALAH and APOGEE. Then the stars with $\sigma_U, \sigma_V \le 20\,\kms$ and with good quality flags were select. For GALAH stars with good data quality flags were included: $flag\_cannon=0$ and $flag\_x\_fe=0$, where where X is a chemical element, and for APOGEE the following quality flags were used: $X\_FE\_FLAG = 0$, where X is a chemical element. This left us a sample of 101\,862 and 72\,517 stars for the GALAH and APOGEE surveys, respectively. To select stars that are possible members of the detected kinematic streams we use our kinematic constraints for action space listed in Table~\ref{_tab_groups_2}, meaning that a star must be within a specific range in $L_z$ and $J_r$ (fall into an ellipse around the structure as shown in Fig.~\ref{_fig_sample}).

%--------------------------------------------------------------------
\begin{figure*}
   \centering
   \resizebox{\hsize}{!}{
   \includegraphics[viewport = 0  0 900 600,clip]{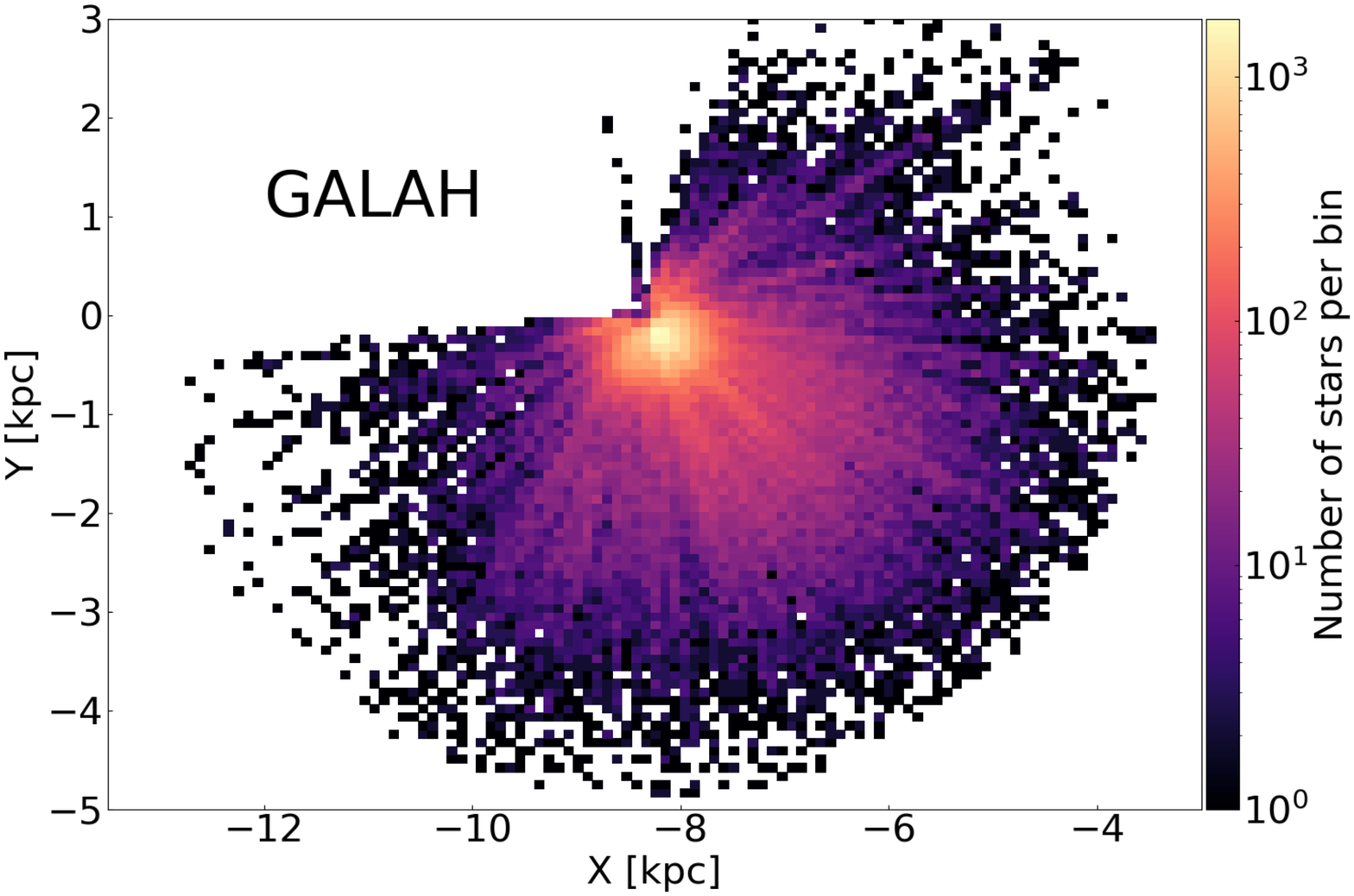}
%   }
%   \resizebox{\hsize}{!}{
   \includegraphics[viewport = 0  0 900 600,clip]{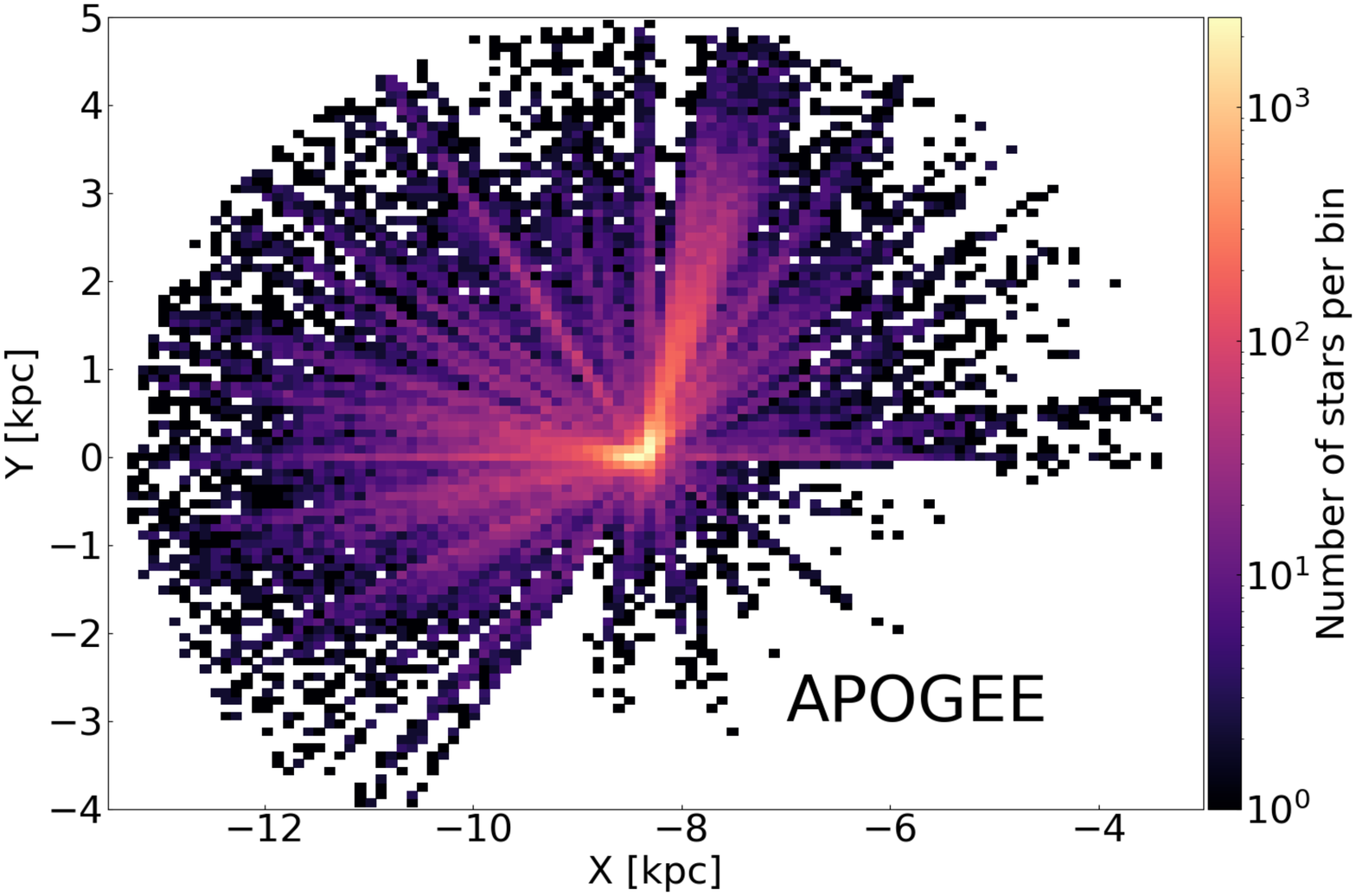}
   }
   \caption{$X-Y$ distributions for GALAH stars (left) and for APOGEE stars (right) with $\sigma_U, \sigma_V \le 20 \kms$ and good quality flags. The bin size is 0.1 kpc for both plots.
   \label{_fig_distr}
   }
\end{figure*}
%--------------------------------------------------------------------

Figure~\ref{_fig_distr} shows the $X-Y$ distributions for the constructed GALAH and APOGEE samples. It is seen that APOGEE covers more stars of the Northern sky and GALAH covers mainly Southern part of the sky.

%--------------------------------------------------------------------
\begin{figure*}
   \centering
   \resizebox{\hsize}{!}{
   \includegraphics[viewport = 0   0 450 335,clip]{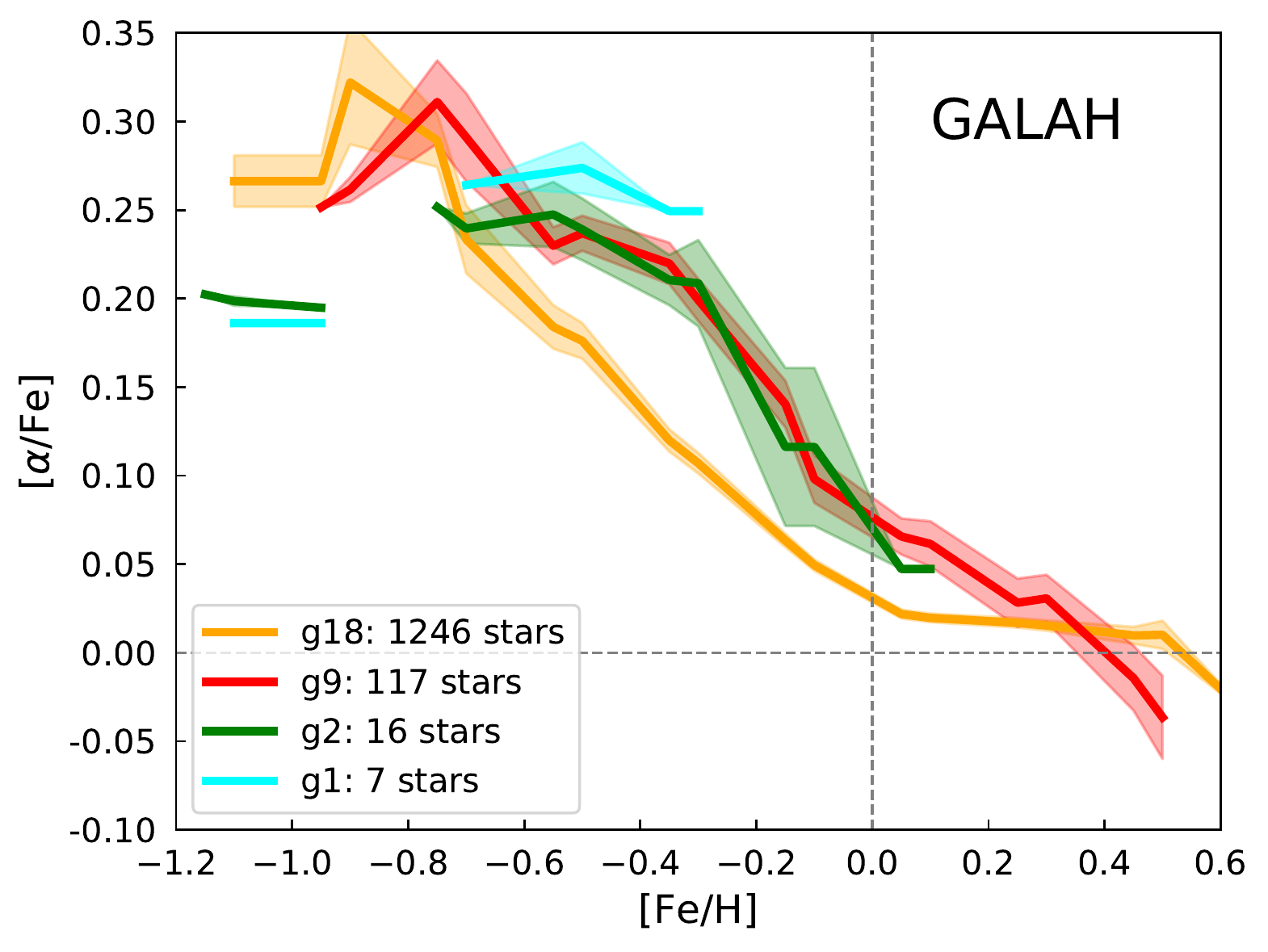}
   \includegraphics[viewport = 0   0 450 335,clip]{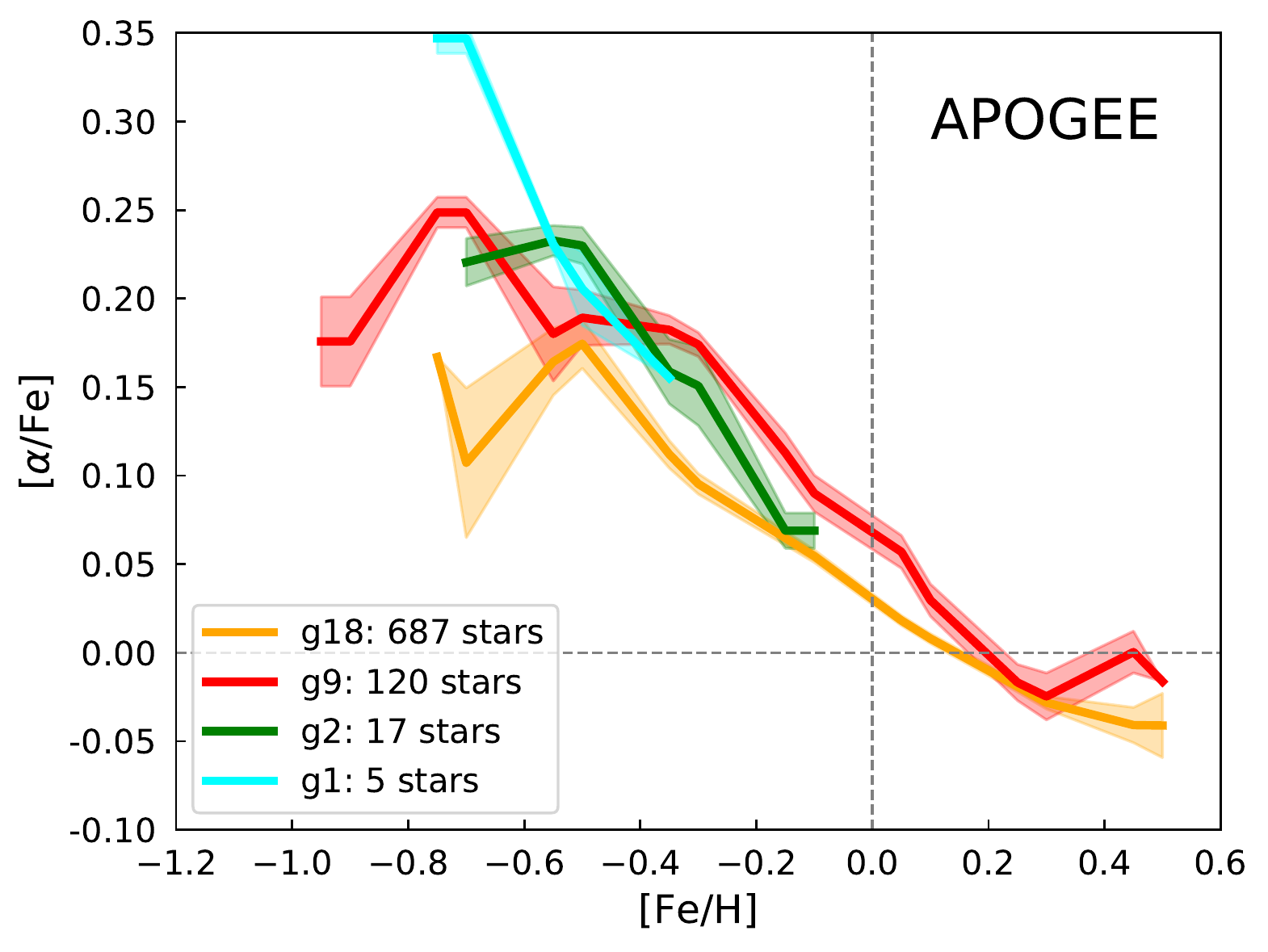}
   }
   \resizebox{\hsize}{!}{
   \includegraphics[viewport = 0   0 450 335,clip]{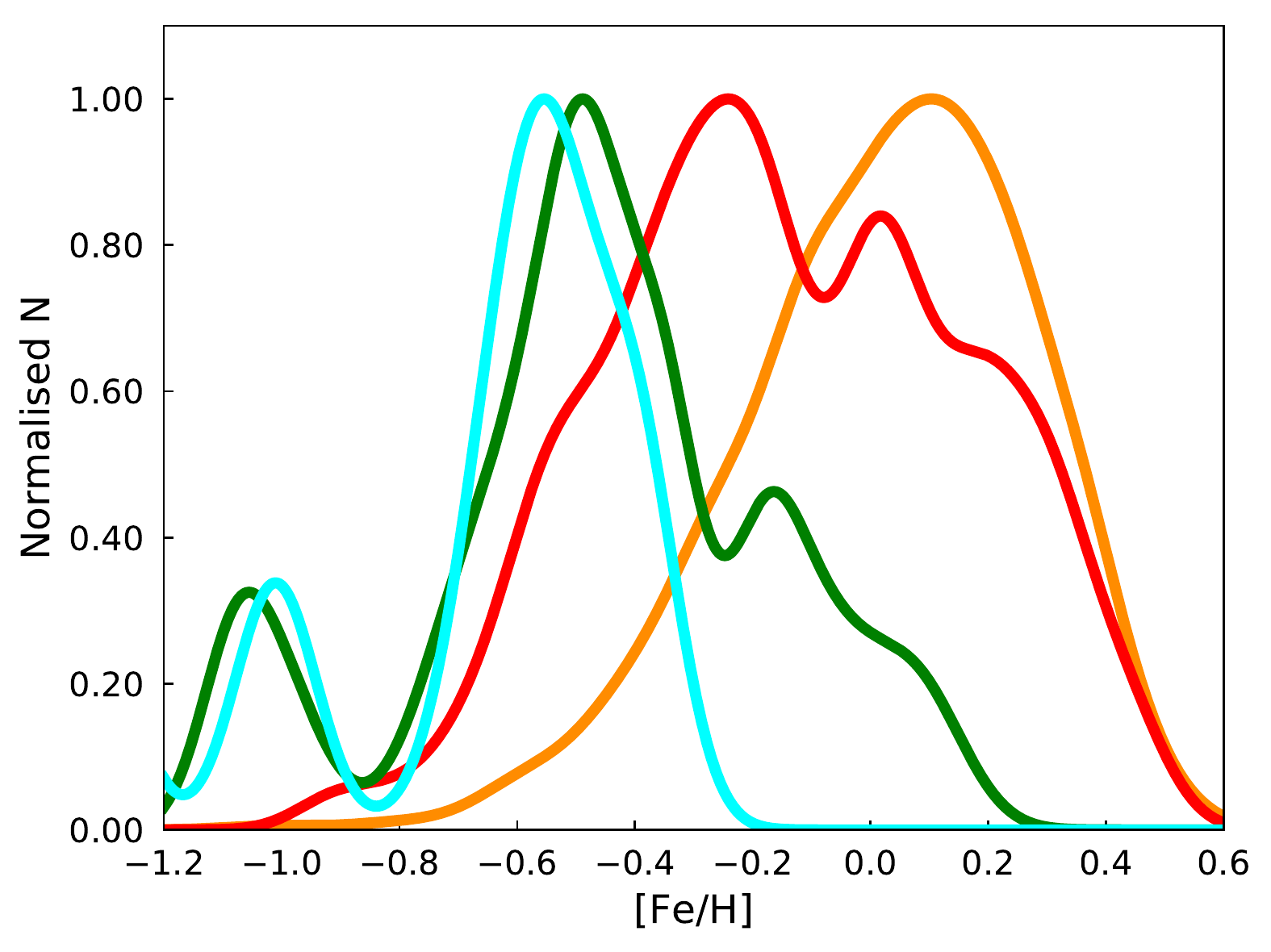}
   \includegraphics[viewport = 0   0 450 335,clip]{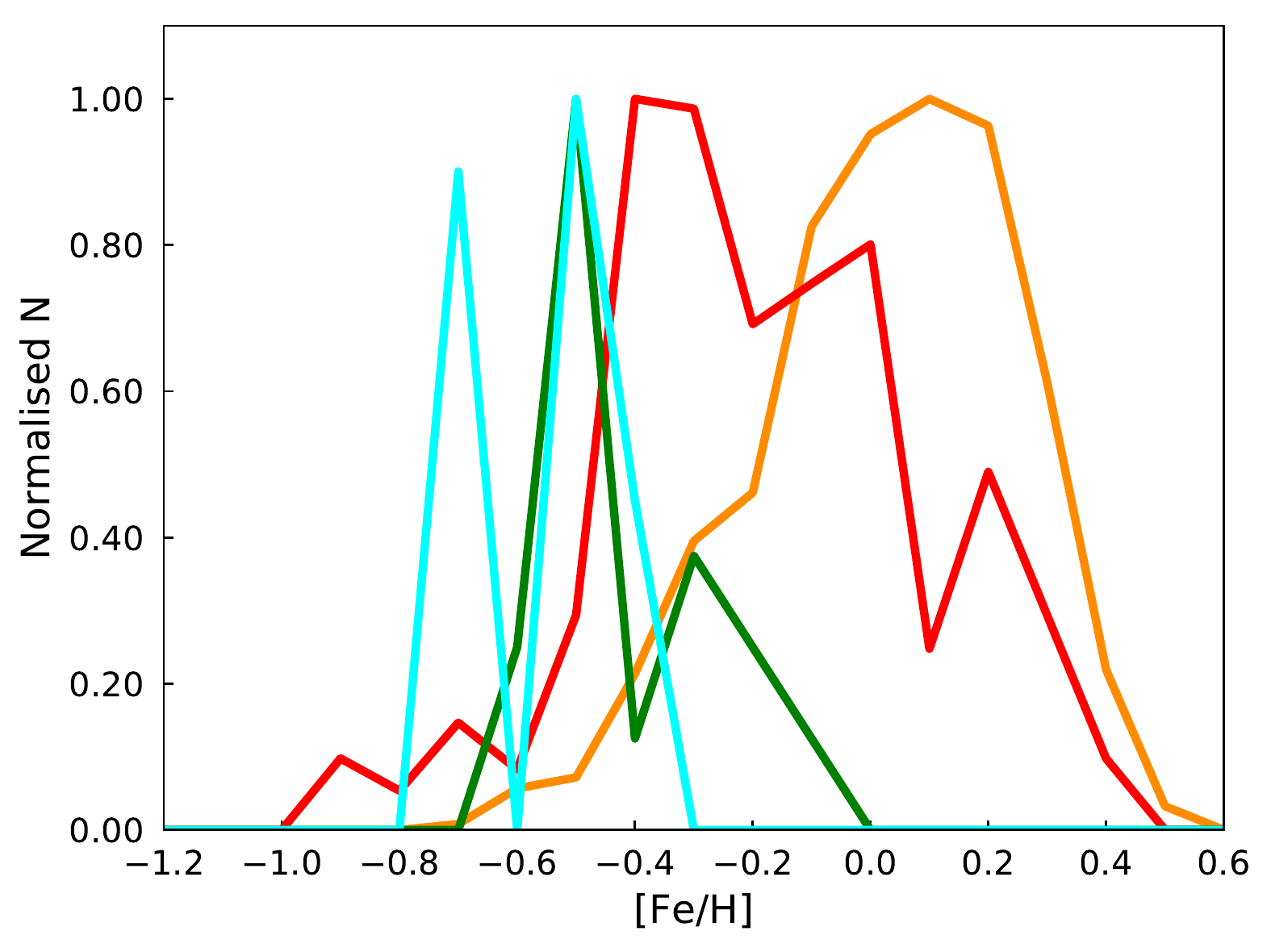}
   }
  \caption{{\it Top plots:} $\rm [Fe/H]-[\alpha/Fe]$ trends for the streams selected from the GALAH and APOGEE data in action space in a square of nine regions defined by regions 01a\_11a, 11a\_21a, 21a\_31a and 31a\_01a. The solid lines show the running means of the [$\alpha$/Fe] distributions in bins of [Fe/H], and the shaded regions give the $1-\sigma$ dispersions of the data around the means. The bin width when calculating running means is 0.05\,dex. %The horizontal lines at $\rm [\alpha/Fe]\,=0.2$ divide the plots into the thin and the thick disks as suggested by \citet{_wojno16}.
  {\it Bottom plots:} Normalised generalised metallicity distributions for the stellar samples shown in the top plots. Here each star is represented by a Gaussian with central peak at the estimated metallicity and a width given by the uncertainty of the metallicity.
   \label{_fig_ab}
   }
\end{figure*}
%--------------------------------------------------------------------

%--------------------------------------------------------------------
\begin{table}[t]
\begin{center}
\caption{Median metallicities and corresponding dispersions of stars members of the groups located in the square defined by nine regions around the central region 00 that were selected from APOGEE and GALAH samples.
\label{_tab_met}}
\footnotesize
\begin{tabular}{c|rr|rr} 
\hline
\hline
\noalign{\smallskip}
\multirow{2}{*}{Group} &  \multicolumn{2}{c}{GALAH} &   \multicolumn{2}{|c}{APOGEE} \\
  & [Fe/H]$_{median}$ & $\sigma_{[Fe/H]}$ & [Fe/H]$_{median}$ & $\sigma_{[Fe/H]}$\\
\noalign{\smallskip}  
\hline 
\noalign{\smallskip}
g18  &    0.0 & 0.2 &   0.0  & 0.2\\
g9   & $-$0.2 & 0.3 &$-$0.2  & 0.3\\
g2   & $-$0.5 & 0.3 &$-$0.5  & 0.2\\
g1   & $-$0.6 & 0.3 &$-$0.5  & 0.1\\
\noalign{\smallskip}
\hline
\end{tabular}
\end{center}
\end{table}
%--------------------------------------------------------------------

The top plots of Fig.~\ref{_fig_ab} shows the $\rm [\alpha/Fe]-[Fe/H]$ diagrams for stars in groups g1, g2, g9 and g18 selected in nine regions around the solar neighbourhood (01a\_11a, 11a, 11a\_21a, 01a, 00, 21a, 31a\_01a, 31a and 21a\_31a) for GALAH and APOGEE samples. We over-plot results for these nine regions simply because there are not enough stars in the low-velocity streams in each region to present them separately. The stars members of the groups were selected based on the properties of the groups in region 00 listed in Table \ref{_tab_groups_2}. Since actions are conserved quantities the groups are expected at the same positions after correcting $L_z$ values for the shift that arises due to difference in Galactocentric radii. The solid lines in these diagrams show the running mean for each stream. The shaded regions around each line show the corresponding $1\sigma$ dispersions around the mean value. %The black horizontal line at $\rm [\alpha/Fe]\,=0.2$ corresponds to the approximate division between the Galactic thin and the thick disks as suggested by \citet{_wojno16}.

The bottom plots of Fig.~\ref{_fig_ab} show generalised metallicity distributions for the same groups as in the upper plots. Median values of the metallicity distributions and corresponding dispersion of groups g18, g9, g2 and g1 are presented in Table~\ref{_tab_met} for the GALAH and APOGEE samples. The low-velocity streams generally have wide metallicity distributions and reaching lower metallicities down to $\rm [Fe/H]\lesssim -1$. The Hercules stream is more metal-rich. These plots shows that the low-velocity streams could be high-alpha thick disk structures, while the Hercules stream is likely a mixture of both the thin and thick disks. 
%In Table~\ref{_tab_met} we present median metallicities of the groups for the same stars and groups used to make Fig.~\ref{_fig_ab}. 
Two-sided Kolmogorov-Smirnov tests are then used to check if the metallicity distribution of any of the streams come from the same distribution. In all cases the $p-$values were infinitesimally small, indicating that the null-hypothesis have to be rejected, meaning that all the distributions are different. Results using APOGEE and GALAH surveys are similar: the low-velocity streams appear to be thick disk structures, while the Hercules stream is a mixture of both the thin and the thick disks, and is a more metal-rich structure.

%===================================================================
%===================================================================
\section{The origin of the Arcturus stream}\label{_sec_arcturus}
%===================================================================
%===================================================================

In this section first we provide a brief overview of the debates around the origin of the Arcturus stream, then summarise the kinematic and chemical characteristics of the Arcturus stream from this work, and based on that we discuss possible origins of the stream.

%%%%%%%%%%%%%%%%%%%%%%%%%%%%%%%%%%%%%%%%%%%%%%%%%%%%%%%%%%%%%%%%%%%%%%%%%
\subsection{Accretion origin: debris of a disrupted satellite}%%%%%%%%%%%%%%
%%%%%%%%%%%%%%%%%%%%%%%%%%%%%%%%%%%%%%%%%%%%%%%%%%%%%%%%%%%%%%%%%%%%%%%%%

In \citet{_eggen96, _eggen98} it was shown that the Arcturus stream (then called a moving group) belongs to the old thick disk population. It was fitted with a 10\,Gyr isochrone and the metallicity of Arcturus was estimated as $\rm [Fe/H] \simeq - 0.6$. Similar properties were observed by \citet{_gilmore02} and \citet{_wyse06} who found a clump of stars at $V\simeq -100 \kms$ that were estimated to be about $10-12$\,Gyr old and metal-poor with $\rm -2.5< [Fe/H] <-0.5$. This is consistent with the properties of the Galactic thick disk. 

One of the first attempts to explain the phenomenon of the Arcturus stream numerically was performed by \citet{_navarro04}. Assuming a merger event that happened $10-12$\,Gyr ago \citet{_navarro04} obtained a structure with similar properties as the Arcturus stream. %This is also the time when, potentially, the thick disk formed \citep[][]{_wyse00}. 
\cite{_navarro04} also estimated the vertical component of the angular momentum of the group to be $L_z \simeq 1000$\,kpc\,$\kms$. 

Other evidence for a possible debris origin for the Arcturus stream comes from \citet{_helmi06}, who found that a satellite galaxy with similar orbital properties as the Arcturus stream can produce three kinematic over-densities. One of the groups was linked to the Arcturus stream and investigated further through a detailed elemental abundances analysis in \citet{_zenoviene14}. They found that the average metallicity of the stream is $\rm [Fe/H]\simeq-0.42$ and that its stars are about 8-12\,Gyr old, which is consistent with the properties of the thick disk. Considering the results from \citet{_helmi06}, the \citet{_zenoviene14} study supported the merger origin for the stream. On the other hand, a comparison with another detailed elemental abundance study by \citet{_ramya12}, who applied different selection criteria for possible Arcturus stream stars, suggested that it is different from the thick disk and the two groups studied in these works are different. 

An alternative approach to search for kinematic over-densities was proposed by \citet{_arifyanto06}, and then followed by \citet{_klement08} and \citet{_zhao14}. Using wavelet transforms they searched for clumps in plane defined by the $\sqrt{U^2+2V^2}$ and $V$ space velocities , where $U$ and $V$ are radial and tangential velocity components, respectively. They detected many kinematic structures in the range $-200\leq V \leq-80 \kms$, including candidates for the Arcturus stream. Following the approach proposed by \citet{_helmi99}, \citet{_klement08} and \citet{_zhao14} studied angular momenta space defined by $\sqrt{L_x^2+L_y^2}$ and $L_z$, the angular momenta components of the stars. They placed the Arcturus stream at $V \simeq -100 \kms$ and $L_z \simeq 1000$ kpc $\kms$. High eccentricity and low metallicity of the low-velocity streams lead to a conclusion that the Arcturus has a merger debris origin.  

The above papers provide evidences that the Arcturus stream and other low-velocity streams can be explained as debris from disrupted satellite galaxies that merged with the Milky Way in the past.

%%%%%%%%%%%%%%%%%%%%%%%%%%%%%%%%%%%%%%%%%%%%%%%%%%%%%%
\subsection{Galactic origin: resonances}%%
%%%%%%%%%%%%%%%%%%%%%%%%%%%%%%%%%%%%%%%%%%%%%%%%%%%%%%

If debris origin for the observed streams is correct, the stars within a stream should have a distinct elemental abundance pattern different from what is observed for the Milky Way disk stars. A detailed chemical analysis of possible members of the Arcturus stream and another group called the AF06 stream, first assigned by \citet{_arifyanto06}, was performed by \citet{_ramya12}. As no unique chemical features were found for either of the groups; their chemical compositions are similar to the background thick disk stars, being metal-poor, alpha-enriched, and have ages between 10-14\,Gyr. This indicates that these structures are likely to have a dynamical origin within the Galaxy.

Another chemical analysis of the Arcturus stream was performed by \citet{_williams09}. The stellar sample was constructed based on results of the N-body simulations, where a satellite was accreted by the Milky Way. It was also here found that the stream stars are chemically inhomogeneous, being similar to the thick disk, and thus, cannot be called a moving group. The authors discuss a possible origin of the group within the Milky Way, being dynamically formed due to Lindblad resonances. At the same time they do not reject the the possibility of a merger origin.

\citet{_bensby14} studied ages and chemical composition of the Galactic disk stars and briefly explored those that potentially could belong to the Arcturus steam ($-115<V<-85 \kms$). They found no chemical signature of a merger event, but rather that a dynamical origin is more probable due to similarities of chemical patterns of the group with the thick disk.

If the discussed spectroscopic studies question an accretion origin for the Arcturus, can the structure be reproduced in a resonant scenario via numerical simulations? Assuming resonances with the Galactic long bar, \citet{_gardner10} simulated a kinematic group which has properties similar to the Arcturus stream. Numerical simulations performed by \citet{_monari13} show similar result; the Galactic long bar can produce a feature that is consistent with the Arcturus stream. 

All these findings lead to a discussion whether the Arcturus stream formed due to resonances or due to a merger event? Simulations assuming either of the hypotheses are able to reproduce a phase-space structure similar to the Arcturus stream. At the same time there is no clear consensus from the detailed elemental abundance studies.

%%%%%%%%%%%%%%%%%%%%%%%%%%%%%%%%%%%%%%%%%%%%%%%%%%%%%
\subsection{Other hypothesis and recent findings} %%%
%%%%%%%%%%%%%%%%%%%%%%%%%%%%%%%%%%%%%%%%%%%%%%%%%%%%%

An alternative opinion on the origin of the Arcturus stream was proposed by \citet{_minchev09}. Assuming the existence of a dynamically unrelaxed population which formed after a merger event \citet{_minchev09} simulated how the distribution of stars in the $U-V$ and $V-\sqrt{U^2+2V^2}$ changes with time. They found ring-like structures that represent a wave with streams appearing almost every $20 \kms$ in $V$, placing the Arcturus stream at $-100 \kms$. Based on the kinematics of the simulated structures the authors state that the Galactic disk was perturbed about 1.9 Gyr ago and match it with the time when the Galactic bar could have formed. 

Another support of the ringing hypothesis came with the {\it Gaia} DR2 \citet{_brown18} release. The analysis of {\it Gaia} DR2 data revealed a rich arch- and ridge-like substructure in the phase-space that is a strong evidence that the Milky Way's disk is far from equilibrium and undergoes phase-mixing \citep{_antoja18, _ramos18, _monari18, _tian18}. This phase-mixing could be a result of external-perturbations due to a passage of the Sagittarius dwarf galaxy \citep[e.g.][]{_antoja18}. At the same time \citet{_hunt18}, \citet{_quillen18} and \citet{_sellwood18} show that the phase-space ridges could be a result of an impact of the Galactic spiral arms, but the simulations do not cover the low-velocity field. The Arcturus stream is one of the arch-like structures seen in \citet{_ramos18}, who performed a deeper study of the substructures with the wavelet analysis. If the Arcturus is a kinematic wave in the Galactic disk, then what triggers the formation of these structures, the bar, spiral arms or a merger event? What is the nature of Arcturus stream and how similar or different is it from the other kinematic structures? 

%%%%%%%%%%%%%%%%%%%%%%%%%%%%%%%%%%%%%%%%%%%%%%%%%%%%%%%%%%%%%%%
\subsection{The origin of the Arcturus stream in this study} %%
%%%%%%%%%%%%%%%%%%%%%%%%%%%%%%%%%%%%%%%%%%%%%%%%%%%%%%%%%%%%%%%

Before discussing possibilities for the nature and origin of the Arcturus stream (g2) we summarise the properties of the kinematic substructures associated with the Arcturus stream that we so far have found:
\begin{itemize}
    \item The rotational velocity of the Arcturus stream is $V\simeq-127\,\kms$, and the vertical component of the angular momentum is $L_z\simeq840$\,kpc\,$\kms$. This is in agreement with the results from \citet{_navarro04} who found the Arcturus stream at $V\simeq-100\,\kms$ and $L_z\simeq$ in the range between 700 and 1100 kpc\,$\kms$, \citet{_klement08} who found it at $V\simeq-120\,\kms$ and $L_z\simeq 1000$\,kpc\,$\kms$, and \citet{_ramya12} who found it at $V\simeq-125\,\kms$ and $L_z\simeq 811$\,kpc\,$\kms$. 
    
    \item The Arcturus stream (g2), as well as the AF06 and KFR08 streams (g1 and g9) are detected not only in the Solar neighbourhood. The structures appear in a larger box defined by regions 01\_12, 22\_12, 22\_32 and 02\_32. The rest of the regions contain significantly less stars as well as distance uncertainties are larger (see Table \ref{_tab_info}). That is why it is not possible to definitely answer if there are no kinematic structures in those regions, or if they are weaker, or if it is a consequence of the properties of the stellar sample. 
    
    \item The Arcturus stream extends to about 2\,kpc vertically from the Galactic plane. At the same time the Hercules stream stars appears to be disappearing at distances greater than $\left|Z\right|\gtrsim1$\,kpc. The streams detected at even lower $V$ velocities extend to even greater distances from the Galactic plane. This is consistent with the results from \citet{_antoja12} who found that the Hercules stream has a lower detection level after $\left|Z\right|\gtrsim0.6$\,kpc. 
    
    \item The Arcturus, KFR08, and AF06 streams are alpha-enhanced in the $\rm [\alpha/Fe]-[Fe/H]$ diagram, and show similarities to the Galactic thick disk. They are clearly different from the Hercules stream which appears to be a mixture of the thin and thick disks. This is in agreement with, for example, \citet{_bensby07, _williams09, _bensby14, _ramya12}.
    
    \item The Arcturus stream has a wide metallicity distribution spanning the interval $\rm -1.2\le [Fe/H]\le0.2$ and peaking at $\simeq-$0.5, which is not different from what has been found in other studies. \citet{_eggen96} found $\rm [Fe/H]\simeq-0.6$, and \citet{_ramya12} found a more metal-poor range for the Arcturus stream, $\rm -1.4\le[Fe/H]\le-0.37$, peaking at $\rm [Fe/H]\simeq-0.7$.
\end{itemize}

Considering the chemical properties of the Arcturus stream (g2) we see that it is not a chemically homogeneous structure. This indicates that the Arcturus stream is not a moving group, similarly to what was found by \citet{_williams09,_ramya12,_bensby14}. Also, the metallicity distributions for the Arcturus, KFR08, and AF06 streams (g2, g1 and g9 respectively) are too wide for them to be called moving groups. At the same time, the mean metallicities are different for each stream, so, they appear to be independent structures, which is different from the conclusion by \citet{_zhao14} that the Arcturus and AF06 streams should be regarded as one structure. It was also shown that a two-sided K-S tests rejected the hypothesis that any of the groups could come from the same population. 

Groups g1, g2 and g9 found in this work appears to be, at least chemically, thick disk structures. Thus, we question the debris origin for the streams, in which case their chemical compositions should be different from the thick disk stars. This is different from the \citet{_liu15} conclusion on a merger origin for the KFR08 stream. It is clearly seen that KFR08 has a wide metallicity distribution and its kinematics and chemistry are consistent with what is seen for the stars in the Galactic thick disk.  

The two remaining possibilities for the origin of the Arcturus stream and the neighbouring KFR08 and AF06 streams (g2, g1 and g9 respectively) are the external-perturbation origin or being due to the resonances with the spiral arms or the Galactic bar. The results we have presented so far are consistent with the results from \citet{_minchev09}, who simulated the velocity distribution in the $V-\sqrt{U^2+2V^2}$ and $U-V$ planes assuming a merger event that perturbed the Galactic disk and caused an ongoing phase-mixing, inducing kinematic over-densities to be placed on the $V$ axis every 20\,$\kms$. This is essentially what we observe observe in this work. Arch-like structures are easily recognisable at Fig.~\ref{_fig_00_2} and similar to what was found in for example \citet{_katz18k,_antoja18} and \citet{_ramos18}. The patterns discussed in \citet{_minchev09} are observed at the $V-\sqrt{U^2+2V^2}$ and action space, the arches observed in the $U-V$ plane become clumps. These clumps and arches are observed as lines in the angular momentum space. Within the uncertainties these features (arches, clumps and lines) show up 20-30 $\kms$ apart in the $V$ velocity component. According to \citet{_minchev09}, this could be due to on-going mixing in the disk after a merger event. The fact that the low-velocity groups extend to higher $\left|Z\right|$ than the bar-originated structures inclines us to conclude on a external-perturbation (phase-mixing) origin for the Arcturus stream and its neighbour KFR08 and AF06 streams is a possible scenario.

%===================================================================
%===================================================================
\section{Summary}\label{_sec_the_end}
%===================================================================
%===================================================================

In order to resolve the nature of the Arcturus stream we analysed the velocity and angular momenta distributions of the {\it Gaia} DR2 stars at different Galactocentric radii. The analysis revealed the following: 

\begin{itemize}
\item The analysis of four spaces defined by velocity, angular momenta and action components in 65 smaller volumes allowed to detect the previously well-studied Sirius, Pleiades, Hyades and Hercules streams; a few low-velocity structures that are associated with the AF06, Arcturus and KFR08 streams; many unknown clumps that might be a part of larger streams.

\item The picture observed in the velocity space is consistent with the results from \citet{_minchev09}. Their model predicts kinematic structures to be placed every $20-30 \kms$ in $V$. This is similar to what was observed with the $Gaia$ DR2 sample: starting with the Sirius at 0 $\kms$ and ending with the KFR08 stream at $\simeq-160 \kms$ we do observe kinematic structures every $V\simeq20-30\kms$ (taking into account velocity uncertainties and sizes of the structures).

\item The arches observed in the $U-V$ plane are observed as clumps in the $V-\sqrt{U^2+2V^2}$ and action space and lines in the angular momentum space. \citet{_minchev09} as well as \citet{_antoja18} link these arches to the on-going phase-mixing in the Milky Way's disk due to a strong disk perturbation likely by a merger event.
    
\item The low-velocity streams are observed further more from the disk at higher $\left|Z\right|$ compared to the Hercules stream which is currently considered as a structure caused by the Galactic bar \citep[e.g.][]{_antoja14, _perez17}. The KFR08 stream, which has the lowest values of $V$ and $L_z$ extends at least 2 kpc further from the Galactic disk, while the Hercules is located closer to the Galactic disk $\left|Z\right|<0.7$ kpc. 

\item The majority of stars from the sample analysed in this work are within 1 kpc in $\left|Z\right|$ and according to \citep{_monari13} the Galactic bar can influence stellar motion up to $\left|Z\right|\simeq1$ kpc in the thin disk and up to $\left|Z\right|\simeq2$ kpc. 
    
\item The lower velocity groups are present mainly in the nearby regions. It is consistent with \citet{_ramos18}, where one of the discovered arches, that the authors associate with the Arcturus group, is located mainly in the Solar neighbourhood and within the Solar circle. At the same time this result can be a consequence of the fact that nearby regions contain more stars and distance uncertainties are smaller. 

\item \citet{_ramos18} discuss the negative gradients of the rotational velocity of the structures with the Galactocentric radii. This gradient should be positive for the Cartesian velocities $V$. We do observe similar gradients in the $U-V$ plane. 
    
\item The analysis of the chemical abundances of stars that are members of the groups taken from the APOGEE and GALAH spectroscopic surveys confirmed that the AF06, Arcturus, and KFR08 resemble the thick disk chemical patterns. The groups cover wider metallicity ranges compared to the Hercules. The latest one appears to be a mixture of the thin and the thick disk stars, which is in agreement with the literature \citep[e.g.][]{_bensby14, _ramya16}. Estimated median metallicity of the Hercules in this work is [Fe/H]\,$\simeq 0.0$ , for example, \citet{_ramya16} obtained [Fe/H]\,$\simeq 0.15$.

\item The two-sided K-S test performed for different combinations of the the groups (Hercules, AF06, Arcturus, KFR08) rejected the possibility for all of them to be drawn from the same distribution.

\end{itemize}

The the Arcturus, KFR08 and AF06 are kinematic structures that have rotational $V$ velocities separated with a fixed step, that they extend farther from the Galactic plane compared to other over-densities such as the Hercules stream, and that they have chemical compositions consistent with the properties of the Galactic thick disk, points towards an origin for the structures related to the ongoing kinematic mixing or ringing in the disk, as was suggested in \citep{_minchev09}. The recently discovered ridges and arches in the phase-space \citep{_antoja18, _ramos18} with the {\it Gaia} DR2 including the Arcturus arch is another evidence that the low-velocity kinematic structures such as the Arcturus stream, could be a result of the external-perturbation process and were formed inside the Milky Way. But the question of origin of these phase-space warps is unresolved. 

Though there has been done a lot of efforts to understand the nature of these phase-space waves with the {\it Gaia} DR2 data, we are still far from an unambiguous answer. Is it a merger origin as was originally proposed by \citep{_minchev09}, or is it an influence of the spiral arms as suggested by, for example, \citet{_quillen18, _hunt18} and \citet{_sellwood18}? Numerical simulations together with the chemical abundances from spectroscopic surveys like the {\it Gaia}-ESO survey \citep{gilmore2012}, WEAVE \citep{dalton2014}, and 4MOST \citep{_dejong2019}, in combination with upcoming {\it Gaia} data releases will broaden the opportunities for us to better understand the formation of the phase-space warps and might give a definite answer about the origin of kinematic structures like the Arcturus stream. Current observational evidence is however pointing towards a phase-space mixing origin.   

%===================================================================
%===================================================================
\begin{acknowledgements}
We thank Prof. F.~Murtagh for making available for us the MR software packages and Dr. P.\,J. McMillan for a valuable help when needed. T.B. was funded by the project grant "The New Milky Way" from the Knut and Alice Wallenberg foundation.
\end{acknowledgements}
%===================================================================
%===================================================================

\bibliographystyle{aa}
\bibliography{references}

\begin{thebibliography}{71}
\expandafter\ifx\csname natexlab\endcsname\relax\def\natexlab#1{#1}\fi

\bibitem[{{Antoja} {et~al.}(2008){Antoja}, {Figueras}, {Fern{\'a}ndez}, \&
  {Torra}}]{_antoja08}
{Antoja}, T., {Figueras}, F., {Fern{\'a}ndez}, D., \& {Torra}, J. 2008, \aap,
  490, 135

\bibitem[{{Antoja} {et~al.}(2012){Antoja}, {Helmi}, {Bienayme},
  {Bland-Hawthorn}, {Famaey}, {Freeman}, {Gibson}, {Gilmore}, {Grebel},
  {Minchev}, {Munari}, {Navarro}, {Parker}, {Reid}, {Seabroke}, {Siebert},
  {Siviero}, {Steinmetz}, {Williams}, {Wyse}, \& {Zwitter}}]{_antoja12}
{Antoja}, T., {Helmi}, A., {Bienayme}, O., {et~al.} 2012, \mnras, 426, L1

\bibitem[{{Antoja} {et~al.}(2014){Antoja}, {Helmi}, {Dehnen}, {Bienaym{\'e}},
  {Bland-Hawthorn}, {Famaey}, {Freeman}, {Gibson}, {Gilmore}, {Grebel},
  {Kordopatis}, {Kunder}, {Minchev}, {Munari}, {Navarro}, {Parker}, {Reid},
  {Seabroke}, {Siebert}, {Steinmetz}, {Watson}, {Wyse}, \&
  {Zwitter}}]{_antoja14}
{Antoja}, T., {Helmi}, A., {Dehnen}, W., {et~al.} 2014, \aap, 563, A60

\bibitem[{{Antoja} {et~al.}(2018){Antoja}, {Helmi}, {Romero-G{\'o}mez}, {Katz},
  {Babusiaux}, {Drimmel}, {Evans}, {Figueras}, {Poggio}, {Reyl{\'e}}, {Robin},
  {Seabroke}, \& {Soubiran}}]{_antoja18}
{Antoja}, T., {Helmi}, A., {Romero-G{\'o}mez}, M., {et~al.} 2018, \nat, 561,
  360

\bibitem[{{Arenou} {et~al.}(2018){Arenou}, {Luri}, {Babusiaux}, {Fabricius},
  {Helmi}, {Muraveva}, {Robin}, {Spoto}, {Vallenari}, {Antoja},
  {Cantat-Gaudin}, {Jordi}, {Leclerc}, {Reyl{\'e}}, {Romero-G{\'o}mez}, {Shih},
  {Soria}, {Barache}, {Bossini}, {Bragaglia}, {Breddels}, {Fabrizio},
  {Lambert}, {Marrese}, {Massari}, {Moitinho}, {Robichon}, {Ruiz-Dern},
  {Sordo}, {Veljanoski}, {Eyer}, {Jasniewicz}, {Pancino}, {Soubiran}, {Spagna},
  {Tanga}, {Turon}, \& {Zurbach}}]{_arenou18}
{Arenou}, F., {Luri}, X., {Babusiaux}, C., {et~al.} 2018, \aap, 616, A17

\bibitem[{{Arifyanto} \& {Fuchs}(2006)}]{_arifyanto06}
{Arifyanto}, M.~I. \& {Fuchs}, B. 2006, \aap, 449, 533

\bibitem[{{Bailer-Jones}(2015)}]{_bailerjones15}
{Bailer-Jones}, C. A.~L. 2015, Publications of the Astronomical Society of the
  Pacific, 127, 994

\bibitem[{{Bensby} {et~al.}(2014){Bensby}, {Feltzing}, \& {Oey}}]{_bensby14}
{Bensby}, T., {Feltzing}, S., \& {Oey}, M.~S. 2014, \aap, 562, A71

\bibitem[{{Bensby} {et~al.}(2007){Bensby}, {Oey}, {Feltzing}, \&
  {Gustafsson}}]{_bensby07}
{Bensby}, T., {Oey}, M.~S., {Feltzing}, S., \& {Gustafsson}, B. 2007, \apjl,
  655, L89

\bibitem[{{Bobylev} \& {Bajkova}(2016)}]{_bobylev16}
{Bobylev}, V.~V. \& {Bajkova}, A.~T. 2016, Astronomy Letters, 42, 90

\bibitem[{{Bovy}(2015)}]{_bovy15}
{Bovy}, J. 2015, The Astrophysical Journal Supplement Series, 216, 29

\bibitem[{{Bovy}(2016)}]{_bovy16}
{Bovy}, J. 2016, \apj, 817, 49

\bibitem[{{Buder} {et~al.}(2018){Buder}, {Asplund}, {Duong}, {Kos}, {Lind},
  {Ness}, {Sharma}, {Bland-Hawthorn}, {Casey}, {De Silva}, {D'Orazi},
  {Freeman}, {Lewis}, {Lin}, {Martell}, {Schlesinger}, {Simpson}, {Zucker},
  {Zwitter}, {Amarsi}, {Anguiano}, {Carollo}, {Casagrande}, {{\v{C}}otar},
  {Cottrell}, {Da Costa}, {Gao}, {Hayden}, {Horner}, {Ireland}, {Kafle},
  {Munari}, {Nataf}, {Nordlander}, {Stello}, {Ting}, {Traven}, {Watson},
  {Wittenmyer}, {Wyse}, {Yong}, {Zinn}, \& {{\v{Z}}erjal}}]{_buder18}
{Buder}, S., {Asplund}, M., {Duong}, L., {et~al.} 2018, \mnras, 478, 4513

\bibitem[{{Chakrabarty}(2007)}]{_chakrabarty07}
{Chakrabarty}, D. 2007, \aap, 467, 145

\bibitem[{{Dalton} {et~al.}(2014){Dalton}, {Trager}, {Abrams}, {Bonifacio},
  {L{\'o}pez Aguerri}, {Middleton}, {Benn}, {Dee}, {Say{\`e}de}, {Lewis},
  {Pragt}, {Pico}, {Walton}, {Rey}, {Allende Prieto}, {Pe{\~n}ate}, {Lhome},
  {Ag{\'o}cs}, {Alonso}, {Terrett}, {Brock}, {Gilbert}, {Ridings}, {Guinouard},
  {Verheijen}, {Tosh}, {Rogers}, {Steele}, {Stuik}, {Tromp}, {Jasko}, {Kragt},
  {Lesman}, {Mottram}, {Bates}, {Gribbin}, {Rodriguez}, {Delgado}, {Martin},
  {Cano}, {Navarro}, {Irwin}, {Lewis}, {Gonzalez Solares}, {O'Mahony},
  {Bianco}, {Zurita}, {ter Horst}, {Molinari}, {Lodi}, {Guerra}, {Vallenari},
  \& {Baruffolo}}]{dalton2014}
{Dalton}, G., {Trager}, S., {Abrams}, D.~C., {et~al.} 2014, in SPIE Conf. Ser.,
  Vol. 9147, 91470L

\bibitem[{{de Jong} {et~al.}(2019){de Jong}, {Agertz}, {Berbel}, {Aird},
  {Alexander}, {Amarsi}, {Anders}, {Andrae}, {Ansarinejad}, {Ansorge},
  {Antilogus}, {Anwand -Heerwart}, {Arentsen}, {Arnadottir}, {Asplund},
  {Auger}, {Azais}, {Baade}, {Baker}, {Baker}, {Balbinot}, {Baldry}, {Banerji},
  {Barden}, {Barklem}, {Barth{\'e}l{\'e}my-Mazot}, {Battistini}, {Bauer},
  {Bell}, {Bellido-Tirado}, {Bellstedt}, {Belokurov}, {Bensby}, {Bergemann},
  {Bestenlehner}, {Bielby}, {Bilicki}, {Blake}, {Bland-Hawthorn}, {Boeche},
  {Boland}, {Boller}, {Bongard}, {Bongiorno}, {Bonifacio}, {Boudon}, {Brooks},
  {Brown}, {Brown}, {Br{\"u}ggen}, {Brynnel}, {Brzeski}, {Buchert},
  {Buschkamp}, {Caffau}, {Caillier}, {Carrick}, {Casagrande}, {Case}, {Casey},
  {Cesarini}, {Cescutti}, {Chapuis}, {Chiappini}, {Childress}, {Christlieb},
  {Church}, {Cioni}, {Cluver}, {Colless}, {Collett}, {Comparat}, {Cooper},
  {Couch}, {Courbin}, {Croom}, {Croton}, {Daguis{\'e}}, {Dalton}, {Davies},
  {Davis}, {de Laverny}, {Deason}, {Dionies}, {Disseau}, {Doel}, {D{\"o}scher},
  {Driver}, {Dwelly}, {Eckert}, {Edge}, {Edvardsson}, {Youssoufi}, {Elhaddad},
  {Enke}, {Erfanianfar}, {Farrell}, {Fechner}, {Feiz}, {Feltzing}, {Ferreras},
  {Feuerstein}, {Feuillet}, {Finoguenov}, {Ford}, {Fotopoulou}, {Fouesneau},
  {Frenk}, {Frey}, {Gaessler}, {Geier}, {Fusillo}, {Gerhard}, {Giannantonio},
  {Giannone}, {Gibson}, {Gillingham}, {Gonz{\'a}lez-Fern{\'a}ndez},
  {Gonzalez-Solares}, {Gottloeber}, {Gould}, {Grebel}, {Gueguen}, {Guiglion},
  {Haehnelt}, {Hahn}, {Hansen}, {Hartman}, {Hauptner}, {Hawkins}, {Haynes},
  {Haynes}, {Heiter}, {Helmi}, {Aguayo}, {Hewett}, {Hinton}, {Hobbs}, {Hoenig},
  {Hofman}, {Hook}, {Hopgood}, {Hopkins}, {Hourihane}, {Howes}, {Howlett},
  {Huet}, {Irwin}, {Iwert}, {Jablonka}, {Jahn}, {Jahnke}, {Jarno}, {Jin},
  {Jofre}, {Johl}, {Jones}, {J{\"o}nsson}, {Jordan}, {Karovicova}, {Khalatyan},
  {Kelz}, {Kennicutt}, {King}, {Kitaura}, {Klar}, {Klauser}, {Kneib}, {Koch},
  {Koposov}, {Kordopatis}, {Korn}, {Kosmalski}, {Kotak}, {Kovalev}, {Kreckel},
  {Kripak}, {Krumpe}, {Kuijken}, {Kunder}, {Kushniruk}, {Lam}, {Lamer},
  {Laurent}, {Lawrence}, {Lehmitz}, {Lemasle}, {Lewis}, {Li}, {Lidman}, {Lind},
  {Liske}, {Lizon}, {Loveday}, {Ludwig}, {McDermid}, {Maguire}, {Mainieri},
  {Mali}, {Mandel}, {Mandel}, {Mannering}, {Martell}, {Martinez Delgado},
  {Matijevic}, {McGregor}, {McMahon}, {McMillan}, {Mena}, {Merloni}, {Meyer},
  {Michel}, {Micheva}, {Migniau}, {Minchev}, {Monari}, {Muller}, {Murphy},
  {Muthukrishna}, {Nandra}, {Navarro}, {Ness}, {Nichani}, {Nichol}, {Nicklas},
  {Niederhofer}, {Norberg}, {Obreschkow}, {Oliver}, {Owers}, {Pai},
  {Pankratow}, {Parkinson}, {Paschke}, {Paterson}, {Pecontal}, {Parry},
  {Phillips}, {Pillepich}, {Pinard}, {Pirard}, {Piskunov}, {Plank},
  {Pl{\"u}schke}, {Pons}, {Popesso}, {Power}, {Pragt}, {Pramskiy}, {Pryer},
  {Quattri}, {Queiroz}, {Quirrenbach}, {Rahurkar}, {Raichoor}, {Ramstedt},
  {Rau}, {Recio-Blanco}, {Reiss}, {Renaud}, {Revaz}, {Rhode}, {Richard},
  {Richter}, {Rix}, {Robotham}, {Roelfsema}, {Romaniello}, {Rosario},
  {Rothmaier}, {Roukema}, {Ruchti}, {Rupprecht}, {Rybizki}, {Ryde}, {Saar},
  {Sadler}, {Sahl{\'e}n}, {Salvato}, {Sassolas}, {Saunders}, {Saviauk},
  {Sbordone}, {Schmidt}, {Schnurr}, {Scholz}, {Schwope}, {Seifert}, {Shanks},
  {Sheinis}, {Sivov}, {Sk{\'u}lad{\'o}ttir}, {Smartt}, {Smedley}, {Smith},
  {Smith}, {Sorce}, {Spitler}, {Starkenburg}, {Steinmetz}, {Stilz}, {Storm},
  {Sullivan}, {Sutherland}, {Swann}, {Tamone}, {Taylor}, {Teillon}, {Tempel},
  {ter Horst}, {Thi}, {Tolstoy}, {Trager}, {Traven}, {Tremblay}, {Tresse},
  {Valentini}, {van de Weygaert}, {van den Ancker}, {Veljanoski}, {Venkatesan},
  {Wagner}, {Wagner}, {Walcher}, {Waller}, {Walton}, {Wang}, {Winkler},
  {Wisotzki}, {Worley}, {Worseck}, {Xiang}, {Xu}, {Yong}, {Zhao}, {Zheng},
  {Zscheyge}, \& {Zucker}}]{_dejong2019}
{de Jong}, R.~S., {Agertz}, O., {Berbel}, A.~A., {et~al.} 2019, The Messenger,
  175, 3

\bibitem[{{De Silva} {et~al.}(2007){De Silva}, {Freeman}, {Bland-Hawthorn},
  {Asplund}, \& {Bessell}}]{_desilva07}
{De Silva}, G.~M., {Freeman}, K.~C., {Bland-Hawthorn}, J., {Asplund}, M., \&
  {Bessell}, M.~S. 2007, \aj, 133, 694

\bibitem[{{Dehnen}(1998)}]{_dehnen98}
{Dehnen}, W. 1998, \aj, 115, 2384

\bibitem[{{Dehnen}(2000)}]{_dehnen00}
{Dehnen}, W. 2000, \aj, 119, 800

\bibitem[{{Dekker}(1976)}]{_dekker76}
{Dekker}, E. 1976, \physrep, 24, 315

\bibitem[{{Eggen}(1971)}]{_eggen71}
{Eggen}, O.~J. 1971, \pasp, 83, 762

\bibitem[{{Eggen}(1996)}]{_eggen96}
{Eggen}, O.~J. 1996, \aj, 112, 1595

\bibitem[{{Eggen}(1998)}]{_eggen98}
{Eggen}, O.~J. 1998, \aj, 115, 2397

\bibitem[{{Famaey} {et~al.}(2005){Famaey}, {Jorissen}, {Luri}, {Mayor}, {Udry},
  {Dejonghe}, \& {Turon}}]{_famaey05}
{Famaey}, B., {Jorissen}, A., {Luri}, X., {et~al.} 2005, \aap, 430, 165

\bibitem[{{Feltzing} \& {Holmberg}(2000)}]{_feltzing00}
{Feltzing}, S. \& {Holmberg}, J. 2000, \aap, 357, 153

\bibitem[{{Gaia Collaboration} {et~al.}(2018{\natexlab{a}}){Gaia
  Collaboration}, {Brown}, {Vallenari}, {Prusti}, {de Bruijne}, {Babusiaux},
  {Bailer-Jones}, {Biermann}, {Evans}, {Eyer}, {Jansen}, {Jordi}, {Klioner},
  {Lammers}, {Lindegren}, {Luri}, {Mignard}, {Panem}, {Pourbaix}, {Randich},
  {Sartoretti}, {Siddiqui}, {Soubiran}, {van Leeuwen}, {Walton}, {Arenou},
  {Bastian}, {Cropper}, {Drimmel}, {Katz}, {Lattanzi}, {Bakker}, {Cacciari},
  {Casta{\~n}eda}, {Chaoul}, {Cheek}, {De Angeli}, {Fabricius}, {Guerra},
  {Holl}, {Masana}, {Messineo}, {Mowlavi}, {Nienartowicz}, {Panuzzo},
  {Portell}, {Riello}, {Seabroke}, {Tanga}, {Th{\'e}venin}, {Gracia-Abril},
  {Comoretto}, {Garcia-Reinaldos}, {Teyssier}, {Altmann}, {Andrae}, {Audard},
  {Bellas-Velidis}, {Benson}, {Berthier}, {Blomme}, {Burgess}, {Busso},
  {Carry}, {Cellino}, {Clementini}, {Clotet}, {Creevey}, {Davidson}, {De
  Ridder}, {Delchambre}, {Dell'Oro}, {Ducourant},
  {Fern{\'a}ndez-Hern{\'a}ndez}, {Fouesneau}, {Fr{\'e}mat}, {Galluccio},
  {Garc{\'\i}a-Torres}, {Gonz{\'a}lez-N{\'u}{\~n}ez}, {Gonz{\'a}lez-Vidal},
  {Gosset}, {Guy}, {Halbwachs}, {Hambly}, {Harrison}, {Hern{\'a}ndez},
  {Hestroffer}, {Hodgkin}, {Hutton}, {Jasniewicz}, {Jean-Antoine-Piccolo},
  {Jordan}, {Korn}, {Krone-Martins}, {Lanzafame}, {Lebzelter}, {L{\"o}ffler},
  {Manteiga}, {Marrese}, {Mart{\'\i}n-Fleitas}, {Moitinho}, {Mora}, {Muinonen},
  {Osinde}, {Pancino}, {Pauwels}, {Petit}, {Recio-Blanco}, {Richards},
  {Rimoldini}, {Robin}, {Sarro}, {Siopis}, {Smith}, {Sozzetti}, {S{\"u}veges},
  {Torra}, {van Reeven}, {Abbas}, {Abreu Aramburu}, {Accart}, {Aerts},
  {Altavilla}, {{\'A}lvarez}, {Alvarez}, {Alves}, {Anderson}, {Andrei},
  {Anglada Varela}, {Antiche}, {Antoja}, {Arcay}, {Astraatmadja}, {Bach},
  {Baker}, {Balaguer-N{\'u}{\~n}ez}, {Balm}, {Barache}, {Barata}, {Barbato},
  {Barblan}, {Barklem}, {Barrado}, {Barros}, {Barstow}, {Bartholom{\'e}
  Mu{\~n}oz}, {Bassilana}, {Becciani}, {Bellazzini}, {Berihuete}, {Bertone},
  {Bianchi}, {Bienaym{\'e}}, {Blanco-Cuaresma}, {Boch}, {Boeche}, {Bombrun},
  {Borrachero}, {Bossini}, {Bouquillon}, {Bourda}, {Bragaglia}, {Bramante},
  {Breddels}, {Bressan}, {Brouillet}, {Br{\"u}semeister}, {Brugaletta},
  {Bucciarelli}, {Burlacu}, {Busonero}, {Butkevich}, {Buzzi}, {Caffau},
  {Cancelliere}, {Cannizzaro}, {Cantat-Gaudin}, {Carballo}, {Carlucci},
  {Carrasco}, {Casamiquela}, {Castellani}, {Castro-Ginard}, {Charlot},
  {Chemin}, {Chiavassa}, {Cocozza}, {Costigan}, {Cowell}, {Crifo}, {Crosta},
  {Crowley}, {Cuypers}, {Dafonte}, {Damerdji}, {Dapergolas}, {David}, {David},
  {de Laverny}, {De Luise}, {De March}, {de Martino}, {de Souza}, {de Torres},
  {Debosscher}, {del Pozo}, {Delbo}, {Delgado}, {Delgado}, {Di Matteo},
  {Diakite}, {Diener}, {Distefano}, {Dolding}, {Drazinos}, {Dur{\'a}n},
  {Edvardsson}, {Enke}, {Eriksson}, {Esquej}, {Eynard Bontemps}, {Fabre},
  {Fabrizio}, {Faigler}, {Falc{\~a}o}, {Farr{\`a}s Casas}, {Federici},
  {Fedorets}, {Fernique}, {Figueras}, {Filippi}, {Findeisen}, {Fonti},
  {Fraile}, {Fraser}, {Fr{\'e}zouls}, {Gai}, {Galleti}, {Garabato},
  {Garc{\'\i}a-Sedano}, {Garofalo}, {Garralda}, {Gavel}, {Gavras}, {Gerssen},
  {Geyer}, {Giacobbe}, {Gilmore}, {Girona}, {Giuffrida}, {Glass}, {Gomes},
  {Granvik}, {Gueguen}, {Guerrier}, {Guiraud}, {Guti{\'e}rrez-S{\'a}nchez},
  {Haigron}, {Hatzidimitriou}, {Hauser}, {Haywood}, {Heiter}, {Helmi}, {Heu},
  {Hilger}, {Hobbs}, {Hofmann}, {Holland}, {Huckle}, {Hypki}, {Icardi},
  {Jan{\ss}en}, {Jevardat de Fombelle}, {Jonker}, {Juh{\'a}sz}, {Julbe},
  {Karampelas}, {Kewley}, {Klar}, {Kochoska}, {Kohley}, {Kolenberg},
  {Kontizas}, {Kontizas}, {Koposov}, {Kordopatis}, {Kostrzewa-Rutkowska},
  {Koubsky}, {Lambert}, {Lanza}, {Lasne}, {Lavigne}, {Le Fustec}, {Le
  Poncin-Lafitte}, {Lebreton}, {Leccia}, {Leclerc}, {Lecoeur-Taibi},
  {Lenhardt}, {Leroux}, {Liao}, {Licata}, {Lindstr{\o}m}, {Lister}, {Livanou},
  {Lobel}, {L{\'o}pez}, {Managau}, {Mann}, {Mantelet}, {Marchal}, {Marchant},
  {Marconi}, {Marinoni}, {Marschalk{\'o}}, {Marshall}, {Martino}, {Marton},
  {Mary}, {Massari}, {Matijevi{\v{c}}}, {Mazeh}, {McMillan}, {Messina},
  {Michalik}, {Millar}, {Molina}, {Molinaro}, {Moln{\'a}r}, {Montegriffo},
  {Mor}, {Morbidelli}, {Morel}, {Morris}, {Mulone}, {Muraveva}, {Musella},
  {Nelemans}, {Nicastro}, {Noval}, {O'Mullane}, {Ord{\'e}novic},
  {Ord{\'o}{\~n}ez-Blanco}, {Osborne}, {Pagani}, {Pagano}, {Pailler},
  {Palacin}, {Palaversa}, {Panahi}, {Pawlak}, {Piersimoni}, {Pineau}, {Plachy},
  {Plum}, {Poggio}, {Poujoulet}, {Pr{\v{s}}a}, {Pulone}, {Racero}, {Ragaini},
  {Rambaux}, {Ramos-Lerate}, {Regibo}, {Reyl{\'e}}, {Riclet}, {Ripepi}, {Riva},
  {Rivard}, {Rixon}, {Roegiers}, {Roelens}, {Romero-G{\'o}mez}, {Rowell},
  {Royer}, {Ruiz-Dern}, {Sadowski}, {Sagrist{\`a} Sell{\'e}s}, {Sahlmann},
  {Salgado}, {Salguero}, {Sanna}, {Santana-Ros}, {Sarasso}, {Savietto},
  {Schultheis}, {Sciacca}, {Segol}, {Segovia}, {S{\'e}gransan}, {Shih},
  {Siltala}, {Silva}, {Smart}, {Smith}, {Solano}, {Solitro}, {Sordo}, {Soria
  Nieto}, {Souchay}, {Spagna}, {Spoto}, {Stampa}, {Steele},
  {Steidelm{\"u}ller}, {Stephenson}, {Stoev}, {Suess}, {Surdej}, {Szabados},
  {Szegedi-Elek}, {Tapiador}, {Taris}, {Tauran}, {Taylor}, {Teixeira},
  {Terrett}, {Teyssand ier}, {Thuillot}, {Titarenko}, {Torra Clotet}, {Turon},
  {Ulla}, {Utrilla}, {Uzzi}, {Vaillant}, {Valentini}, {Valette}, {van Elteren},
  {Van Hemelryck}, {van Leeuwen}, {Vaschetto}, {Vecchiato}, {Veljanoski},
  {Viala}, {Vicente}, {Vogt}, {von Essen}, {Voss}, {Votruba}, {Voutsinas},
  {Walmsley}, {Weiler}, {Wertz}, {Wevers}, {Wyrzykowski}, {Yoldas},
  {{\v{Z}}erjal}, {Ziaeepour}, {Zorec}, {Zschocke}, {Zucker}, {Zurbach}, \&
  {Zwitter}}]{_brown18}
{Gaia Collaboration}, {Brown}, A.~G.~A., {Vallenari}, A., {et~al.}
  2018{\natexlab{a}}, \aap, 616, A1

\bibitem[{{Gaia Collaboration} {et~al.}(2018{\natexlab{b}}){Gaia
  Collaboration}, {Katz}, {Antoja}, {Romero-G{\'o}mez}, {Drimmel}, {Reyl{\'e}},
  {Seabroke}, {Soubiran}, {Babusiaux}, {Di Matteo}, {Figueras}, {Poggio},
  {Robin}, {Evans}, {Brown}, {Vallenari}, {Prusti}, {de Bruijne},
  {Bailer-Jones}, {Biermann}, {Eyer}, {Jansen}, {Jordi}, {Klioner}, {Lammers},
  {Lindegren}, {Luri}, {Mignard}, {Panem}, {Pourbaix}, {Randich}, {Sartoretti},
  {Siddiqui}, {van Leeuwen}, {Walton}, {Arenou}, {Bastian}, {Cropper},
  {Lattanzi}, {Bakker}, {Cacciari}, {Casta n}, {Chaoul}, {Cheek}, {De Angeli},
  {Fabricius}, {Guerra}, {Holl}, {Masana}, {Messineo}, {Mowlavi},
  {Nienartowicz}, {Panuzzo}, {Portell}, {Riello}, {Tanga}, {Th{\'e}venin},
  {Gracia-Abril}, {Comoretto}, {Garcia-Reinaldos}, {Teyssier}, {Altmann},
  {Andrae}, {Audard}, {Bellas-Velidis}, {Benson}, {Berthier}, {Blomme},
  {Burgess}, {Busso}, {Carry}, {Cellino}, {Clementini}, {Clotet}, {Creevey},
  {Davidson}, {De Ridder}, {Delchambre}, {Dell'Oro}, {Ducourant},
  {Fern{\'a}ndez-Hern{\'a}ndez}, {Fouesneau}, {Fr{\'e}mat}, {Galluccio},
  {Garc{\'\i}a-Torres}, {Gonz{\'a}lez-N{\'u}{\~n}ez}, {Gonz{\'a}lez-Vidal},
  {Gosset}, {Guy}, {Halbwachs}, {Hambly}, {Harrison}, {Hern{\'a}ndez},
  {Hestroffer}, {Hodgkin}, {Hutton}, {Jasniewicz}, {Jean-Antoine-Piccolo},
  {Jordan}, {Korn}, {Krone-Martins}, {Lanzafame}, {Lebzelter}, {L{\"o}ffler},
  {Manteiga}, {Marrese}, {Mart{\'\i}n-Fleitas}, {Moitinho}, {Mora}, {Muinonen},
  {Osinde}, {Pancino}, {Pauwels}, {Petit}, {Recio-Blanco}, {Richards},
  {Rimoldini}, {Sarro}, {Siopis}, {Smith}, {Sozzetti}, {S{\"u}veges}, {Torra},
  {van Reeven}, {Abbas}, {Abreu Aramburu}, {Accart}, {Aerts}, {Altavilla},
  {{\'A}lvarez}, {Alvarez}, {Alves}, {Anderson}, {Andrei}, {Anglada Varela},
  {Antiche}, {Arcay}, {Astraatmadja}, {Bach}, {Baker},
  {Balaguer-N{\'u}{\~n}ez}, {Balm}, {Barache}, {Barata}, {Barbato}, {Barblan},
  {Barklem}, {Barrado}, {Barros}, {Barstow}, {Bartholom{\'e} Mu{\~n}oz},
  {Bassilana}, {Becciani}, {Bellazzini}, {Berihuete}, {Bertone}, {Bianchi},
  {Bienaym{\'e}}, {Blanco-Cuaresma}, {Boch}, {Boeche}, {Bombrun}, {Borrachero},
  {Bossini}, {Bouquillon}, {Bourda}, {Bragaglia}, {Bramante}, {Breddels},
  {Bressan}, {Brouillet}, {Br{\"u}semeister}, {Brugaletta}, {Bucciarelli},
  {Burlacu}, {Busonero}, {Butkevich}, {Buzzi}, {Caffau}, {Cancelliere},
  {Cannizzaro}, {Cantat-Gaudin}, {Carballo}, {Carlucci}, {Carrasco},
  {Casamiquela}, {Castellani}, {Castro-Ginard}, {Charlot}, {Chemin},
  {Chiavassa}, {Cocozza}, {Costigan}, {Cowell}, {Crifo}, {Crosta}, {Crowley},
  {Cuypers}, {Dafonte}, {Damerdji}, {Dapergolas}, {David}, {David}, {de
  Laverny}, {De Luise}, {De March}, {de Souza}, {de Torres}, {Debosscher}, {del
  Pozo}, {Delbo}, {Delgado}, {Delgado}, {Diakite}, {Diener}, {Distefano},
  {Dolding}, {Drazinos}, {Dur{\'a}n}, {Edvardsson}, {Enke}, {Eriksson},
  {Esquej}, {Eynard Bontemps}, {Fabre}, {Fabrizio}, {Faigler}, {Falc a},
  {Farr{\`a}s Casas}, {Federici}, {Fedorets}, {Fernique}, {Filippi},
  {Findeisen}, {Fonti}, {Fraile}, {Fraser}, {Fr{\'e}zouls}, {Gai}, {Galleti},
  {Garabato}, {Garc{\'\i}a-Sedano}, {Garofalo}, {Garralda}, {Gavel}, {Gavras},
  {Gerssen}, {Geyer}, {Giacobbe}, {Gilmore}, {Girona}, {Giuffrida}, {Glass},
  {Gomes}, {Granvik}, {Gueguen}, {Guerrier}, {Guiraud}, {Guti{\'e}}, {Haigron},
  {Hatzidimitriou}, {Hauser}, {Haywood}, {Heiter}, {Helmi}, {Heu}, {Hilger},
  {Hobbs}, {Hofmann}, {Holland }, {Huckle}, {Hypki}, {Icardi}, {Jan{\ss}en},
  {Jevardat de Fombelle}, {Jonker}, {Juh{\'a}sz}, {Julbe}, {Karampelas},
  {Kewley}, {Klar}, {Kochoska}, {Kohley}, {Kolenberg}, {Kontizas}, {Kontizas},
  {Koposov}, {Kordopatis}, {Kostrzewa-Rutkowska}, {Koubsky}, {Lambert},
  {Lanza}, {Lasne}, {Lavigne}, {Le Fustec}, {Le Poncin-Lafitte}, {Lebreton},
  {Leccia}, {Leclerc}, {Lecoeur-Taibi}, {Lenhardt}, {Leroux}, {Liao}, {Licata},
  {Lindstr{\o}m}, {Lister}, {Livanou}, {Lobel}, {L{\'o}pez}, {Managau}, {Mann},
  {Mantelet}, {Marchal}, {Marchant}, {Marconi}, {Marinoni}, {Marschalk{\'o}},
  {Marshall}, {Martino}, {Marton}, {Mary}, {Massari}, {Matijevi{\v{c}}},
  {Mazeh}, {McMillan}, {Messina}, {Michalik}, {Millar}, {Molina}, {Molinaro},
  {Moln{\'a}r}, {Montegriffo}, {Mor}, {Morbidelli}, {Morel}, {Morris},
  {Mulone}, {Muraveva}, {Musella}, {Nelemans}, {Nicastro}, {Noval},
  {O'Mullane}, {Ord{\'e}novic}, {Ord{\'o}{\~n}ez-Blanco}, {Osborne}, {Pagani},
  {Pagano}, {Pailler}, {Palacin}, {Palaversa}, {Panahi}, {Pawlak},
  {Piersimoni}, {Pineau}, {Plachy}, {Plum}, {Poujoulet}, {Pr{\v{s}}a},
  {Pulone}, {Racero}, {Ragaini}, {Rambaux}, {Ramos-Lerate}, {Regibo}, {Riclet},
  {Ripepi}, {Riva}, {Rivard}, {Rixon}, {Roegiers}, {Roelens}, {Rowell},
  {Royer}, {Ruiz-Dern}, {Sadowski}, {Sagrist{\`a} Sell{\'e}s}, {Sahlmann},
  {Salgado}, {Salguero}, {Sanna}, {Santana-Ros}, {Sarasso}, {Savietto},
  {Schultheis}, {Sciacca}, {Segol}, {Segovia}, {S{\'e}gransan}, {Shih},
  {Siltala}, {Silva}, {Smart}, {Smith}, {Solano}, {Solitro}, {Sordo}, {Soria
  Nieto}, {Souchay}, {Spagna}, {Spoto}, {Stampa}, {Steele},
  {Steidelm{\"u}ller}, {Stephenson}, {Stoev}, {Suess}, {Surdej}, {Szabados},
  {Szegedi-Elek}, {Tapiador}, {Taris}, {Tauran}, {Taylor}, {Teixeira},
  {Terrett}, {Teyssand ier}, {Thuillot}, {Titarenko}, {Torra Clotet}, {Turon},
  {Ulla}, {Utrilla}, {Uzzi}, {Vaillant}, {Valentini}, {Valette}, {van Elteren},
  {Van Hemelryck}, {van Leeuwen}, {Vaschetto}, {Vecchiato}, {Veljanoski},
  {Viala}, {Vicente}, {Vogt}, {von Essen}, {Voss}, {Votruba}, {Voutsinas},
  {Walmsley}, {Weiler}, {Wertz}, {Wevers}, {Wyrzykowski}, {Yoldas},
  {{\v{Z}}erjal}, {Ziaeepour}, {Zorec}, {Zschocke}, {Zucker}, {Zurbach}, \&
  {Zwitter}}]{_katz18k}
{Gaia Collaboration}, {Katz}, D., {Antoja}, T., {et~al.} 2018{\natexlab{b}},
  \aap, 616, A11

\bibitem[{{Gardner} \& {Flynn}(2010)}]{_gardner10}
{Gardner}, E. \& {Flynn}, C. 2010, \mnras, 405, 545

\bibitem[{{Gilmore} {et~al.}(2012){Gilmore}, {Randich}, {Asplund}, {Binney},
  {Bonifacio}, {Drew}, {Feltzing}, {Ferguson}, {Jeffries}, {Micela},
  {Negueruela}, {Prusti}, {Rix}, {Vallenari}, {Alfaro}, {Allende-Prieto},
  {Babusiaux}, {Bensby}, {Blomme}, {Bragaglia}, {Flaccomio}, {Francois},
  {Irwin}, {Koposov}, {Korn}, {Lanzafame}, {Pancino}, {Paunzen},
  {Recio-Blanco}, {Sacco}, {Smiljanic}, {van Eck}, \& {Walton}}]{gilmore2012}
{Gilmore}, G., {Randich}, S., {Asplund}, M., {et~al.} 2012, The Messenger, 147,
  25

\bibitem[{{Gilmore} {et~al.}(2002){Gilmore}, {Wyse}, \& {Norris}}]{_gilmore02}
{Gilmore}, G., {Wyse}, R.~F.~G., \& {Norris}, J.~E. 2002, \apjl, 574, L39

\bibitem[{{Helmi} {et~al.}(2018){Helmi}, {Babusiaux}, {Koppelman}, {Massari},
  {Veljanoski}, \& {Brown}}]{_helmi18}
{Helmi}, A., {Babusiaux}, C., {Koppelman}, H.~H., {et~al.} 2018, \nat, 563, 85

\bibitem[{{Helmi} {et~al.}(2006){Helmi}, {Navarro}, {Nordstr{\"o}m},
  {Holmberg}, {Abadi}, \& {Steinmetz}}]{_helmi06}
{Helmi}, A., {Navarro}, J.~F., {Nordstr{\"o}m}, B., {et~al.} 2006, \mnras, 365,
  1309

\bibitem[{{Helmi} {et~al.}(2017){Helmi}, {Veljanoski}, {Breddels}, {Tian}, \&
  {Sales}}]{_helmi17}
{Helmi}, A., {Veljanoski}, J., {Breddels}, M.~A., {Tian}, H., \& {Sales}, L.~V.
  2017, \aap, 598, A58

\bibitem[{{Helmi} {et~al.}(1999){Helmi}, {White}, {de Zeeuw}, \&
  {Zhao}}]{_helmi99}
{Helmi}, A., {White}, S.~D.~M., {de Zeeuw}, P.~T., \& {Zhao}, H. 1999, \nat,
  402, 53

\bibitem[{{Holtzman} {et~al.}(2018){Holtzman}, {Hasselquist}, {Shetrone},
  {Cunha}, {Allende Prieto}, {Anguiano}, {Bizyaev}, {Bovy}, {Casey},
  {Edvardsson}, {Johnson}, {J{\"o}nsson}, {Meszaros}, {Smith}, {Sobeck},
  {Zamora}, {Chojnowski}, {Fernandez-Trincado}, {Garcia-Hernandez}, {Majewski},
  {Pinsonneault}, {Souto}, {Stringfellow}, {Tayar}, {Troup}, \&
  {Zasowski}}]{_holzman18}
{Holtzman}, J.~A., {Hasselquist}, S., {Shetrone}, M., {et~al.} 2018, \aj, 156,
  125

\bibitem[{{Hunt} {et~al.}(2018){Hunt}, {Hong}, {Bovy}, {Kawata}, \&
  {Grand}}]{_hunt18}
{Hunt}, J. A.~S., {Hong}, J., {Bovy}, J., {Kawata}, D., \& {Grand}, R. J.~J.
  2018, \mnras, 481, 3794

\bibitem[{{Johnson} \& {Soderblom}(1987)}]{_johnson87}
{Johnson}, D.~R.~H. \& {Soderblom}, D.~R. 1987, \aj, 93, 864

\bibitem[{{Katz} {et~al.}(2019){Katz}, {Sartoretti}, {Cropper}, {Panuzzo},
  {Seabroke}, {Viala}, {Benson}, {Blomme}, {Jasniewicz}, {Jean-Antoine},
  {Huckle}, {Smith}, {Baker}, {Crifo}, {Damerdji}, {David}, {Dolding},
  {Fr{\'e}mat}, {Gosset}, {Guerrier}, {Guy}, {Haigron}, {Jan{\ss}en},
  {Marchal}, {Plum}, {Soubiran}, {Th{\'e}venin}, {Ajaj}, {Allende Prieto},
  {Babusiaux}, {Boudreault}, {Chemin}, {Delle Luche}, {Fabre}, {Gueguen},
  {Hambly}, {Lasne}, {Meynadier}, {Pailler}, {Panem}, {Royer}, {Tauran},
  {Zurbach}, {Zwitter}, {Arenou}, {Bossini}, {Gerssen}, {G{\'o}mez},
  {Lemaitre}, {Leclerc}, {Morel}, {Munari}, {Turon}, {Vallenari}, \&
  {{\v{Z}}erjal}}]{_katz18}
{Katz}, D., {Sartoretti}, P., {Cropper}, M., {et~al.} 2019, \aap, 622, A205

\bibitem[{{Klement} {et~al.}(2008){Klement}, {Fuchs}, \& {Rix}}]{_klement08}
{Klement}, R., {Fuchs}, B., \& {Rix}, H.-W. 2008, \apj, 685, 261

\bibitem[{{Koppelman} {et~al.}(2018){Koppelman}, {Helmi}, \&
  {Veljanoski}}]{_koppelman18}
{Koppelman}, H., {Helmi}, A., \& {Veljanoski}, J. 2018, \apj, 860, L11

\bibitem[{{Kushniruk} {et~al.}(2017){Kushniruk}, {Schirmer}, \&
  {Bensby}}]{_kushniruk17}
{Kushniruk}, I., {Schirmer}, T., \& {Bensby}, T. 2017, \aap, 608, A73

\bibitem[{Lindegren(2018)}]{_lindegren18n}
Lindegren, L. 2018, Gaia Technical Note: GAIA-C3-TN-LU-LL-124-01

\bibitem[{{Liu} {et~al.}(2015){Liu}, {Feltzing}, \& {Ruchti}}]{_liu15}
{Liu}, C., {Feltzing}, S., \& {Ruchti}, G. 2015, \aap, 580, A111

\bibitem[{{McMillan}(2018)}]{_mcmillan18}
{McMillan}, P.~J. 2018, Research Notes of the American Astronomical Society, 2,
  51

\bibitem[{{Minchev} {et~al.}(2010){Minchev}, {Boily}, {Siebert}, \&
  {Bienayme}}]{_minchev10}
{Minchev}, I., {Boily}, C., {Siebert}, A., \& {Bienayme}, O. 2010, \mnras, 407,
  2122

\bibitem[{{Minchev} {et~al.}(2007){Minchev}, {Nordhaus}, \&
  {Quillen}}]{_minchev07}
{Minchev}, I., {Nordhaus}, J., \& {Quillen}, A.~C. 2007, \apj, 664, L31

\bibitem[{{Minchev} {et~al.}(2009){Minchev}, {Quillen}, {Williams}, {Freeman},
  {Nordhaus}, {Siebert}, \& {Bienaym{\'e}}}]{_minchev09}
{Minchev}, I., {Quillen}, A.~C., {Williams}, M., {et~al.} 2009, \mnras, 396,
  L56

\bibitem[{{Monari} {et~al.}(2013){Monari}, {Antoja}, \& {Helmi}}]{_monari13}
{Monari}, G., {Antoja}, T., \& {Helmi}, A. 2013, ArXiv e-prints
  [\eprint[arXiv]{1306.2632}]

\bibitem[{{Monari} {et~al.}(2018){Monari}, {Famaey}, {Minchev}, {Antoja},
  {Bienaym{\'e}}, {Gibson}, {Grebel}, {Kordopatis}, {McMillan}, {Navarro},
  {Parker}, {Quillen}, {Reid}, {Seabroke}, {Siebert}, {Steinmetz}, {Wyse}, \&
  {Zwitter}}]{_monari18}
{Monari}, G., {Famaey}, B., {Minchev}, I., {et~al.} 2018, Research Notes of the
  American Astronomical Society, 2, 32

\bibitem[{{Monari} {et~al.}(2019){Monari}, {Famaey}, {Siebert}, {Bienaym{\'e}},
  {Ibata}, {Wegg}, \& {Gerhard}}]{_monari19}
{Monari}, G., {Famaey}, B., {Siebert}, A., {et~al.} 2019, arXiv e-prints,
  arXiv:1908.01318

\bibitem[{{Monari} {et~al.}(2017){Monari}, {Kawata}, {Hunt}, \&
  {Famaey}}]{_monari17}
{Monari}, G., {Kawata}, D., {Hunt}, J. A.~S., \& {Famaey}, B. 2017, \mnras,
  466, L113

\bibitem[{{Navarro} {et~al.}(2004){Navarro}, {Helmi}, \&
  {Freeman}}]{_navarro04}
{Navarro}, J.~F., {Helmi}, A., \& {Freeman}, K.~C. 2004, \apjl, 601, L43

\bibitem[{{P{\'e}rez-Villegas} {et~al.}(2017){P{\'e}rez-Villegas}, {Portail},
  {Wegg}, \& {Gerhard}}]{_perez17}
{P{\'e}rez-Villegas}, A., {Portail}, M., {Wegg}, C., \& {Gerhard}, O. 2017,
  \apjl, 840, L2

\bibitem[{{Quillen} {et~al.}(2018){Quillen}, {Carrillo}, {Anders}, {McMillan},
  {Hilmi}, {Monari}, {Minchev}, {Chiappini}, {Khalatyan}, \&
  {Steinmetz}}]{_quillen18}
{Quillen}, A.~C., {Carrillo}, I., {Anders}, F., {et~al.} 2018, \mnras, 480,
  3132

\bibitem[{{Ramos} {et~al.}(2018){Ramos}, {Antoja}, \& {Figueras}}]{_ramos18}
{Ramos}, P., {Antoja}, T., \& {Figueras}, F. 2018, \aap, 619, A72

\bibitem[{{Ramya} {et~al.}(2012){Ramya}, {Reddy}, \& {Lambert}}]{_ramya12}
{Ramya}, P., {Reddy}, B.~E., \& {Lambert}, D.~L. 2012, \mnras, 425, 3188

\bibitem[{{Ramya} {et~al.}(2016){Ramya}, {Reddy}, {Lambert}, \&
  {Musthafa}}]{_ramya16}
{Ramya}, P., {Reddy}, B.~E., {Lambert}, D.~L., \& {Musthafa}, M.~M. 2016,
  \mnras, 460, 1356

\bibitem[{{Sch{\"o}nrich} {et~al.}(2010){Sch{\"o}nrich}, {Binney}, \&
  {Dehnen}}]{_schonrich10}
{Sch{\"o}nrich}, R., {Binney}, J., \& {Dehnen}, W. 2010, \mnras, 403, 1829

\bibitem[{{Sellwood}(2010)}]{_sellwood10}
{Sellwood}, J.~A. 2010, \mnras, 409, 145

\bibitem[{{Sellwood} {et~al.}(2019){Sellwood}, {Trick}, {Carlberg}, {Coronado},
  \& {Rix}}]{_sellwood18}
{Sellwood}, J.~A., {Trick}, W.~H., {Carlberg}, R.~G., {Coronado}, J., \& {Rix},
  H.-W. 2019, \mnras, 484, 3154

\bibitem[{{Skuljan} {et~al.}(1999){Skuljan}, {Hearnshaw}, \&
  {Cottrell}}]{_skuljan99}
{Skuljan}, J., {Hearnshaw}, J.~B., \& {Cottrell}, P.~L. 1999, \mnras, 308, 731

\bibitem[{Starck \& Murtagh(2002)}]{starck_astronomical_2002}
Starck, J.-L. \& Murtagh, F. 2002, Astronomical image and data analysis
  (Berlin; New York: Springer), oCLC: 679368657

\bibitem[{{Starck} {et~al.}(1998){Starck}, {Murtagh}, \& {Bijaoui}}]{_starck98}
{Starck}, J.-L., {Murtagh}, F.~D., \& {Bijaoui}, A. 1998, {Image Processing and
  Data Analysis}, 297

\bibitem[{{Tian} {et~al.}(2018){Tian}, {Liu}, {Wu}, {Xiang}, \&
  {Zhang}}]{_tian18}
{Tian}, H.-J., {Liu}, C., {Wu}, Y., {Xiang}, M.-S., \& {Zhang}, Y. 2018, \apj,
  865, L19

\bibitem[{{Trick} {et~al.}(2019){Trick}, {Coronado}, \& {Rix}}]{_trick18}
{Trick}, W.~H., {Coronado}, J., \& {Rix}, H.-W. 2019, \mnras, 484, 3291

\bibitem[{{{\v Z}enovien{\.e}} {et~al.}(2014){{\v Z}enovien{\.e}}, {Tautvai{\v
  s}ien{\.e}}, {Nordstr{\"o}m}, \& {Stonkut{\.e}}}]{_zenoviene14}
{{\v Z}enovien{\.e}}, R., {Tautvai{\v s}ien{\.e}}, G., {Nordstr{\"o}m}, B., \&
  {Stonkut{\.e}}, E. 2014, \aap, 563, A53

\bibitem[{van~der Walt {et~al.}(2014)van~der Walt, {S}ch\"onberger,
  {Nunez-Iglesias}, {B}oulogne, {W}arner, {Y}ager, {G}ouillart, {Y}u, \& the
  scikit-image contributors}]{_sc_image}
van~der Walt, S., {S}ch\"onberger, J.~L., {Nunez-Iglesias}, J., {et~al.} 2014,
  PeerJ, 2, e453

\bibitem[{{Wegg} {et~al.}(2015){Wegg}, {Gerhard}, \& {Portail}}]{_wegg15}
{Wegg}, C., {Gerhard}, O., \& {Portail}, M. 2015, \mnras, 450, 4050

\bibitem[{{Williams} {et~al.}(2009){Williams}, {Freeman}, {Helmi}, \& {RAVE
  Collaboration}}]{_williams09}
{Williams}, M.~E.~K., {Freeman}, K.~C., {Helmi}, A., \& {RAVE Collaboration}.
  2009, in IAU Symposium, Vol. 254, The Galaxy Disk in Cosmological Context,
  ed. J.~{Andersen}, {Nordstr{\"o}ara}, B.~{m}, \& J.~{Bland-Hawthorn},
  139--144

\bibitem[{{Wyse} {et~al.}(2006){Wyse}, {Gilmore}, {Norris}, {Wilkinson},
  {Kleyna}, {Koch}, {Evans}, \& {Grebel}}]{_wyse06}
{Wyse}, R.~F.~G., {Gilmore}, G., {Norris}, J.~E., {et~al.} 2006, \apjl, 639,
  L13

\bibitem[{{Zhao} {et~al.}(2014){Zhao}, {Zhao}, {Chen}, {Oswalt}, {Tan}, \&
  {Zhang}}]{_zhao14}
{Zhao}, J.~K., {Zhao}, G., {Chen}, Y.~Q., {et~al.} 2014, \apj, 787, 31

\end{thebibliography}

\begin{appendix}

\section{Tables}

%--------------------------------------------------------------------
\begin{table}[h]
\begin{center}
\caption{The number of stars, median distance and median distance uncertainty for the stars located in the 65 regions (as defined in Fig.~\ref{_fig_regions}).
\label{_tab_info}}
\resizebox{.25\textwidth}{!}{%
\begin{tabular}{crrrr} 
\hline
\hline
\noalign{\smallskip}
N & Region  & N stars &  $D_{median}$ & $\sigma_{D_{median}}$\\
  &         &         &  [pc]         & [pc]\\
\noalign{\smallskip}
\hline 
\noalign{\smallskip}
1	&00	     &666505 &212	&2\\
2	&01	     &48759	 &851	&45\\
3	&02	     &13647	 &1689	&181\\
4	&03	     &6332	 &2483	&386\\
5	&04	     &3041	 &3285	&677\\
6	&05	     &1621	 &4083	&932\\
7	&11	     &70747	 &861	&38\\
8	&12	     &22070	 &1778	&164\\
9	&13	     &8991	 &2636	&375\\
10	&14	     &3521	 &3502	&669\\
11	&15	     &1685	 &4338	&918\\
12	&21	     &81838	 &777	&38\\
13	&22	     &24823	 &1602	&171\\
14	&23	     &10146	 &2392	&410\\
15	&24	     &3307	 &3204	&803\\
16	&25	     &1386	 &4011	&1171\\
17	&31	     &64875	 &868	&38\\
18	&32	     &19806	 &1778	&164\\
19	&33	     &10747	 &2624	&365\\
20	&34	     &4307	 &3489	&655\\
21	&35	     &1873	 &4329	&826\\
22	&22\_11	 &18596	 &1798	&209\\
23	&22\_12	 &13098	 &2258	&306\\
24	&22\_13	 &8223	 &2853	&495\\
25	&22\_14	 &4132	 &3522	&724\\
26	&22\_15	 &2197	 &4216	&898\\
27	&22\_31	 &27838	 &1780	&194\\
28	&22\_32	 &18135	 &2242	&284\\
29	&22\_33	 &9545	 &2839	&458\\
30	&22\_34	 &5646	 &3507	&690\\
31	&22\_35	 &2832	 &4200	&922\\
32	&02\_11	 &11447	 &1942	&215\\
33	&02\_12	 &8189	 &2546	&349\\
34	&02\_13	 &3219	 &3321	&608\\
35	&02\_14	 &2267	 &4145	&852\\
36	&02\_15	 &308    &4897	&954\\
37	&02\_31	 &12179	 &1939	&232\\
38	&02\_32	 &9594	 &2544	&372\\
39	&02\_33	 &4786	 &3306	&602\\
40	&02\_34	 &1887	 &4147	&868\\
41	&02\_35	 &263	 &4881	&903\\
42	&12\_01	 &13658	 &2042	&217\\
43	&12\_03	 &4287	 &3177	&589\\
44	&12\_04	 &1942	 &3893	&847\\
45	&12\_05	 &1125	 &4605	&1024\\
46	&12\_21	 &21401	 &1864	&189\\
47	&12\_23	 &6129	 &2823	&538\\
48	&12\_24	 &2440	 &3500	&896\\
49	&12\_25	 &1195	 &4206	&1173\\
50	&32\_01	 &13717	 &2044	&219\\
51	&32\_03	 &3988	 &3176	&609\\
52	&32\_04	 &2172	 &3871	&829\\
53	&32\_05	 &1290	 &4592	&1081\\
54	&32\_21	 &22441	 &1856	&186\\
55	&32\_23	 &3331	 &3496	&855\\
56	&32\_24	 &11387	 &2806	&495\\
57	&32\_25	 &1447	 &4209	&1166\\
58	&01a	 &211846 &453	&12\\
59	&11a	 &271059 &444	&10\\
60	&21a	 &317397 &387	&9\\
61	&31a	 &256489 &448	&10\\
62	&01a\_11a &127033 &630	&22\\
63	&11a\_21a &175045 &568	&18\\
64	&21a\_31a &184213 &574	&18\\
65	&31a\_01a &127733 &636	&22\\
\hline
\end{tabular}
}
\end{center}
\end{table}
%--------------------------------------------------------------------

\begin{table*}[h]
\begin{center}
\caption{Kinematic structures found in region 00 in the $U-V$ (Plane 1), $V-\sqrt{U^2+2V^2}$ (Plane 2), $L_z-\sqrt{L_x^2+L_y^2}$ (Plane 3) and $L_z-\sqrt{J_r}$ (Plane 4) planes at scale $J=2$. First column is a line number in the table; the second one denotes the plane; names of the groups as in Figure \ref{_fig_00_2} are given in column 3 and names of the groups as in the literature are provided in column 4; number of stars in each group is given in column 5; median U, V velocities, median angular momentum $L_z$ and median value of square root of radial action per group is given in columns 6-9; columns 8-13 are standard deviations of the same quantities as in columns 6-9.
\label{_tab_groups_2}
}
\resizebox{0.7\textwidth}{!}{%
\begin{tabular}{rrrrrrrrrrrrr} 
\hline
\hline 
\noalign{\smallskip}
N &Plane &Group &Name &N stars &$U$ &$V$ &$L_z$ &$\sqrt{J_r}$ &$\sigma_U$ &$\sigma_V$ &$\sigma_{L_z}$ & $\sigma_{\sqrt{J_r}}$ \\
\noalign{\smallskip}
& & & & & [$\kms$] & [$\kms$] & [kpc $\kms$] & [kpc $\kms$] & [$\kms$] & [$\kms$] &  [kpc $\kms$] &  [kpc $\kms$] \\ 
\noalign{\smallskip}
\hline 
\noalign{\smallskip}
1    &  1  &  g1   &  A1/A2          & 173     &     39  &     37  &  2158  &    12  &   2  &  2 &  30 & 0.6 \\
2    &  1  &  g2   & $\gamma$Leo     & 2643    &     34  &      8  &  1926  &     7  &   2  &  2 &  28 & 0.5 \\
3    &  1  &  g3   &  Sirius         & 6966    &  $-$13  &      6  &  1918  &   3.6  &   2  &  2 &  31 & 0.5 \\
4    &  1  &  g4   &  Sirius         & 1659    &  $-$53  &      4  &  1900  &   6.3  &   2  &  2 &  30 & 0.5 \\
5    &  1  &  g5   & $\gamma$Leo     & 2277    &     47  &      3  &  1894  &   7.9  &   2  &  2 &  31 & 0.5 \\
6    &  1  &  g6   &  Sirius         & 10303   &     12  &      2  &  1888  &   4.1  &   2  &  2 &  30 & 0.5 \\
7    &  1  &  g7   &  Sirius         & 1416    &  $-$61  &      2  &  1887  &   6.9  &   2  &  2 &  30 & 0.5 \\
8    &  1  &  g8   & $\gamma$Leo     & 1080    &     65  &      1  &  1872  &   9.9  &   2  &  2 &  27 & 0.5 \\
9    &  1  &  g9   &  Sirius         & 871     &  $-$79  &      0  &  1858  &   8.8  &   2  &  2 &  31 & 0.5 \\
10   &  1  &  g10  & $\gamma$Leo     & 785     &     73  &   $-$2  &  1844  &  10.6  &   2  &  2 &  28 & 0.5 \\
11   &  1  &  g11  & $\gamma$Leo     & 1881    &     57  &   $-$5  &  1819  &   8.5  &   2  &  2 &  32 & 0.5 \\
12   &  1  &  g12  & Coma Berenices  & 12046   &  $-$15  &   $-$7  &  1808  &   1.3  &   2  &  2 &  29 & 0.5 \\
13   &  1  &  g13  & Bobylev16       & 180     & $-$118  &  $-$10  &  1776  &  13.1  &   2  &  2 &  31 & 0.5 \\
14   &  1  &  g14  & Coma Berenices  & 8439    &      8  &  $-$14  &  1745  &   2.4  &   2  &  2 &  30 & 0.5 \\
15   &  1  &  g15  & Pleiades/Hyades & 960     &  $-$81  &  $-$14  &  1744  &   8.3  &   2  &  2 &  29 & 0.5 \\
16   &  1  &  g16  & Pleiades/Hyades & 14380   &  $-$33  &  $-$16  &  1733  &   2.7  &   2  &  2 &  28 & 0.5 \\
17   &  1  &  g17  & Pleiades/Hyades & 11749   &  $-$42  &  $-$19  &  1710  &   3.8  &   2  &  2 &  25 & 0.4 \\
18   &  1  &  g18  & Antoja12(12)    & 302     &     98  &  $-$21  &  1695  &    13  &   2  &  2 &  29 & 0.5 \\
19   &  1  &  g19  & Wolf 630        & 6104    &     21  &  $-$22  &  1678  &   4.1  &   2  &  2 &  29 & 0.5 \\
20   &  1  &  g20  & Dehnen98        & 4129    &     41  &  $-$25  &  1664  &   6.3  &   2  &  2 &  26 & 0.4 \\
21   &  1  &  g21  & Pleiades/Hyades & 13084   &  $-$10  &  $-$25  &  1657  &   2.2  &   2  &  2 &  26 & 0.4 \\
22   &  1  &  g22  & Hercules        & 2604    &  $-$58  &  $-$37  &  1562  &   6.5  &   2  &  2 &  24 & 0.4 \\
23   &  1  &  g23  & Pleiades/Hyades & 3428    &     25  &  $-$37  &  1561  &   5.7  &   2  &  2 &  27 & 0.4 \\
24   &  1  &  g24  & Hercules        & 1380    &  $-$82  &  $-$47  &  1473  &   9.5  &   2  &  2 &  27 & 0.4 \\
25   &  1  &  g25  & Hercules        & 3529    &  $-$15  &  $-$48  &  1468  &   5.8  &   2  &  2 &  25 & 0.4 \\
26   &  1  &  g26  & Hercules        & 4626    &  $-$37  &  $-$49  &  1463  &   6.6  &   2  &  2 &  22 & 0.4 \\
27   &  1  &  g27  & Hercules        & 2816    &      0  &  $-$50  &  1455  &   6.2  &   2  &  2 &  24 & 0.5 \\
28   &  1  &  g28  & Hercules        & 1593    &  $-$70  &  $-$51  &  1450  &   8.9  &   2  &  2 &  25 & 0.5 \\
29   &  1  &  g29  & Antoja12(15)    & 633     &     65  &  $-$52  &  1441  &  10.2  &   2  &  2 &  27 & 0.4 \\
30   &  1  &  g30  & HR1614          & 1286    &      9  &  $-$64  &  1342  &   8.2  &   2  &  2 &  25 & 0.5 \\
31   &  1  &  g31  & HR1614          & 1989    &  $-$16  &  $-$64  &  1343  &   8.1  &   2  &  2 &  25 & 0.4 \\
32   &  1  &  g32  & HR1614          & 1619    &  $-$26  &  $-$67  &  1321  &   8.5  &   2  &  2 &  24 & 0.5 \\
33   &  1  &  g33  & HR1614          & 1345    &  $-$36  &  $-$68  &  1313  &   8.9  &   2  &  2 &  25 & 0.5 \\
34   &  1  &  g34  & $\epsilon$Ind   & 498     &  $-$83  &  $-$75  &  1252  &    12  &   2  &  2 &  25 & 0.4 \\
35   &  1  &  g35  & $\epsilon$Ind   & 626     &  $-$73  &  $-$76  &  1247  &  11.5  &   2  &  2 &  24 & 0.5 \\
36   &  1  &  g36  & Arcturus        & 358     &  $-$13  &  $-$92  &  1118  &  11.6  &   2  &  2 &  23 & 0.6 \\
\noalign{\smallskip}
\hline 
\noalign{\smallskip}
37   &  2  &  g1   &  Arcturus        & 1840    &  $-$11  &  $-$91  &  1125  &  11.7  &  25  &  2 &  22 & 0.6 \\
38   &  2  &  g2   &  AF06            & 2013    &  $-$12  &  $-$86  &  1168  &  11.1  &  26  &  2 &  24 & 0.7 \\
39   &  2  &  g3   &  Hercules        & 1483    &  $-$85  &  $-$54  &  1431  &  10.4  &  57  &  2 &  26 & 0.8 \\
40   &  2  &  g4   &  HR1614          & 7629    &  $-$10  &  $-$65  &  1337  &   8.3  &  20  &  1 &  23 & 0.5 \\
41   &  2  &  g5   &  HR1614          & 8490    &   $-$9  &  $-$61  &  1366  &   7.8  &  19  &  1 &  24 & 0.5 \\
42   &  2  &  g6   &  Hercules        & 14860   &  $-$10  &  $-$46  &  1482  &   5.8  &  19  &  2 &  27 & 0.5 \\
43   &  2  &  g7   &  Sirius          & 2459    &  $-$60  &      1  &  1876  &     8  &  64  &  2 &  29 & 1.4 \\
44   &  2  &  g8   &  Hercules        & 15404   &   $-$5  &  $-$40  &  1539  &   4.9  &  21  &  2 &  25 & 0.7 \\
45   &  2  &  g9   &  A1/A2           & 1129    &   $-$5  &     38  &  2173  &  10.6  &  17  &  2 &  30 & 0.6 \\
46   &  2  &  g10  &  Pleiades/Hyades & 16788   &  $-$41  &  $-$19  &  1707  &   3.9  &  34  &  2 &  26 & 1.1 \\
47   &  2  &  g11  &  $\gamma$Leo     & 4112    &     45  &      3  &  1893  &   7.4  &  49  &  2 &  31 & 1.3 \\
48   &  2  &  g12  &  Pleiades/Hyades & 38181   &   $-$7  &  $-$25  &  1660  &   2.5  &  14  &  2 &  27 & 0.7 \\
49   &  2  &  g13  &  $\gamma$Leo     & 5460    &     30  &      7  &  1921  &   6.3  &  34  &  2 &  29 & 1.1 \\
50   &  2  &  g14  &  A1/A2           & 3361    &   $-$5  &     22  &  2044  &     7  &  13  &  1 &  32 & 0.6 \\
51   &  2  &  g15  &  Coma Berenices  & 26737   &   $-$2  &  $-$12  &  1766  &   1.2  &  10  &  2 &  29 &   1 \\
52   &  2  &  g16  &  Sirius          & 19311   &      0  &      8  &  1926  &   4.3  &   9  &  2 &  27 & 0.5 \\
53   &  2  &  g17  &  Sirius          & 19358   &      0  &      0  &  1872  &   2.9  &   5  &  2 &  31 & 0.5 \\
\noalign{\smallskip}
\hline 
\noalign{\smallskip}
54   &  3  &  g1   &  Pleiades/Hyades & 81      &  $-$13  &  $-$23  &  1661  &   3.2  &  40  &  2 &   2 & 2.8 \\
55   &  3  &  g2   &  Hercules        & 47      &  $-$12  &  $-$52  &  1439  &   6.6  &  47  &  2 &   2 & 2.3 \\
56   &  3  &  g3   &  Hercules        & 66      &  $-$18  &  $-$41  &  1529  &   5.7  &  47  &  2 &   2 & 2.3 \\
57   &  3  &  g4   &  Sirius          & 86      &   $-$6  &      4  &  1895  &   4.4  &  33  &  3 &   2 & 1.9 \\
58   &  3  &  g5   &  HR1614          & 41      &  $-$17  &  $-$67  &  1325  &   8.7  &  45  &  2 &   1 & 1.9 \\
59   &  3  &  g6   &  Pleiades/Hyades & 171     &  $-$12  &  $-$20  &  1698  &   3.5  &  34  &  2 &   2 & 2.1 \\
60   &  3  &  g7   &  Hercules        & 120     &  $-$27  &  $-$51  &  1452  &   6.7  &  34  &  2 &   2 & 1.3 \\
61   &  3  &  g8   &  Sirius          & 165     &   $-$3  &      2  &  1890  &   4.2  &  35  &  3 &   2 & 2.2 \\
62   &  3  &  g9   &  Pleiades/Hyades & 264     &  $-$19  &  $-$20  &  1698  &   2.8  &  36  &  2 &   2 & 2.5 \\
63   &  3  &  g10  &  AF06            & 17      &  $-$22  &  $-$94  &  1109  &  12.4  &  45  &  2 &   2 & 1.5 \\
64   &  3  &  g11  &  A1/A2           & 41      &  $-$10  &     24  &  2064  &   8.1  &  38  &  3 &   2 &   2 \\
65   &  3  &  g12  &  Sirius          & 263     &   $-$2  &      7  &  1926  &   4.6  &  29  &  3 &   2 & 1.7 \\
66   &  3  &  g13  &  A1/A2           & 55      &  $-$13  &     26  &  2063  &   7.9  &  30  &  3 &   2 & 1.5 \\
67   &  3  &  g14  &  HR1614          & 119     &  $-$11  &  $-$62  &  1350  &   8.4  &  38  &  2 &   2 & 1.3 \\
68   &  3  &  g15  &  Hercules        & 339     &  $-$20  &  $-$39  &  1547  &   5.4  &  39  &  2 &   2 & 1.7 \\
69   &  3  &  g16  &  Sirius          & 456     &      0  &      6  &  1914  &   4.3  &  28  &  3 &   2 & 1.7 \\
70   &  3  &  g17  &  A1/A2           & 54      &  $-$16  &     24  &  2057  &   7.5  &  28  &  3 &   2 & 1.5 \\
71   &  3  &  g18  &  HR1614          & 108     &  $-$17  &  $-$62  &  1349  &   8.4  &  39  &  2 &   2 & 1.6 \\
72   &  3  &  g19  &  Hercules        & 379     &  $-$32  &  $-$48  &  1470  &   6.5  &  31  &  2 &   2 & 1.4 \\
73   &  3  &  g20  &  Sirius          & 427     &      3  &      5  &  1919  &   4.2  &  23  &  3 &   2 & 1.5 \\
74   &  3  &  g21  &  Arcturus        & 8       &  $-$44  & $-$111  &   967  &  15.4  &  60  &  1 &   3 & 0.6 \\
75   &  3  &  g22  &  Coma Berenices  & 644     &   $-$9  &   $-$7  &  1806  &     2  &  27  &  2 &   2 & 2.3 \\
76   &  3  &  g23  &  Pleiades/Hyades & 865     &  $-$21  &  $-$19  &  1703  &   2.5  &  25  &  2 &   2 & 1.7 \\
77   &  3  &  g24  &  HR1614          & 90      &  $-$18  &  $-$61  &  1362  &   8.3  &  39  &  2 &   2 & 1.5 \\
\noalign{\smallskip}
\hline 
\noalign{\smallskip}
78   &  4  &  g1   &  KFR08           & 107     &  $-$13  & $-$160  &   575  &  19.5  &  49  &  2 &  23 & 0.2 \\
79   &  4  &  g2   &  Arcturus        & 303     &   $-$6  & $-$127  &   841  &  16.1  &  40  &  3 &  23 & 0.2 \\
80   &  4  &  g3   &  Antoja12(12)    & 122     &    103  &  $-$19  &  1707  &  15.1  & 125  &  3 &  22 & 0.3 \\
81   &  4  &  g4   &  A1/A2           & 132     &     31  &     43  &  2223  &  13.3  &  42  &  4 &  22 & 0.2 \\
82   &  4  &  g5   &  Hercules        & 326     & $-$111  &  $-$50  &  1460  &    13  & 103  &  3 &  22 & 0.2 \\
83   &  4  &  g6   &  A1/A2           & 255     &     27  &     30  &  2110  &  12.8  &  69  &  4 &  22 & 0.2 \\
84   &  4  &  g7   &  Antoja12(12)    & 579     &     92  &  $-$19  &  1710  &  12.9  & 105  &  4 &  23 & 0.3 \\
85   &  4  &  g8   &  $\gamma$Leo     & 405     &     84  &   $-$7  &  1805  &  12.7  & 102  &  4 &  22 & 0.2 \\
86   &  4  &  g9   &  AF06            & 1836    &  $-$18  &  $-$92  &  1113  &  11.9  &  27  &  2 &  21 & 0.2 \\
87   &  4  &  g10  &  $\gamma$Leo     & 803     &     77  &   $-$4  &  1828  &  11.4  &  84  &  4 &  23 & 0.3 \\
88   &  4  &  g11  &  A1/A2           & 1026    &   $-$7  &     39  &  2177  &  10.9  &  26  &  3 &  20 & 0.3 \\
89   &  4  &  g12  &  $\gamma$Leo     & 1765    &     64  &      0  &  1867  &    10  &  70  &  4 &  23 & 0.3 \\
90   &  4  &  g13  &  Hercules        & 3649    &  $-$72  &  $-$50  &  1460  &   9.4  &  64  &  4 &  23 & 0.3 \\
91   &  4  &  g14  &  $\gamma$Leo     & 3393    &     48  &      1  &  1878  &   8.3  &  57  &  4 &  24 & 0.2 \\
92   &  4  &  g15  &  HR1614          & 8691    &  $-$15  &  $-$64  &  1342  &   8.3  &  22  &  2 &  21 & 0.3 \\
93   &  4  &  g16  &  A1/A2           & 2922    &  $-$13  &     25  &  2064  &   7.7  &  20  &  3 &  20 & 0.3 \\
94   &  4  &  g17  &  $\gamma$Leo     & 5271    &     33  &      6  &  1910  &     7  &  44  &  4 &  23 & 0.3 \\
95   &  4  &  g18  &  Hercules        & 16701   &  $-$21  &  $-$49  &  1463  &   6.2  &  19  &  2 &  20 & 0.3 \\
96   &  4  &  g19  &  Dehnen98        & 7245    &     36  &  $-$23  &  1679  &   6.1  &  47  &  4 &  22 & 0.3 \\
97   &  4  &  g20  &  Pleiades/Hyades & 15246   &  $-$15  &  $-$38  &  1551  &   4.6  &  19  &  3 &  22 & 0.3 \\
98   &  4  &  g21  &  Sirius          & 22970   &      0  &      6  &  1916  &   4.1  &  15  &  3 &  22 & 0.3 \\
99   &  4  &  g22  &  Pleiades/Hyades & 21418   &  $-$39  &  $-$20  &  1699  &   3.9  &  29  &  4 &  22 & 0.3 \\
100  &  4  &  g23  &  Pleiades/Hyades & 32987   &  $-$13  &  $-$25  &  1664  &   2.4  &  11  &  3 &  22 & 0.2 \\
101  &  4  &  g24  &  Coma Berenices  & 21624   &  $-$11  &  $-$11  &  1769  &   0.9  &   6  &  3 &  23 & 0.2 \\
\noalign{\smallskip}
\hline 
\end{tabular}   
}
\end{center}
\end{table*}
%------------------------------------------------------------

\begin{table*}[h]
\begin{center}
\caption{Kinematic structures found in region 00 in the $U-V$ (Plane 1), $V-\sqrt{U^2+2V^2}$ (Plane 2), $L_z-\sqrt{L_x^2+L_y^2}$ (Plane 3) and $L_z-\sqrt{J_r}$ (Plane 4) planes at scale $J=3$. First column is a line number in the table; the second one denotes the plane; names of the groups as in Figure \ref{_fig_00_3} are given in column 3 and names of the groups as in the literature are provided in column 4; number of stars in each group is given in column 5; median U, V velocities, median angular momentum $L_z$ and median value of square root of radial action per group is given in columns 6-9; columns 8-13 are standard deviations of the same quantities as in columns 6-9.
\label{_tab_groups_3}
}
\resizebox{\textwidth}{!}{%
\begin{tabular}{rrrrrrrrrrrrr} 
\hline
\hline 
\noalign{\smallskip}
N &Plane &Group & Name &N stars &$U$ &$V$ &$L_z$ &$\sqrt{J_r}$ &$\sigma_U$ &$\sigma_V$ &$\sigma_{L_z}$ & $\sigma_{\sqrt{J_r}}$ \\
\noalign{\smallskip}
& & & & & [$\kms$] & [$\kms$] & [kpc $\kms$] & [kpc $\kms$] & [$\kms$] & [$\kms$] &  [kpc $\kms$] &  [kpc $\kms$] \\ 
\noalign{\smallskip}
\hline 
\noalign{\smallskip}
1	 & 1	& g1	& A1/A2           &44       &$-$122	  &17	    &2003		&15.9	&2	    &2	&34	&0.5\\
2	 & 1	& g2	& Sirius          &10806	&     9	  &2	    &1890	    & 3.9	&2	    &2	&30	&0.5\\
3	 & 1	& g3	& $\gamma$Leo     &2325	    &    49	  &0	    &1872		& 7.9	&2	    &2	&32	&0.5\\
4	 & 1	& g4	& Bobylev16       &824	    & $-$83	  &$-$3	    &1840	 	& 9.1	&2	    &2	&30	&0.5\\
5	 & 1	& g5	& Coma Berenices  &12027	& $-$11	  &$-$6	    &1814	    & 1.3	&2	    &2	&29	&0.5\\
6	 & 1	& g6	& Bobylev16       &194	    &$-$118	  &$-$13	&1750		&12.9	&2	    &2	&29	&0.4\\
7	 & 1	& g7	& Pleiades/Hyades &14437	& $-$33	  &$-$17	&1726	    & 2.7	&2	    &2	&28	&0.5\\
8	 & 1	& g8	& Antoja12(12)    &332	    &    96	  &$-$22	&1686		&12.8	&2	    &2	&30	&0.5\\
9	 & 1	& g9	& Pleiades/Hyades &14069	& $-$14	  &$-$23	&1672	    &   2	&2	    &2	&28	&0.4\\
10	 & 1	& g10	& Wolf 630        &5913   	&    21	  &$-$24	&1665	    & 4.2	&2	    &2	&28	&0.5\\
11	 & 1	& g11	& Dehnen98        &4086	    &    41	  &$-$25	&1659		& 6.4	&2	    &2	&26	&0.5\\
12	 & 1	& g12	& Hercules        &4670	    & $-$37	  &$-$49	&1463	    & 6.5	&2	    &2	&22	&0.4\\
13	 & 1	& g13	& Hercules        &1394	    & $-$81	  &$-$49	&1464		& 9.6	&2	    &2	&25	&0.4\\
14	 & 1	& g14	& Hercules        &2400	    &     1	  &$-$53	&1438		& 6.5	&2	    &2	&26	&0.5\\
15	 & 1	& g15	& HR1614          &1769	    & $-$13	  &$-$66	&1331		& 8.2	&2	    &2	&25	&0.5\\
16	 & 1	& g16   & $\epsilon$Ind	  &544	    & $-$76	  &$-$77	&1239		&11.7	&2	    &2	&23	&0.5\\
\noalign{\smallskip}
\hline 
\noalign{\smallskip}
17	 & 2	& g1	& Arcturus        &1837	    & $-$13	  &$-$92	&1113		&11.9	&28	    &2	&22	&0.7\\
18	 & 2	& g2	& HR1614          &7724	    & $-$14	  &$-$66	&1330		& 8.5	&24	    &2	&24	&0.5\\
19	 & 2	& g3	& Hercules        &15998	& $-$20	  &$-$49	&1463		& 6.3	&24	    &2	&25	&0.5\\
20	 & 2	& g4	& Hercules        &15702	& $-$12	  &$-$42	&1524		& 5.4	&25	    &2	&29	&0.7\\
21	 & 2	& g5	& A1/A2           &1270	    &  $-$6	  &37	    &2167		&10.5	&21	    &2	&32	&0.6\\
22	 & 2	& g6	& A1/A2           &1835	    &  $-$7	  &30	    &2108		& 8.8	&17	    &2	&36	&0.6\\
23	 & 2	& g7	& Pleiades/Hyades &40542	& $-$11	  &$-$21	&1696	    &   2	&17	    &2	&29	&0.9\\
24	 & 2	& g8	& Sirius          &21306	&     6	  &0	    &1865	   	&   3	&10	    &2	&33	&0.8\\
\noalign{\smallskip}
\hline 
\noalign{\smallskip}
25	 & 3	& g1	&    HR1614           &35	    & $-$16	  &$-$68	&1315		& 8.7	&40	    &2	&2	&1.7\\
26	 & 3	& g2	&    Pleiades/Hyades  &190  	& $-$11	  &$-$21	&1692		& 3.3	&39	    &2	&2	&2.6\\
27	 & 3	& g3	&    Hercules         &125  	& $-$26	  &$-$50	&1457		& 6.6	&42	    &2	&2	&1.8\\
28	 & 3	& g4	&    Sirius           &165 	    &  $-$3	  &2	    &1890		& 4.2	&35	    &3	&2	&2.2\\
29	 & 3	& g5	&    HR1614           &85       & $-$33	  &$-$68	&1308	    & 9.2	&37	    &2	&2	&1.5\\
30	 & 3	& g6	&    Sirius           &522	    &     6	  &4	    &1901	    & 4.1	&27	    &3	&2	&1.7\\
31	 & 3	& g7	&    Hercules         &383    	& $-$31	  &$-$49	&1464	    & 6.6	&33	    &2	&2	&1.4\\
32	 & 3	& g8	&    Pleiades/Hyades  &939	    & $-$20	  &$-$20	&1697	    & 2.8	&28	    &2	&2	&1.8\\
33	 & 3	& g9	&    Arcturus         &8	    & $-$49	  &$-$106	&1000	    &14.2	&38	    &1	&2	&0.8\\
\noalign{\smallskip}
\hline 
\noalign{\smallskip}
34	 & 4	& g1	&    KFR08            &94   	& $-$14	  &$-$159	&586	&19.3	&51	    &2	&21	&0.2\\
35	 & 4	& g2	&    Antoja12(12)     &585	    &    91	  &$-$18	&1718	&12.8	&104	&4	&23	&0.3\\
36	 & 4	& g3	&    A1/A2            &1081	    &  $-$6	  &38	    &2172	&10.8	&27	    &3	&22	&0.3\\
37	 & 4	& g4	&    Hercules         &2939	    & $-$77	  &$-$49	&1463	&9.7	&70	    &4	&23	&0.2\\
38	 & 4	& g5	&    $\gamma$Leo      &3455	    &    49	  &0	    &1869	&8.3	&58	    &4	&24	&0.2\\
39	 & 4	& g6	&    A1/A2            &3076	    & $-$14	  &24	    &2058	&7.6	&21	    &3	&21	&0.3\\
40	 & 4	& g7	&    $\gamma$Leo      &4959     &    35	  &6	    &1910	&7.2	&45	    &4	&24	&0.2\\
41	 & 4	& g8	&    Hercules         &17305	& $-$24	  &$-$49	&1466	&6.3	&21	    &3	&21	&0.3\\
42	 & 4	& g9	&    Sirius           &23880	&     0	  &5	    &1909	&  4	&15	    &3	&22	&0.3\\
43	 & 4	& g10   &    Pleiades/Hyades  &35424	& $-$17	  &$-$22	&1687	    &2.3	&13	    &3	&24	&0.2\\
\noalign{\smallskip}
\hline 
\end{tabular}   
}
\end{center}
\end{table*}
%------------------------------------------------------------

\end{appendix}

\end{document}